\documentclass[pdflatex,sn-basic]{sn-jnl}

\usepackage{graphicx}%
\usepackage{multirow}%
\usepackage{amsmath,amssymb,amsfonts}%
\usepackage{amsthm}%
\usepackage{mathrsfs}%
\usepackage[title]{appendix}%
\usepackage{xcolor}%
\usepackage{textcomp}%
\usepackage{manyfoot}%
\usepackage{booktabs}%
\usepackage{array}
\usepackage{algorithm}%
\usepackage{algorithmicx}%
\usepackage{algpseudocode}%
\usepackage{listings}%
\usepackage{orcidlink}

\definecolor{bla}{rgb}{0,0,0} 
\definecolor{red}{rgb}{1,0,0} 
\definecolor{gre}{rgb}{0,0.6,0} 
\definecolor{blu}{rgb}{0,0,1} 
\definecolor{ora}{rgb}{1,0.65,0} 
\definecolor{yel}{rgb}{0.7,0.7,0} 
\definecolor{pur}{rgb}{0.5,0,0.5} 
\definecolor{cya}{rgb}{0,1,1} 

\newcommand\bla{\color{bla}}

\newcommand\gre{\color{gre}}

\newcommand{\bs}[1]{\textcolor{bla}{#1}}

\newcommand{\mwb}[1]{\textcolor{bla}{#1}}

\newcommand{\coa}[1]{\textcolor{bla}{#1}}

\graphicspath{ {./figures/} }

\begin{document}

\title[Occultations by TNOs]{Stellar occultations by Trans-Neptunian Objects} 


\author*[1]{\fnm{Bruno} \sur{Sicardy}\,\orcidlink{0000-0003-1995-0842}}\email{bruno.sicardy@obspm.fr}

\author[2]{\fnm{Felipe} \sur{Braga-Ribas}\,\orcidlink{0000-0003-2311-24382}}\email{fribas@utfpr.edu.br}
\equalcont{These authors contributed equally to this work.}

\author[3]{\fnm{Marc W.} \sur{Buie}\,\orcidlink{0000-0003-0854-745X}}\email{buie@boulder.swri.edu}
\equalcont{These authors contributed equally to this work.}

\author[4]{\fnm{Jos\'e Luis} \sur{Ortiz}\,\orcidlink{0000-0002-8690-2413}}\email{ortiz@iaa.es}
\equalcont{These authors contributed equally to this work.}

\author[5]{\fnm{Fran\c{c}oise} \sur{Roques},\orcidlink{0000-0001-9900-7760}}\email{francoise.roques@obspm.fr}
\equalcont{%
These authors contributed equally to this work. \\
Accepted for publication in  \it The Astronomy and Astrophysics Review.
}

\affil*[1]{\orgdiv{IMCCE}, \orgname{Observatoire de Paris, PSL Universit\'e, Sorbonne Universit\'e, Universit\'e Lille 1, CNRS UMR 8028}, \orgaddress{\street{77, avenue Denfert-Rochereau}, \city{Paris}, \postcode{75014}, \country{France}}}

\affil[2]{\orgdiv{UTFPR-Curitiba/PPGFA}, \orgname{Federal University of Technology - Paran\'a}, \orgaddress{\street{Rua Sete de Setembro, 3165}, \city{Curitiba}, \postcode{CEP 80230-901}, \state{Paran\'a}, \country{Brazil}}}



\affil[3]{\orgname{Southwest Research Institute}, \orgaddress{\street{1050 Walnut Street, Suite 300}, \city{Boulder}, \postcode{80302}, \state{Colorado}, \country{USA}}}

\affil[4]{\orgdiv{IAA-CSIC}, \orgname{Instituto de Astrof\'{\i}sica de Andaluc\'{\i}a}, \orgaddress{\street{Glorieta de la Astronom\'{\i}a s/n}, \city{Granada}, \postcode{18008}, \country{Spain}}}

\affil[5]{\orgdiv{LESIA}, \orgname{Observatoire de Paris, Universit\'e PSL, CNRS, UPMC, Sorbonne Universit\'e, Univ. Paris Diderot, Sorbonne Paris Cit\'e}, \orgaddress{\street{5 place Jules Janssen}, \city{Meudon}, \postcode{92195}, \country{France}}}


\abstract{%
Stellar occultations provide a powerful tool to explore objects of the outer solar system. 
The Gaia mission now provides milli-arcsec accuracy on the predictions of these events and makes possible observations that were previously unthinkable.
Occultations return kilometric accuracies on the three-dimensional shape of bodies irrespective of their geocentric distances, 
with the potential of detecting 
topographic features along the limb. From the shape, accurate values of albedo can be derived, and if the mass is known, the  bulk density is pinned down, thus constraining the internal structure and equilibrium state of the object. 
Occultations are also extremely sensitive to tenuous atmospheres, down to the nanobar level. They allowed the monitoring of Pluto's and Triton's atmospheres in the last three decades, constraining their seasonal evolution. They may unveil in the near future atmospheres around other remote bodies of the solar system.
Since 2013, occultations have led to the surprising discovery of ring systems around the Centaur object Chariklo, the dwarf planet Haumea and the large trans-Neptunian object Quaoar, while revealing dense material around the Centaur Chiron. This suggests that rings are probably much more common features than previously thought. Meanwhile, they have raised new dynamical questions concerning the confining effect of resonances forced by irregular objects on ring particles.
Serendipitous occultations by km-sized trans-Neptunian or Oort objects has the potential to provide the size distribution of a population that suffered \bs{few} collisions until now, thus constraining the history of primordial planetesimals in the 1--100~km range. 
%
}%


\maketitle

\setcounter{tocdepth}{3}
\tableofcontents




\section{Introduction}
\label{sec_intro}

Several techniques are used to explore the outer solar system, each with its own advantages and caveats. Among them, direct imaging reveals sizes, shapes, as well as surface and atmospheric features, while spectroscopy constrains compositions. For a limited list of objects, space missions return unique data at high resolution, and in some cases, in situ measurements.

In that context, Earth-based stellar occultations\footnote{Occultations occur when a celestial body comes in front of another celestial body. This differs from eclipses, that occur when a body casts its shadow over another body, a topic not covered here.}
allow results that are impossible to obtain with any other techniques, thanks to their extremely high spatial resolution and high sensitivity to tenuous atmospheres.
Various reviews on occultations are available, see for instance the papers by
\cite{elli79} and \cite{elli96}, mainly dedicated to asteroids, planets, large satellites and Pluto.
A more recent review by \cite{orti20a} focuses on occultations by TNOs, 
while \cite{sica23} provides details on the role of refraction in stellar occultations.

Occultations have led to major discoveries in the solar system, among which are
the narrow and dense rings of Uranus in 1977 \citep{elli77,mill77,bhat77},
Neptune's ring arcs in 1984 \citep{hubb86},
Pluto's atmosphere in 1985 \citep{bros95} and 1988 \citep{hubb88,elli89},
Titan's atmospheric super-rotation \citep{hubb93} and gravity waves \citep{stro97},
the drastic increase of the Pluto atmospheric pressure since 1988 \citep{elli03a,sica03},
the dense rings around 
the Centaur object Chariklo in 2013 \citep{brag14}, 
the trans-Neptunian dwarf planet Haumea in 2017 \citep{orti17}, 
the large TNO Quaoar in 2018-2022 \citep{morg23,pere23} and 
\mwb{ephemeral} dense material around the Centaur object Chiron \citep{bus96,elli95,rupr15,sick20,orti23}.
Finally, an occultation revealed the binarity of the TNO 2014 WC$_{510}$ \citep{leiv20}.

We focus in this review on recent progresses made on Trans-Neptunian Objects (TNOs), a topic that has greatly expanded in the past three decades. This review includes Kuiper Belt Objects (KBOs), a population that extends beyond Neptune up to some 50~au, and Oort cloud objects, that orbits much farther out, up to tens of thousands~au. 
We will also consider the Trojans of Jupiter and the Centaurs objects that orbit between Jupiter and Neptune since both populations are thought to be TNOs that have been implanted in the giant planet region.

The discovery of the first TNO 
(other than Pluto), 
(15760) Albion, dates back to 1992 \citep{jewi93}. Currently,
the number of observed TNOs is around five thousand,
\bs{with an estimated total of more than one hundred thousand bodies}
with sizes larger than 100~km \citep{pria20}.
The orbital distribution, sizes, topographic features, compositions, and surroundings (including satellites, atmospheres, and rings) of these many objects provide precious hints on how our solar system formed and evolved.

Due to their remoteness, these bodies are difficult to study. They are both faint and angularly very small, thus requiring, in general, large telescopes to be observed. Contrarily to other techniques, the performances of occultations are essentially independent of the geocentric distances. They can be recorded by small telescopes and are accessible to a large community of amateur astronomers.
As explained below, the ultimate resolving power of occultations is imposed by diffraction and apparent stellar diameters, resulting in spatial resolutions down to the kilometer level. This is several orders of magnitudes better than the resolving power of any Earth-based instruments, for objects orbit at distances greater than a few astronomical units.

This said, occultations have until recently suffered from two main limitations. One was the difficulty to predict the events. In practice, detecting an occultation by a small remote body in the outer solar system was mainly a matter of luck. The other limitation was the high cost of the sensitive and high-speed cameras that are necessary to monitor occultations, thus limiting the widespread distribution of this equipment.

These difficulties have now been overcome. Since 2018, the Gaia mission of the European Space Agency (ESA) has provided astrometric accuracies at the milli-arcsec (mas) level for the stars to be occulted. 
On the other hand, the prices of sensitive cameras have drastically dropped in the last decade, allowing a large community of professional and amateur astronomers to equip their telescopes and to participate in large-scale campaigns. 
%
%
The occultation technique provides key information concerning a very large sample of TNOs, Trojans, and Centaurs. In particular, it can provide
\begin{itemize}
\item The size and shape of objects at kilometric accuracy
\item The topography and the transition between shape and topographic features
\item The accurate value of albedo and its connection with surface composition
\item The discovery of satellites down to contact binaries, that are impossible to image directly 
\item The improved astrometry of the occulting body down to mas-accuracy level
\item The accurate bulk density, if the mass is known, e.g., through the motion of a satellite
\item The satellite/primary mass ratio from the barycentric wobble
\item The equilibrium status, in particular, whether the body is in hydrostatic equilibrium, homogeneous, differentiated, or with a more complex behavior
\item The detection of atmospheres down to the nanobar level, with their thermal structures and temporal evolutions
\item The discovery of rings, with the accurate determination of their orbits and optical depth profiles
\end{itemize}

\section{The limits of stellar occultations}
\label{sec_limits}

Assuming a ``perfect experiment", i.e., an arbitrarily large acquisition rate and an infinite signal-to-noise ratio (SNR), there are two ultimate physical limits to the resolving power of stellar occultations observed from Earth:
diffraction and finite stellar diameters. These two effects are discussed in turn.

\subsection{Diffraction}

As a plane wave with wavelength $\lambda$ hits the edge of an opaque body or a semi-transparent ring, diffraction blurs the shadow profile seen by an observer at a distance $\Delta$. This blurring occurs over a characteristic length called the Fresnel scale, given by
\begin{equation}
\lambda_{\rm F} \sim \sqrt{\frac{\lambda\Delta}{2}}.
\label{eq_Fresnel_Scale}
\end{equation}

\begin{figure}[ht]
\centering
\includegraphics[width=0.8\textwidth]{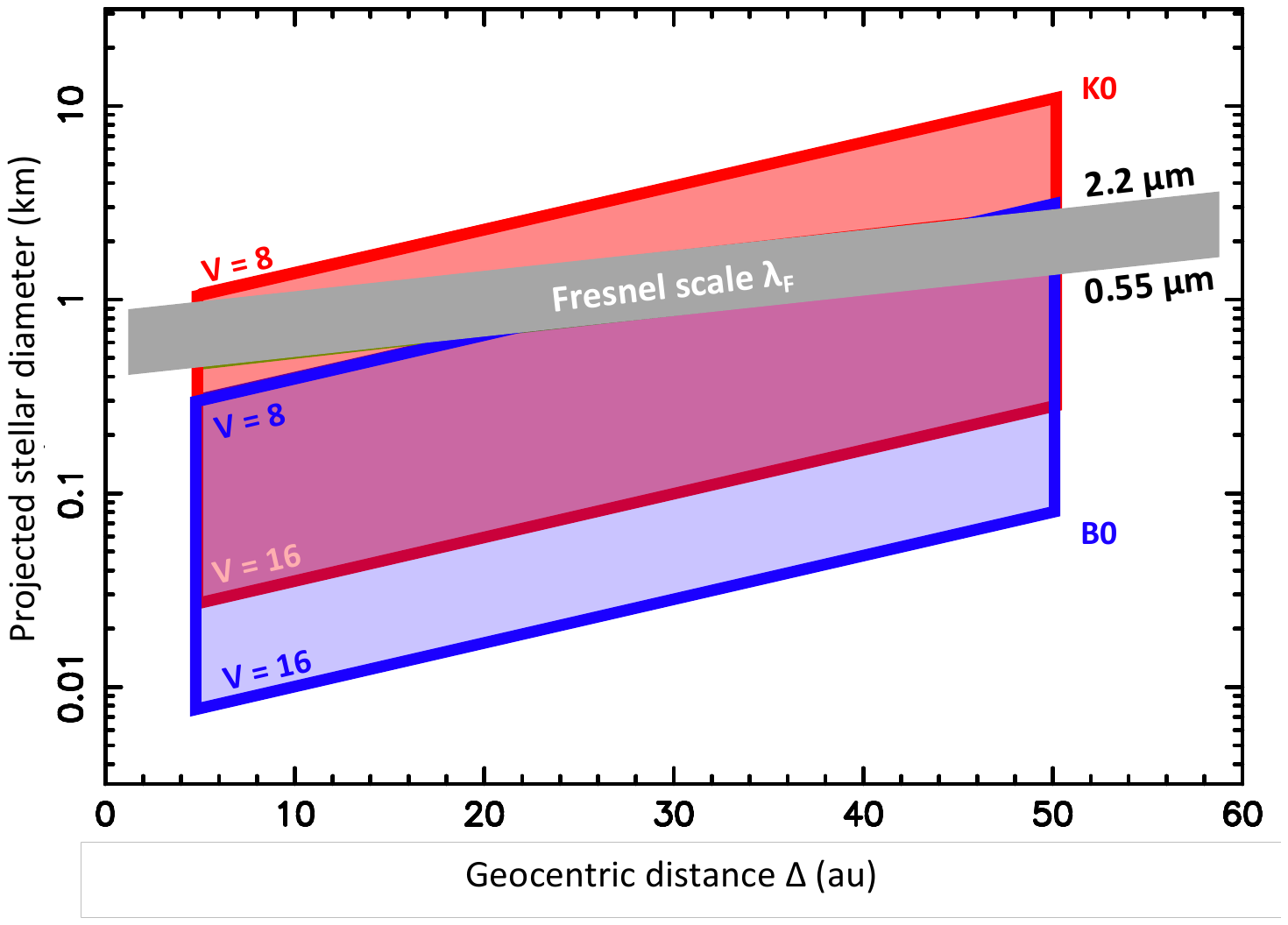}
\caption{%
The Fresnel scale $\lambda_{\rm F}$ (Eq.~\eqref{eq_Fresnel_Scale}) is plotted as a gray box. 
The upper bound of the box corresponds to visible wavelength (typically 0.55~$\mu$m),
while the lower bound corresponds to near-infrared (typically 2.2~$\mu$m).
The stellar diameters projected at objects with geocentric distances $\Delta$, 
are plotted in the colored boxes according to Eq.~\eqref{eq_star_diam_km}.
Two  extreme stellar types are plotted, the cooler K0-type stars with $V-K \sim 1.85$ (red box)  
and the hotter B0-type (blue) with $V-K \sim -0.6$ (blue box).
The upper bound of each box corresponds to bright stars with apparent magnitude $V=8$,
while the lower bound corresponds to fainter stars with $V=16$.
}%
\label{fig_Delta_diam}
\end{figure}

The value of $\lambda_{\rm F}$ is plotted versus the geocentric distance $\Delta$ in Fig.~\ref{fig_Delta_diam}. The Fresnel scale increases with both $\lambda$ and $\Delta$. For Trojan objects, it is typically a fraction of a kilometer, and for remote TNOs at 50~au, it is typically a few kilometers.

\subsection{Stellar diameters}

Stars appear in the sky plane as small disks with finite angular diameters $\theta_{\star,\rm ang}$.
This diameter depends on both the apparent visual magnitude V and its color V-K, where K is the apparent magnitude in the near infrared (2.2 $\mu$m). Typically, we have
$$
\theta_{\star,\rm ang} \sim 4.7 \times 10^{-0.2V} \times10^{+0.223(V-K)}~{\rm mas},
$$
see \citealt{vanb99}. More refined calculations, that account in particular for limb darkening, 
are provided by \citealt{kerv04}. 
The angular diameter $\theta_{\star,\rm ang}$ can be projected at the planet geocentric distance in astronomical units, $\Delta_{\rm au}$, and re-expressed as a linear diameter in kilometer, $\theta_\star$, providing
\begin{equation}
\theta_\star \sim 3.4 \times \Delta_{\rm au} \times 10^{-0.2V} \times10^{+0.223(V-K)}~{\rm km}.
\label{eq_star_diam_km}
\end{equation}
We see that redder stars (with larger V-K) have a lower surface brightness and thus a larger apparent diameter for a given magnitude V. Adopting extreme stellar spectral types from K0 (red) and B0 (blue), there is a factor of about 3.5 in diameters between the reddest and bluest stars. Results for typical hot and cool stars are presented in Fig.~\ref{fig_Delta_diam}.


The Gaia catalog also provides estimates of stellar diameters, using their individual physical properties, such as apparent G magnitude, parallax, and effective temperature. The stellar radius and the luminosity can be calculated using the \textit{Apsis-Priam-FLAME} packages\footnote{For the basic equations, see \cite{and18}, and \cite{fou23} for the latest discussions and implementations using Gaia DR3.} and then converted to $\theta_\star$, using the equation
\begin{equation}
\theta_\star = 2\pi \arctan\left( \frac{R_{\star}}{2d}\right)\times \Delta_{\rm au}\times
\left( \frac{1.49597870691\times10^8}{6.48\times10^8} \right)~{\rm km},
\label{eq_star_diam_proj}	
\end{equation}
where $R_\star$ is the physical stellar radius and $d$ is the distance of the star\footnote{Caution should be taken when estimating distances using the inversion of the parallax ($d=1/\tan(\alpha/2)$), see the thorough discussion by \cite{Bailer18}.}.

In Fig.~\ref{fig_Delta_diam}, we consider stars as bright as $V=8$, 
a rare occurrence in Earth-based occultations by remote bodies, 
down to stars fainter than $V=16$ that are mainly accessible to large professional telescopes.
Considering typical cases of stars in the range $V \sim$~12-16, this provides typical ranges of 
$\theta_\star \sim$0.01-0.05~km for B0 stars at geocentric distances $\Delta=4$~au (typical of Trojans) and $\theta_\star \sim$0.03-0.2~km for K0 stars, again at Trojan distances.
As the diameter increases linearly with $\Delta$ (Eq.~\eqref{eq_star_diam_km}), the range
of $\theta_\star$ is typically 0.1-0.5~km at $\Delta=50$~au for blue stars and
0.3-2~km for red stars at that same distance.

As seen in Fig.~\ref{fig_Delta_diam}, for Trojans observed from Earth 
the smoothing of the shadow mainly comes from Fresnel diffraction, except for the (rare)
occasions where red stars with apparent magnitudes brighter than $V \sim 8$ are involved.
For objects as far as 50~au, the stellar diameters of red stars may compare 
with diffraction in terms of shadows smoothing.
Blue stars, in contrast, generally cause a negligible blurring when compared to diffraction.
Whatever the cause of blurring is, it is always larger than about one kilometer. 
This is the limiting factor for detecting small objects using serendipitous occultations, 
see Sect.~\ref{sec_serendip}.

In any instance, the limits quoted here are well beyond -- by several orders of magnitudes -- the resolving power of any current Earth-based imaging systems that reach at best some 50~km at the level of the Trojans, or about 500~km at the level of Pluto.This  explains why occultations are and will continue to be used even in the era of large ground-based or space-borne telescopes.

\subsection{Predictions}
\label{sec_prediction}

The approach to predict, plan and observe stellar occultations by small bodies is detailed in Appendix~\ref{app_deployment}. It consists \bs{of} a bootstrap method. 
First, the star position provided by Gaia, combined with classical astrometric measurements of the occulting body, is used to attempt an occultation event. 
Second, once at least one positive ``chord"\footnote{An occulting chord is defined as the segment delimited by the disappearance and reappearance of the star behind the body, as observed in the sky plane.}
has been obtained, a much more accurate ephemeris of the body is derived, pinning down the accuracy of subsequent predictions. 
Third, as more successful campaigns with more positive chords are achieved, the ultimate Gaia limit, typically the mas-level or less, can be reached, remembering that 1~mas corresponds to about 3~km for Trojans, 10~km for Centaurs, 20~km for Pluto and 40~km for objects at 50~au.

\begin{figure}[ht]
\centering
\includegraphics[width=0.75\textwidth]{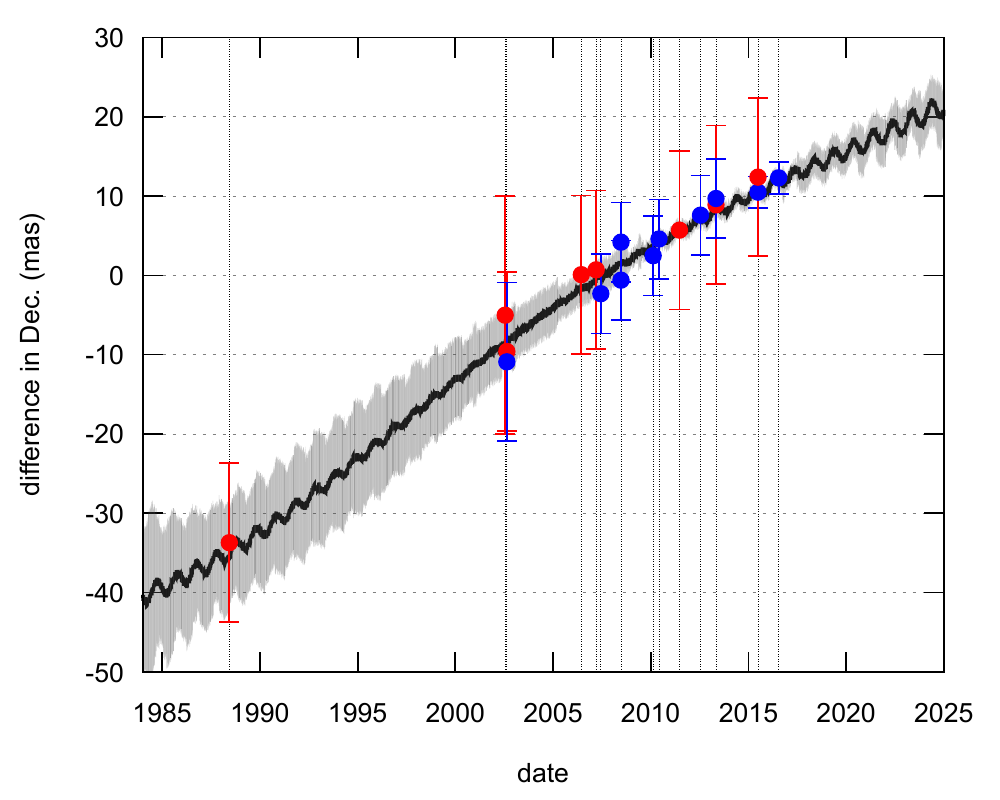}
\caption{%
The black line displays the offsets in declination between the NIMAv8 and the JPL DE436  ephemerides of Pluto’s system barycenter. The blue points are based on the stellar occultation results of \cite{meza19}, using Gaia star positions, while the red points are positions obtained from other observations. The gray area marks the 1$\sigma$ uncertainty of the NIMA ephemeris.
One can note the increasing accuracy of the ephemeris as more observations are made, reaching a few mas in 2016. Subsequent occultations were used to pin down even further the accuracy of NIMA see \url{https://lesia.obspm.fr/lucky-star/obj.php?p=818}.
Image reproduced with permission from \citealt{desm19}, copyright by the author(s).
}%
\label{fig_ephem_pluto}
\end{figure}

An example of the bootstrap approach is shown in Fig.~\ref{fig_ephem_pluto}. The uncertainty on Pluto's ephemeris so obtained (the mas-level) is much smaller than the angular diameter of Pluto's atmosphere, about 150~mas. This means that it is now possible to plan observations of specific regions of this atmosphere, for instance the winter vs. summer hemispheres, or the morning vs. evening limbs.
It is now also possible to plan a fine scanning of the central flash region, that extends by not more than about 100~km around the shadow center, corresponding to less than 5~mas when projected at Pluto's distance.

Another example of the efficiency of this approach is provided by the occultation of 8 August 2020 by the large TNO 2002 MS$_4$ that subtends about 25~mas as seen from Earth. A double-chord event observed two weeks before the event (on 26 July 2020) pinned down the accuracy of the prediction to the 8-mas level, allowing an efficient planning of a large campaign in Europe with more than 60 positive chords on August 8 \citep{romm23}.

Besides its utility for pinning down predictions, the bootstrap approach can fill an important role in an initial survey of the system, by revealing a more complex situation before ever getting to the tightly coordinated efforts. It can for instance reveal the presence of an unseen body when combining multiple occultations through the detection of a wobble when fitting an orbital solution.
In fact, once three occultations are observed, 
\bs{one may sometimes put limits on unseen bodies that will be very constraining, as discussed further in Sect.~\ref{sec_trojans} and Appendix~\ref{app_deployment}.
}

\section{Sizes, albedos, shapes and topographic features}
\label{sec_shape_retrieval}

The measurements made possible by occultations are very well suited to in-depth studies of small bodies in the Solar System.
The accuracy on the length of each chord is limited by diffraction and stellar diameter, that is at the kilometer level or so, see the previous section and Fig.~\ref{fig_Delta_diam}. Thus, these limitations become a problem only for very small or very distant objects.

%
We note that the faintness of the objects under consideration is an advantage rather than a limitation.  
The reflected light from the occulting body is actually a source of noise in the recorded flux, 
which sets the minimum useful brightness for a star that can be considered.  
As a result, fainter (and more numerous) stars can be used for occultations by faint bodies, 
with the further advantage of reducing the confounding effect of stellar diameters
(Fig.~\ref{fig_Delta_diam}).

By combining multiple 
occultations and accounting for the rotation phase of the object (using photometric light curves, for instance), it is possible to retrieve the 3D shape and size of the object.
Once the size and shape are accurately determined, the visual albedo can be derived, using the equation
\begin{equation}
p_{\rm V}  = 10^{0.4 (H_{\odot,\rm V}-H_{\rm V})}  /R^2_{\rm eq},
\label{eq_albedo}
\end{equation}
where $R_{\rm eq}$ is the equivalent radius (i.e., the square root of the projected area divided by $\pi$) measured in astronomical units, $H_{\odot,\rm V}= -26.74$ 
is the magnitude of the Sun at 1~au, and $H_{\rm V}$ is the absolute magnitude of the body at the moment of the occultation.
This usually provides $p_{\rm V}$ at a level of precision of a few percent. This is more accurate than one obtains from measurements of thermal radiation from the surface, which have typical errors of 10--20\%, see Sect.~\ref{sec_centaurs_tnos}.

The size and shape also provide an accurate volume determination. 
If the mass of the object is known, for instance, from a satellite motion or deflection 
of other objects, the average bulk density can be calculated.
Also, besides the general shape, an occultation may reveal interesting topographic
features along the limb of the object, such as craters, depressions, chasms, or mountains.

The size, shape, topography, albedo, and density provide, in turn, important diagnostic information about the processes that have been at work on the body. These processes may concern the initial formation phase of the object, or later modifications caused by collisions, or even processes that are entirely described by its current dynamical and thermal state, e.g. the YORP\footnote{Yarkovsky--O'Keefe--Radzievskii--Paddack}-modified effect in the case of near-Earth asteroids.  
 
Differences of tens of percent in the bulk density can explain the object's initial composition (water/olivine ratio) and, thus, its internal stratification. For example, an ``Orcus-like" object, with a density of 1.53~g~cm$^{-3}$ \citep{forn13} will have a very small fraction of serpentine in its interior at the end of its formation phase, whereas a ``Pluto-like" object, with 1.86 g~cm$^{-3}$ \citep{ster15}, will have a modest quantity of serpentine. Meanwhile, an ``Eris-like" object, with 2.4 g~cm$^{-3}$ \citep{sica11} will end up with almost no water in its interior, which will be made by a significant ratio of serpentine over non-reactive rock (see Fig.~4 of \citealt{fark23}).

\subsection{Limb fitting}
\label{subsec_limb_fit}

In the earliest days of occultation studies, obtaining a circular equivalent size was usually the
best one could hope for, due to the prevailing uncertainties in the event predictions 
leading to a small number of chords.
%
However, this was a unique way to pin down albedo and density measurements, 
that were otherwise poorly known for remote objects, 
see for instance \cite{elli10} or \cite{sica11}.
Another important by-product of even simple circular limb fitting is the drastic improvement of the ephemeris of the object, leading to more accurate predictions for subsequent occultations (Sect.~\ref{sec_prediction}).

The next step, elliptical limb fitting, provides important information on the internal state of the object, especially when comparing it with equilibrium solutions. This is mainly relevant for larger bodies, while for smaller bodies, the shapes are controlled by the internal strength of the object (Sec. \ref{subsubsec_topography}). 

\begin{figure}[ht]
\centering
\includegraphics[width=0.6\textwidth]{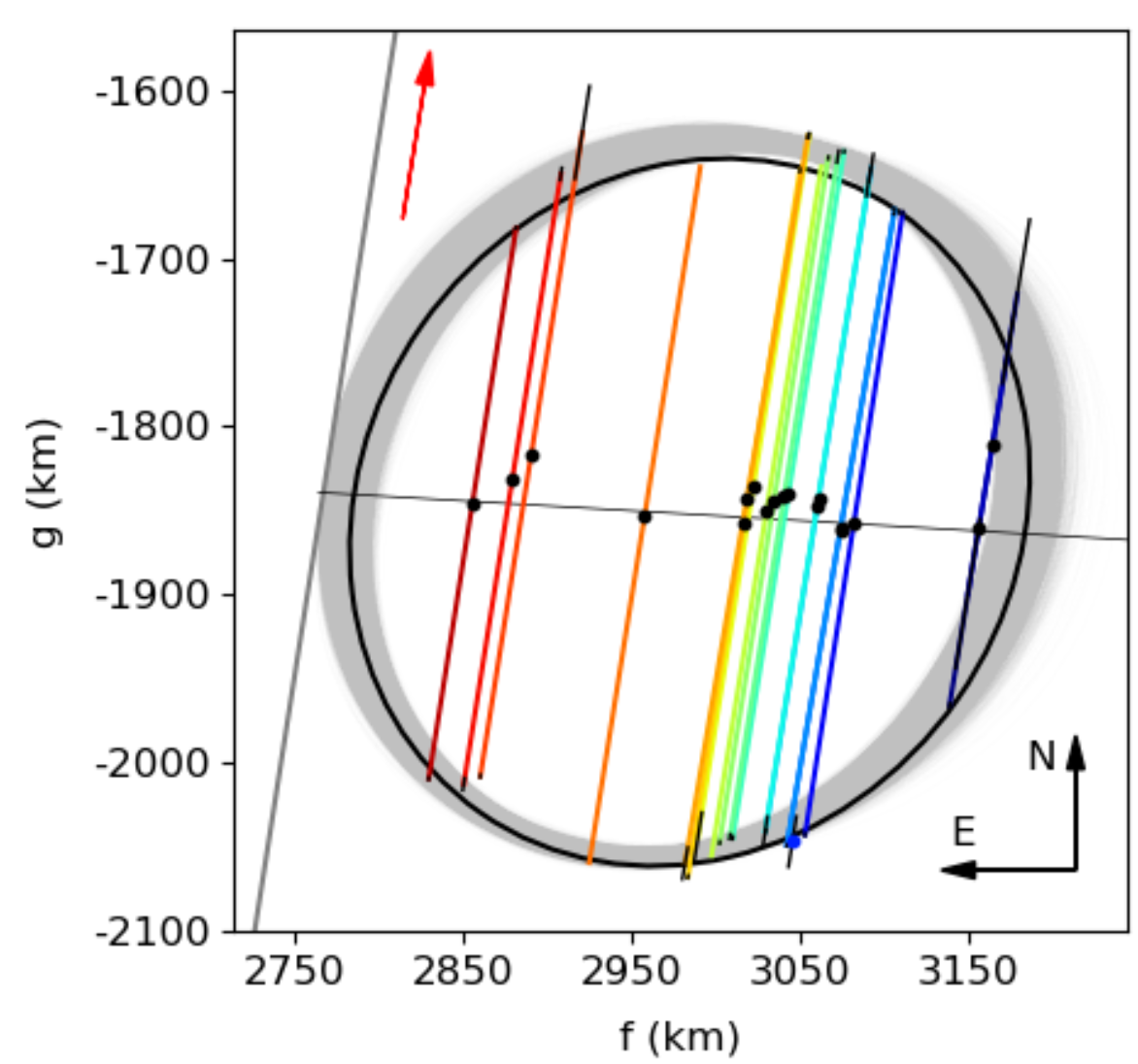}
\caption{%
The best fitting elliptical limb (in black) obtained from a stellar occultation by the TNO (38628) Huya observed on 18 March 2019, using 19 chords (colored segments). The gray region indicates the 1$\sigma$ uncertainty region for the elliptical fit. The black dots mark the center of each chord and the red arrow shows the direction of motion of the star with respect to Huya.
Image reprodcued with permission from \cite{sant22}, copyright by the author(s).
}%
\label{fig_fit_limb_Huya}
\end{figure}

An elliptical limb is defined by five parameters: 
the offset ($f, g$) to apply to the center of the body with respect to the prediction (counted positively towards the East- and North-celestial directions, respectively),  
the apparent semi-major axis $a'$ and apparent semi-minor axis $b'$ (or equivalently, the apparent oblateness $\epsilon=(a'-b')/a'$), and 
the position angle $P$ of the apparent semi-minor axis $b'$ (counted positively from the North to the East directions). Therefore, at least five points are needed to fit an elliptical profile. 
This can be achieved by three well-spaced observers who provide three chords, i.e. six data points along the limb. An example of a limb fitting using chords obtained during an occultation by the TNO Huya is displayed in Fig.~\ref{fig_fit_limb_Huya}.

\subsection{Retrieval of 3D shape}
\label{subsec_3D_shape}


From an instantaneous apparent limb provided by an occultation, the three-dimensional shape of an object can be obtained by either combining several stellar occultation events observed at different rotational phases \citep{morg21}, or by combining the occultation results with the photometric rotational light curve of the object. 

We can adopt for the body the simple model of a triaxial ellipsoid with semi-axes $a > b > c$, resulting in six unknown parameters: $a$, $b$, $c$, the coordinates of the pole position and the rotational phase, with only three observables: $a'$, $b'$ and $P$, see the previous subsection.

The rotational phase at the time of occultation can be derived from rotational light curves close in time to the event if the rotational period is well determined, which decreases the degeneracy of the model \citep{sant21}, but still leaving five unknowns. 
Further constraints on the axial ratios $\beta= b/a$ and $\gamma= c/a$ are provided by the amplitude $\Delta m$ of the rotational light curve, given by  
\begin{equation}
\Delta m = 2.5\cdot\log_{10}\left(\frac{1}{\beta}\right)-1.25\cdot\log_{10}\left[ \frac{1+\gamma^2\tan^2\theta}{1+(\gamma/\beta)^2\tan^2\theta} \right],
\label{eq_deltam}
\end{equation}
where $\theta$ is the aspect angle, i.e. the angle between the  $c$-axis and the line of sight. This decreases further the degeneracy of the model, but more information is still needed to come up with a unique solution, namely the pole position.

The latter can be constrained based on the change of $\Delta m$ with time. However, given the long orbital periods of these bodies, this is not as easy as for asteroids, as detailed by \cite{carry24}, but could be feasible for Trojans and Centaurs.
For objects surrounded by rings, we can safely assume that the ring is equatorial, which provides the spin axis orientation of the body, resulting in a unique solution \citep{orti17}. Similarly, for TNOs with large satellites of known orbits, the satellite plane can be assumed to be equatorial, which again provides the pole orientation.
\bs{However, care should be taken when adopting this a priori assumption, which is not always true. For instance Namaka (one of Haumea's satellites) has an inclination of almost 13~degrees with respect to Haumea's equator \citep{prou24}.}

As an example, let us consider the case where the pole (and therefore $\theta$) is known and where the occultation occurs at the maximum brightness, meaning that the apparent semi-major axis $a'$ coincides with $a$. Then it can be shown that \citep{bene19}
\begin{equation}
\gamma=\sqrt{\frac{\beta'^2-\beta^2\cos^2\theta}{\sin^2\theta}}.
\label{eq_gamma}
\end{equation}
For any rotation phases, more complex expression can be used, see for instance the equations~25-27 of \cite{mag86}.

If no information in the rotational phase is available, then the light curve amplitude assuming an equilibrium figure (see next subsection) can be a first approximation to obtain the 3D shape of an object and its density \citep{sica11,brag13,brag23}.

\subsection{Equilibrium solutions: MacLaurin and Jacobi}
\label{subsec_equil_fig}

The simplest case beyond the homogeneous spherical body is a homogeneous object of density $\rho$ in hydrostatic equilibrium under the effect of its rotation with spin frequency $\omega$. 
In the low angular momentum regime, the body assumes an oblate spheroidal shape with equatorial and polar radii $a$ and $c$, respectively, known as the Maclaurin solution \citep{macl1741}. It is given by \citep{plum1919}
\begin{equation}
\frac{\omega^{\sf 2}a^{\sf 3}}{GM}= 
\frac{3\left\{2\psi\left[2+\cos(2\psi)\right] - 3\sin(2\psi)\right\}}{4\sin^{\sf 3}(\psi)},
\label{eq_maclaurin}
\end{equation}
where $G$ is the gravitational constant, $M$ is the mass of the body, and $\cos(\psi)$=$c/a$.
Note that the density $\rho$ can appear explicitly in the equation above by writing $\rho= 3M/[4\pi \cos(\psi) a^3]$.

Once $a$ and $c$ are derived from occultation data, it is possible to derive $\rho$, assuming a Maclaurin spheroid. Conversely, if $M$ is known independently (for instance, from the motion of a satellite), it is possible to check whether the body is indeed a Maclaurin spheroid.

In the high angular momentum regime, the Maclaurin solution becomes unstable and bifurcates towards the triaxial homogeneous Jacobi  solution \citep{jaco1834} with semi-axes $a > b > c$ given by
\begin{equation}
\beta^{\sf 2}       \int_{\sf 0}^\infty \frac{du}{(1+u)(\beta^{\sf 2}+u)\Delta(\beta,\gamma,u)}= 
\gamma^{\sf 2} \int_{\sf 0}^\infty \frac{du}{(\gamma^{\sf 2}+u)\Delta(\beta,\gamma,u)}
\label{eq_jacobi1}
\end{equation}
where
\begin{equation}
\frac{\omega^{\sf 2}a^{\sf 3}}{GM} =  \frac{3}{2}  \int_{\sf 0}^\infty \frac{udu}{(1+u)(\beta^{\sf 2}+u)\Delta(\beta,\gamma,u)},
\label{eq_jacobi2}
\end{equation}
where $\Delta(\beta,\gamma,u)$=$[(1+u)(\beta^{\sf 2}+u)(\gamma^{\sf 2}+u)]^{\sf 1/2}$ \citep{lace07}.
If for instance $\beta$=$b/a$ is given, Eq.~\eqref{eq_jacobi1} yields $\gamma$, which in turn is related to the spin rate $\omega$ and the mass $M$ through Eq.~\eqref{eq_jacobi2}. Again, the density $\rho$ may appear explicitly in this equation by noting that $\rho= 3M/[4\pi \beta \gamma a^3]$.

Once $(a,b,c)$ is known, it is possible to check whether Eqs.~\ref{eq_jacobi1} and \ref{eq_jacobi2} are satisfied, that is, whether the shape of the body is consistent with the Jacobi solution. Nevertheless, this condition is necessary but not sufficient. For instance, Haumea's shape derived from the 21 January 2017 occultation is close to the Jacobi solution in terms of the ratios $\beta$ and $\gamma$. However, Haumea's spin rate $\omega$ is well known thanks to its rotational light curve, while its mass $M$ is well determined from the motions of the satellites Namaka and Hi'iaka.  It then turns out that Haumea's mean density $\rho=1885 \pm 80$~kg~m$^{-3}$ is well below the value expected from the Jacobi solution, $2530 < \rho < 3340$~kg~m$^{-3}$ \citep{orti17}. With such low density, a homogeneous body should fly apart due to the centrifugal effect of its own rotation.

It is then necessary to consider more complex internal structures than provided by the Jacobi solution. For instance, Haumea's shape might be explained by a differentiated object in hydrostatic equilibrium with a denser hydrated silicate core and  thick ice crust \citep{dunh19}.

\subsection{Topography}
\label{subsubsec_topography}

For objects with diameters above $\sim$400~km, given that they had time to achieve hydrostatic equilibrium, it is reasonable to consider that their global shape does not deviate from an ellipsoidal body \citep{thom89,line10}. Its shape will depend on its density, \bs{rotational} spin, internal strength, and friction that the material on the crust can support \citep{hols01}.

\begin{figure}[ht]
\centering
\includegraphics[width=0.75\textwidth]{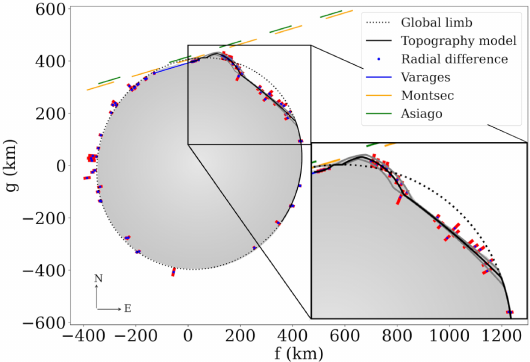}
\caption{%
A large depression was revealed along the limb of the TNO 2002~MS$_4$ from a stellar occultation with more than 60~chords. The red segments around each data point (in blue) indicate their uncertainties in the radial directions. The dotted line is the global elliptical fit to the limb. A large depression of depth $41.5\pm1.5$~km and length $322\pm39$~km appears in the top-right part of the body. 
Image reproduced with permission from \citealt{romm23}, copyright by the author(s).
}%
\label{fig_depression_MS4}
\end{figure}

Besides providing the global shape, the occultation chords can reveal local topographic features, 
or ``roughness" of the surface. Little is known about the roughness of small objects in the outer solar system. 
High-resolution images of Saturn's, Uranus', and Pluto's icy satellites and 
Arrokoth, obtained with the Cassini, Voyager II, and the New Horizons spacecraft, respectively, are the only direct measurements of global roughness of TNOs or similar objects \citep{sche18,sche20,nimm17,spe20}.
This said, a stellar occultation by the TNO 2003~AZ$_{84}$ showed a gradual fading during the star disappearance. 
\bs{This gradual fading could be caused by an extremely low-angle (grazing) occultation geometry in which the limb of the body partially occulted over several data points the finite stellar disk projected at the object distance. In the present case, it} 
could be interpreted either as the presence of an 8~km deep and 23~km long chasm or a 13~km deep and 80~km long depression along the limb \bs{of 2003~AZ$_{84}$} \citep{dias17}.  In the same vein, a multi-chord occultation revealed a large and deep depression along the limb of the large TNO 2002~MS$_4$, see Fig.~\ref{fig_depression_MS4}.

Even if individual topographic features cannot be resolved, the scatter in the residuals of a smooth limb fit can be used to set a lower limit to the global roughness of the body, as is the case for the Uranian satellite Umbriel \citep{assa23a}. The theoretical limit on the topography supported by a large icy body is estimated from \cite{john73},
\begin{equation}
    h_{\rm topo} \leq \frac{3k S}{4\pi \rho^2 G R},
    \label{eq_topography}
\end{equation}
where $R$ is the radius of the object and 
$k$ is a dimensionless parameter related to the strength $S$ of the material;
it is generally close to unity 
with extreme values of up to three.
The value of $S$ for icy material is $S_{\rm ice} \sim 3 \times 10^6$~N~m$^{-2}$ \citep{nimm17}. 
%


\bla

\subsection{Small-body characterization}
\label{subsec_smallbody_char}

Determining the surface area and volume of small bodies (D $\lesssim$ 100 km) can be challenging using only stellar occultations. Even multi-chord observations can only show a picture of the object's instantaneous profile \citep{buie20}. In this case, no model like Maclaurin of Jacobi solutions proposes reliable templates for these objects; see more discussion of this point in Sect.~\ref{sec_trojans}. Note also that although more precise than classical astrometry, the astrometry obtained from an occultation with a few chords can be wrong by an object's radius (see Appendix \ref{app_deployment}).

Thus, stellar occultations by small bodies should be associated with techniques capable of obtaining the 3D shape of the object. High-resolution images obtained with adaptive optics can be used for the biggest main-belt objects \citep{vern21}, and spacecraft images are available for a limited number of objects. The inversion technique, made with rotation light curves  \citep{kaa01}, can be an option for the Trojans and a few Centaur objects \citep{mot24,rous21}. As it needs observations of the rotation light curve along a good fraction of its orbital period, i.e., at different aspect angles, this can not be applied to the TNOs at this time 
without a statistical analysis of a very large population sample.

\begin{figure}[ht]
\centering
\includegraphics[width=0.6\textwidth]{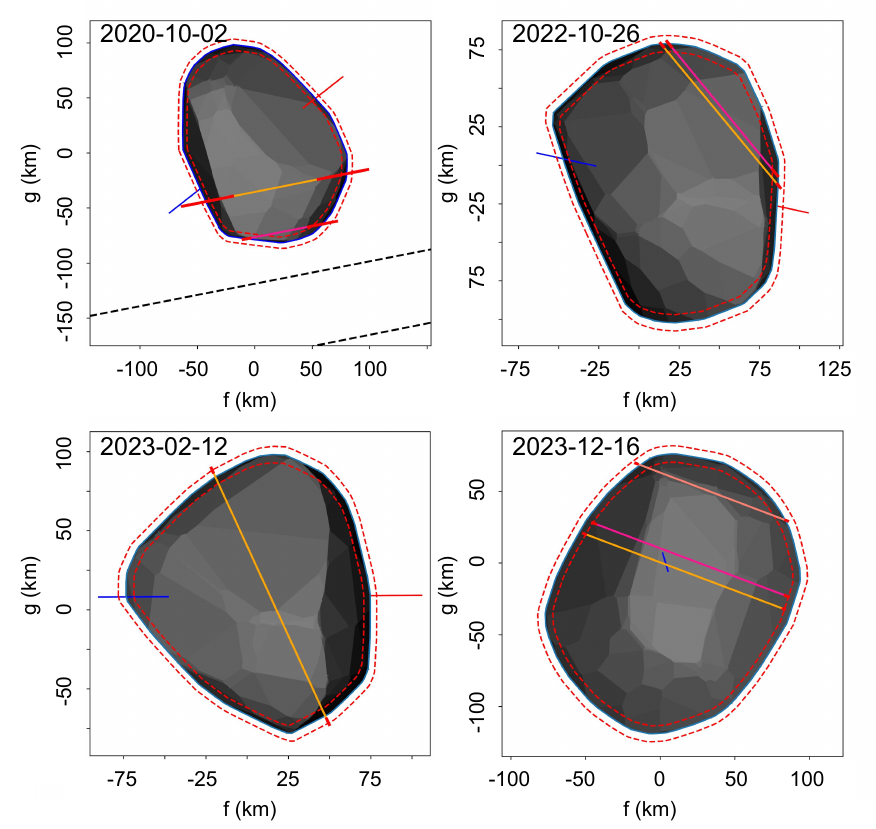}
\caption{%
An example of a fit of four stellar occultations by the Trojan (911) Agamemnon with the 3D model from Database of Asteroid Models from Inversion Techniques (\href{https://astro.troja.mff.cuni.cz/projects/damit/}{DAMIT}), from \cite{dur10}. Considering the available rotation period, the model is rotated to the occultation dates 
and then scaled to the chords (colored segments). The dashed lines represent the model uncertainty, and the rotation pole direction is shown in the blue and red segments. All these parameters can be fitted jointly, considering all the available chords \citep{gome22}.
}%
\label{fig_trojan_3Dfit}
\end{figure}

The 3D models obtained with the inversion technique can be used with occultation chords to scale the model and, thus, the volume of the object. If a few occultations are available, then the scale, pole position, spin rate, phase orientation, center, and even the model uncertainty can be refined (Fig.~\ref{fig_trojan_3Dfit}) as done for the satellite \bs{Phoebe} (Saturn XI, see \citealt{gome20}) and the Centaur (60558) 174P/Echeclus \citep{pere24}. The best solution is to use all available sources, such as rotation light curves, high-resolution images, thermal measurements, and occultation detections to fit a non-convex 3D shape, such as made using ADAM software \citep{viik15}.




\section{Atmospheres}
\label{sec_atmo}

Due to their low surface temperatures (typically 30-60~K), Centaurs or TNOs can maintain tenuous atmospheres that are essentially controlled by the vapor pressure equilibrium with volatile ices such as N$_2$, CO and CH$_4$, see the detailed review of \cite{youn20}.

As explained below, Earth-based occultations represent a unique and sensitive tool to probe atmospheres down to pressures levels of a few nanobars.
They provide a high spatial resolution, with details as small as a kilometer for the retrieved molecular density $n(r)$, the pressure $P(r)$ and the temperature $T(r)$ of the gas, as a function of the distance $r$ to the body center.
When the star, the body and the observer are aligned, a central flash can be seen.
The flash structure is very sensitive to the shape of the atmosphere, and can thus
constrain the zonal wind regime of the layer responsible for the flash.

Atmospheric occultations cause a gradual loss of the stellar flux that lasts for several seconds or minutes,
in contrast to the rapid, sub-second disappearance caused by a sharp opaque limb.
This gradual loss has two different origins: refraction, that deviates the stellar rays, and extinction or scattering due to hazes or molecules that reduce the stellar flux received by the observer. 

It is important to note that the drop due to refraction occurs 
\textit{even if the atmosphere is transparent}, as it follows from 
the defocusing of the stellar rays, as sketched in Fig.~\ref{fig_dp_dz}.
Conversely, the limb curvature has the opposite effect,
as it focuses the stellar rays toward the observer, resulting in a strong central flash 
near the shadow center (Fig.~\ref{fig_Triton_05oct17}).

Meanwhile, absorption provides the atmospheric extinction 
along the line of sight.
If this extinction is observed in various bands, it may reveal chromatic effects 
that constrain in turn the properties of the haze particles responsible for the extinction, 
such as the size distribution or the nature of the aerosols (e.g. spheres vs. fractal aggregates).

This section mainly presents
Earth-based stellar occultations, knowing that 
occultations have also been performed from spacecraft, a topic not reviewed here.
In practice, spacecraft can record radio occultations (typically at a few-cm wavelengths)
that cause the phase shift of a coherent radio signal emitted from or sent to the spacecraft.
Mathematically, the phase shift is equivalent to refraction, 
and thus also provides the $n(r)$, $P(r)$ and $T(r)$ profiles.
Space missions can also monitor multi-wavelength absorption occultations of the Sun or 
bright stars by atmospheres, from the UV to near IR. In doing so, they provide
density profiles of several minor species and haze particles.

Earth-based occultations offer a competitive approach compared to space exploration. An example of such complementarity is given by Earth-based occultations by Pluto. They can provide accurate atmospheric profiles between altitude levels of 5 and 380~km \citep{meza19}, while the New Horizons radio science occultations retrieved profiles between the surface and the $\sim$30~km altitude level, see \cite{hins17} and Sect.~\ref{subsec_atmo_waves}.

Another advantage of Earth-based observations is that they allow a long-term monitoring of an
atmosphere, thus revealing seasonal effects, something which is not possible with space flybys.
Moreover, Earth-based occultations \bs{make it possible} to search for atmospheres around many objects, 
while space exploration is targeted towards a handful of objects.

\subsection{The physics of atmospheric occultations}

\begin{figure}[ht]
\centering
\includegraphics[width=0.8\textwidth,trim=0 0 0 0]{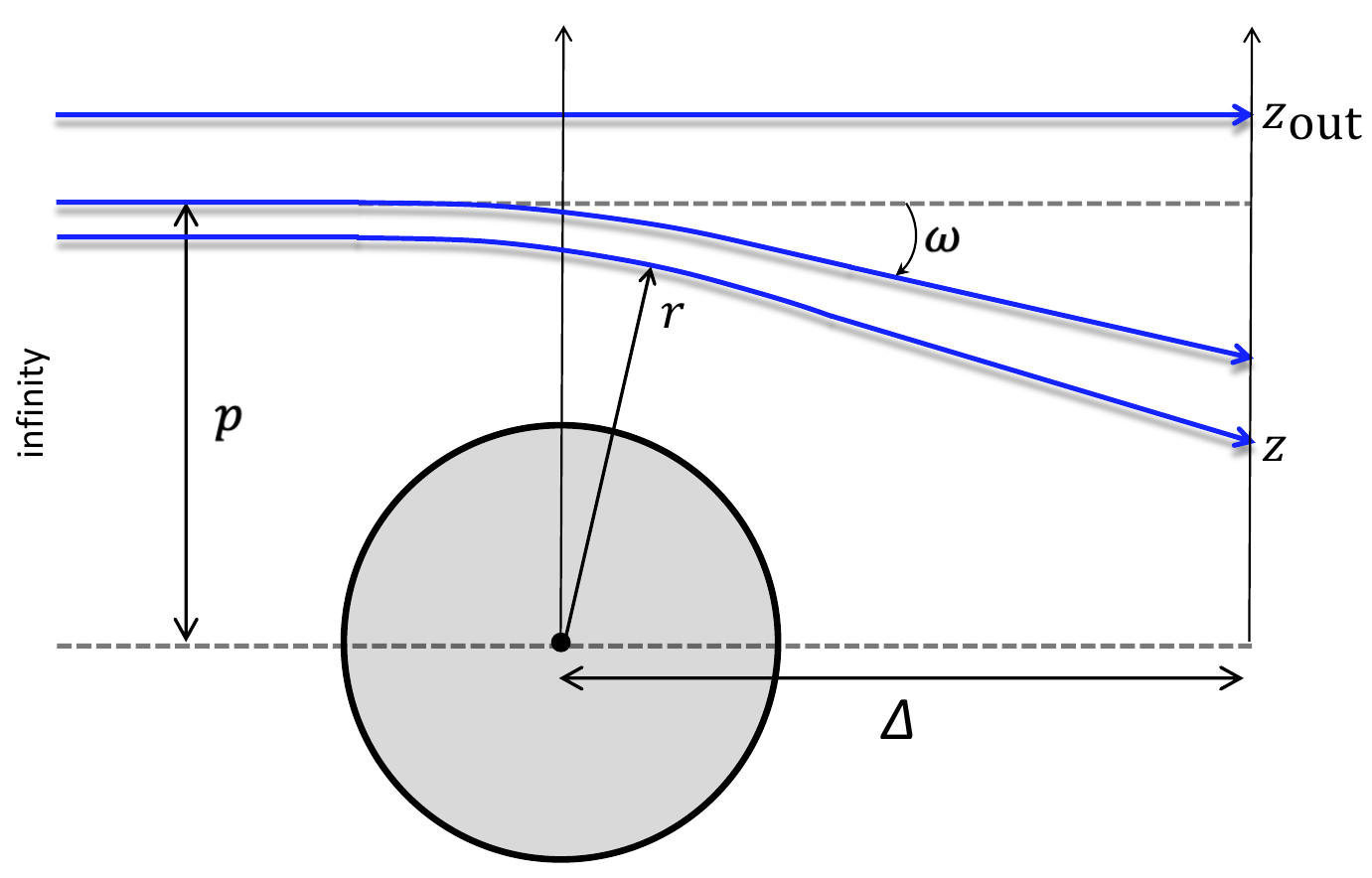}
\caption{%
The principle of refractive occultations. 
Stellar rays coming from infinity with impact parameter $p$ are refracted 
by an angle $\omega$ (taken here as negative) as they traverse the atmospheric layers, 
coming to closest approach $r$ to the planet center, 
before reaching the observer located at distance $\Delta$ from the body and
at ordinate $z$ in the shadow plane.
As deeper layers are probed, the angle $\omega$ increases in absolute value, 
causing the divergence of the rays. 
Energy conservation (in a transparent atmosphere) results in a decrease of 
the stellar flux received at $z$. 
Due to the exponential decrease of molecular density with altitude, 
the ray deviations rapidly becomes negligible above a distance $z_{\rm out}$
from the shadow center, where the stellar flux reaches its unocculted level. 
}%
\label{fig_dp_dz}
\end{figure}

The simplest version of atmospheric occultations is illustrated in Fig.~\ref{fig_dp_dz}.
It considers stellar rays coming from infinity
with impact parameter $p$ and traversing a spherical and transparent atmosphere.

\begin{figure}[ht]
\centerline{
\includegraphics[width=1.\textwidth,trim=0 0 0 0]{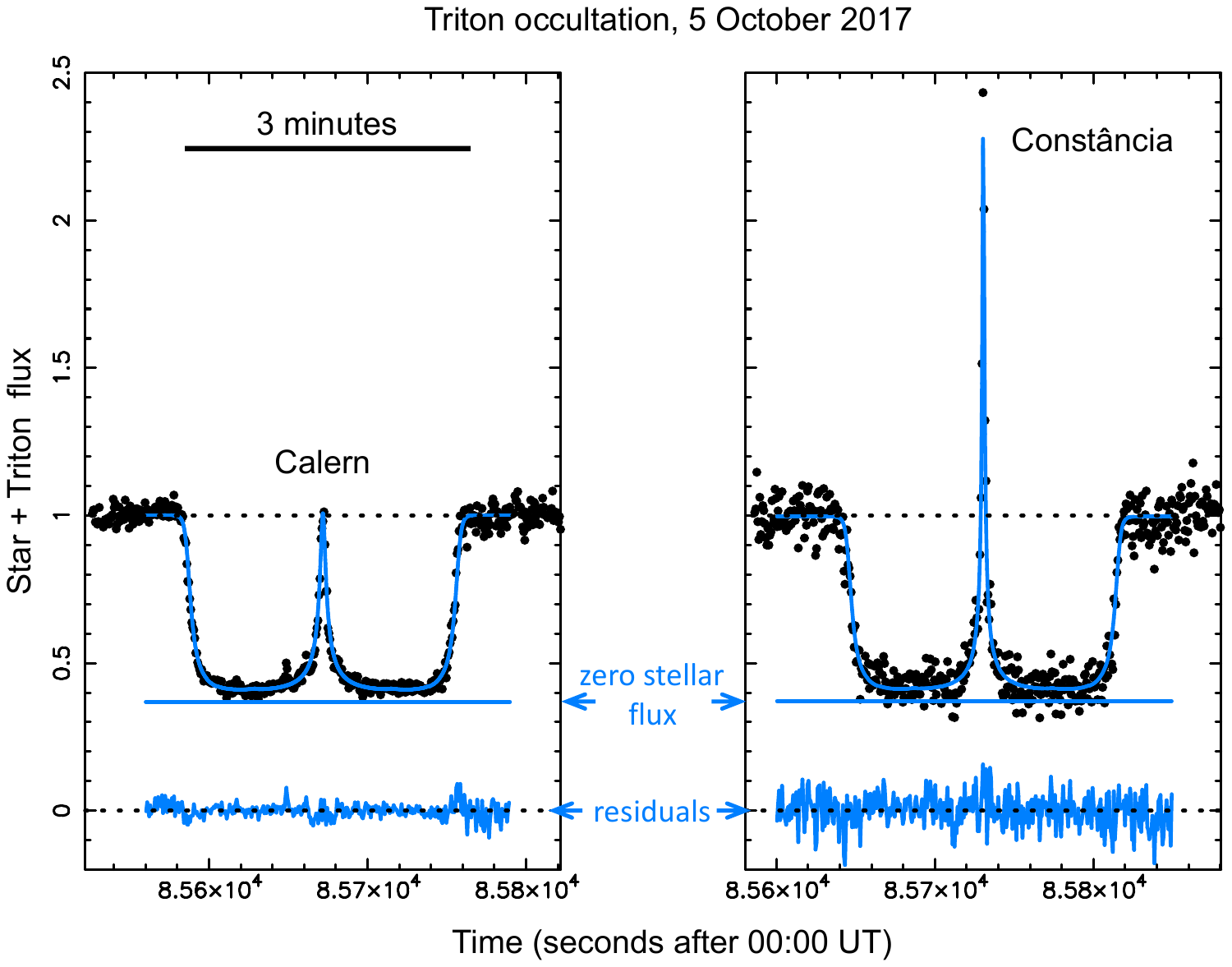}
}
\caption{%
Two light curves (with the data points plotted as dots) observed at the Calern and Const\^ancia stations
during the Triton occultation of 5 October 2017. 
The flux of the star plus Triton, normalized to its unocculted value, is plotted vs. time.
The blue curves on top of the data points are best-fitting models,
while lower blue curve shows the residuals of the fit.
The gradual drop at the edge of the occultation is caused by the divergence of the 
refracted rays (Fig.~\ref{fig_dp_dz}).
Note that the stellar flux never drops to zero. It reaches a minimal value of about 7\% of its unocculted 
level due to the refraction of the stellar rays around Triton's limb.
The central flash stems from the focusing effect of Triton's limb curvature, 
caused by the factor $r/\mid z \mid$ in Eq.~\eqref{eq_phi_focusing}.
For Triton $r \approx 1360$~km, while at closest approach, 
the Calern station was at $\mid z \mid = 29$~km from the shadow center, 
causing a flux amplification of $r/\mid z \mid = 40$.
For Const\^ancia ($\mid z \mid \sim 8$~km at closest approach), 
this amplification factor was $r/\mid z \mid = 145$, causing a much higher flash.
Image adapted from \cite{marq22}.
%
}
\label{fig_Triton_05oct17}
\end{figure}

Using the nomenclature of Fig.~\ref{fig_dp_dz}, the combined effects of 
refraction, limb curvature and haze absorption 
results in a normalized stellar flux of
\begin{equation}
\phi = 
\left(  \frac{1}{\displaystyle 1 + \Delta \frac{d \omega}{d r}}  \right) \left( \frac{r}{\mid z \mid} \right) \exp(-\tau).
\label{eq_phi_focusing}
\end{equation}
The first factor in parentheses comes from the defocusing of the stellar rays 
caused by differential refraction, see Fig.~\ref{fig_dp_dz}.
The second factor stems from the flux amplification caused by the limb curvature
(assumed to be circular for simplicity) that focuses the stellar rays.
This causes a surge of flux for small $z$'s, a phenomenon known as central flash, see Fig.~\ref{fig_Triton_05oct17}.
The last factor describes the extinction caused by hazes of optical depth $\tau$ along the line of sight. 

The mathematical formalism involved in refractive occultations has been developed for a century or so, 
see the review by \cite{sica23}.
Using an Abel inversion, the observed light curve provides the refractivity profile 
$\nu(r)= n_{\rm r}(r)-1$, where $n_{\rm r}$ is the refraction index.
The $\nu(r)$ profile then provides the molecular density of the gas
through $n = K/\nu$, where $K$ is the molecular refractivity.
Finally, the ideal gas and the hydrostatic equations provide
the pressure $P(r)$ and temperature $T(r)$ profiles of the atmosphere,
\begin{equation}
    P(r) =  n(r) k_B T(r) {\rm ~~and~~}
\frac{1}{T} \frac{dT}{dr} =  
-\left[\frac{\mu g(r)}{k_B T} + \frac{1}{n} \frac{dn}{dr} \right],
\label{eq_ideal_gas_eq}
\end{equation}
where $k_B$, $\mu$ and $g$ are respectively 
the Boltzmann constant, the molecular weight and the acceleration of gravity.
Note that the differential equation in $T$ is of first order kind.
Consequently, it requires a boundary condition $T_0(r_0)$ at some prescribed radius $r_0$.
This introduces a further \bs{source} of uncertainty (besides the photometric noise in the light curve)
that cannot be solved, unless an independent knowledge of $T_0(r_0)$ is available,
for instance from a space mission. 
This initial condition problem is illustrated in Sect.~\ref{subsec_atmo_waves}.

%
%

Examples of occultation light curves and their ray-tracing modeling 
are displayed in Fig.~\ref{fig_Triton_05oct17}.
A simplified equation can be used to estimate the effect of an atmosphere during an occultation.
It assumes that the atmosphere has a constant scale height $H$ that is much smaller
than the radius, $H \ll r$. It was derived by \cite{baum53} and provides the flux $\phi$
observed at position $z$ in the shadow plane through an implicit equation, 
\begin{equation}
\displaystyle
\left( \frac{1}{\phi} - 2 \right) + \ln \left( \frac{1}{\phi} - 1 \right) = -\frac{z - z_{1/2}}{H},
\label{eq_BC}
\end{equation}
where $z_{1/2}$ is the location in the shadow where the stellar flux reaches half of its unocculted value,
known as the ``half-light level".
It can be shown (e.g. \citealt{sica23}) that the half-light levels $z_{1/2}$ in the shadow plane
and in the atmosphere are related by 
$$
z_{1/2} = r_{1/2} - H.
$$
In other words, the ``half-light rays" are deviated by one scale height before reaching the observer.

\subsection{Detecting tenuous atmospheres}

As $\phi$ approaches unity, i.e., when the atmosphere gets more and more tenuous, 
the Baum and Code equation provides
\begin{equation}
\phi \approx 1 -  \exp \left(-\frac{\Delta z}{H}\right),
\label{eq_phi_edge}
\end{equation}
where $\Delta z = z - z_{1/2}$. 
Thus, as $\Delta z$ increases, the flux $\phi$ approaches very rapidly (exponentially) its unocculted level. 
For instance, the equation above shows that the 1\% drop occurs at $\Delta z= -H \log (0.01) \approx 4.6H$. 
This means in practice that a light curve with a noise level of 1\% can detect the atmosphere 
up to but no more than about 4.6 scale heights above the half-light level $r_{1/2}$.
Assuming an isothermal atmosphere and $H \ll r$, the molecular density corresponding to 
the half-light level is given by \citep{sica23}
\begin{equation}
n_{1/2} = \frac{1}{K} \sqrt{\frac{H^3}{2\pi r_{1/2}\Delta^2}}.
\label{eq_n_half}
\end{equation}
Because of the large value of the geocentric distance $\Delta$, 
$n_{1/2}$ is very small, explaining the high sensitivity of Earth-based occultations.

At the edge of the occultation, $\phi \approx 1$ (Eq.~\eqref{eq_phi_edge}),
which corresponds to a molecular density of \citep{dias15}
\begin{equation}
n_{\rm top} = (1-\phi) n_{1/2}.
\label{eq_n_top}
\end{equation}
Let $\sigma_\phi$ be the noise level of the normalized light curve.
Then, the signature of the atmosphere is lost in the noise for $\phi >\sim 1 - \sigma_\phi$. 
According to Eq.~\eqref{eq_n_top}, this corresponds to a density level 
$n_{\rm top} \sim \sigma_\phi n_{1/2}$, i.e. a typical pressure level of
\begin{equation}
P_{\rm top} \sim \sigma_\phi P_{1/2}.
\label{eq_p_top}
\end{equation}

\begin{table}[ht]
\caption{Half-light pressure level $P_{1/2}$ probed during Earth-based stellar occultations.}
\label{tab_p_half}
\begin{tabular}{@{}lllll@{}}
\toprule
Object      & $r_{1/2}$ & $\Delta$ 	& $H$ 	& $P_{1/2}$     \\
	        & (km)      & \bs{(au)}	& (km) 	& ($\mu$bar)    \\
\midrule
Jupiter	    & 71840     &  4.2  & 25    &  1            \\
Saturn   	& 61000     &  8.5  & 55    &  2            \\ 
Uranus   	& 25900     & 18.2  & 65    &  2            \\
Neptune 	& 25100     & 29.0  & 50    &  1            \\

Titan       & 3070      &  8.5 & 45     & 3             \\

Triton      & 1440      & 29.0 & 25     & 0.2           \\
Pluto       & 1300      & 32.0 & 65     & 2             \\
\bottomrule
\end{tabular}
\footnotetext{%
The atmospheres of the giant planets are assumed to be composed of 90\% H$_2$ and 10\% H$_{\rm e}$, with 
molecular mass of $\mu= 0.37 \times 10^{-26}$~kg and
molecular refractivity $K= 0.479 \times 10^{-29}$ m$^3$ molecule$^{-1}$.
The atmospheres of Titan, Triton and Pluto are assumed to be composed of pure nitrogen N$_2$, with
$\mu= 4.7 \times 10^{-26}$~kg and
$K= 1.11 \times 10^{-29}$ m$^3$ molecule$^{-1}$.
}%
\end{table}
 
Using Eq.~\eqref{eq_n_half}, 
Table~\ref{tab_p_half} lists the typical half-light pressures $P_{1/2} = n_{1/2} k_{\rm B} T$ 
applicable to Earth-based occultations by various bodies. 
They all have typical values ranging from a fraction of to a few microbars.
For these pressures, the occultations cause a significant drop of the stellar flux.
However, pushing the method to its limit, Eq.~\eqref{eq_p_top} shows that fainter atmospheres can be detected.
For instance, for a notional noise level of 1\% ($\sigma_\phi = 0.01$), 
pressure levels as low as 10 nanobars can be reached by Earth-based occultations.

\begin{figure}[ht]
\centering
\includegraphics[width=1.\textwidth,trim=0 0 0 0]{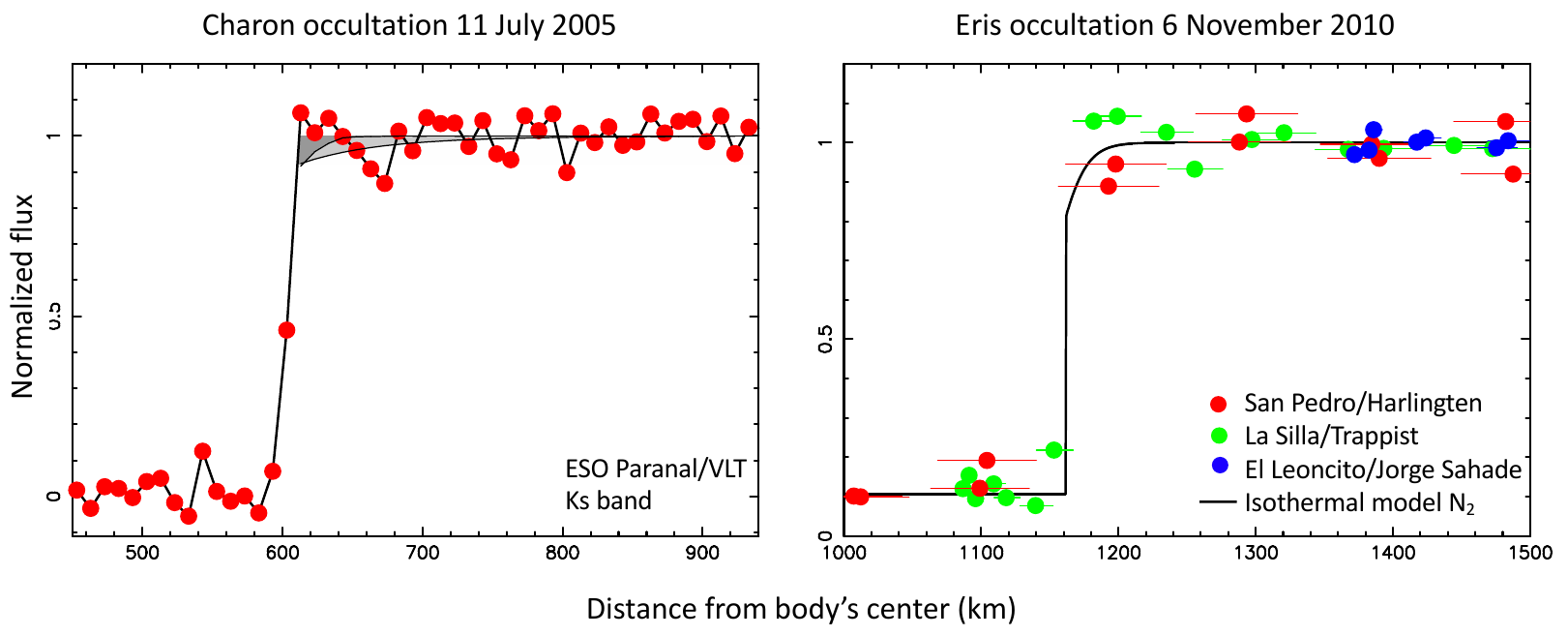}
\caption{%
Left panel -  
An occultation by Charon observed on 11 July 2005 \citep{sica06}.
The light curve has been re-plotted as a function of the projected distance to Charon's center, 
instead of time.
The lighter grey drop is a model using an isothermal N$_2$ atmosphere at 56~K. 
It marks the 3$\sigma$ upper limit of 110~nbar for the surface pressure.
The darker grey drop corresponds to the 3$\sigma$ upper limit of 15~nbar for a CH$_4$ atmosphere.
Right panel - 
The atmospheric limit for Eris from an occultation observed 
on 6 November 2010 from various stations in Argentina and Chile \citep{sica11}.
The horizontal bars correspond to the finite radial resolution associated 
with the finite integration intervals. 
The black solid line is the 3$\sigma$ upper limit of 
a pure isothermal N$_2$ atmosphere with surface pressure of 2.9 nbar.
}%
\label{fig_limit_atmo_Charon_Eris}
\end{figure}

Example of detection thresholds for Pluto's main satellite Charon and for 
the dwarf planet Eris are shown in Fig.~\ref{fig_limit_atmo_Charon_Eris}. 
These limits, together with those of other bodies, are listed in Table~\ref{tab_atmo_upper_limits}. 
They stay in an interval of several nanobars to a few microbars,
depending on the gas composition and the quality of the data.
This said, the SNR of the light curves shown 
in Fig.~\ref{fig_limit_atmo_Charon_Eris} and corresponding to the results of Table~\ref{tab_atmo_upper_limits}
can certainly be improved in the future. 
Thus, one may expect that the atmospheric limits will be pushed to the nanobar level in the near future.

\begin{table}[ht]
\caption{%
Upper limits of global atmospheres deduced from Earth-based stellar occultations
by Centaurs and TNOs.}
\label{tab_atmo_upper_limits}
\begin{tabular}{@{}lll@{}}
\toprule
Body            & Upper limit\footnotemark[1]  & Reference     \\
	              & (nbar)                       &               \\ 
\midrule
2002 TC$_{302}$ &  100 (N$_2$ or CH$_4$)                      & \cite{orti20b}\\
2002 TX$_{300}$ & 5000 (Xe)                                   & \cite{elli10} \\
2003 VS$_2$     &  200 (N$_2$)                                & \cite{bene19} \\
Charon	        &   40 (N$_2$), 5 (CH$_4$)                    & \cite{sica06} \\
Chiron	        & 5000-9000 (N$_2$, CH$_4$, CO, CO$_2$,       & \cite{sick23} \\
     	        &            H$_2$O or H$_2$)                 &               \\
Eris            &    1 (N$_2$, CH$_4$ or Ar)                  & \cite{sica11} \\
Haumea          &    5 (N$_2$), 17 (CH$_4$)                   & \cite{orti17} \\
Huya            &   10 (N$_2$)                                & \cite{sant22} \\

Ixion           & 80-600 (N$_2$, CH$_4$, CO, CO$_2$, H$_2$O,  & \cite{levn21} \\
                &         Ar, Kr, Ne or Xe)                   &               \\

Makemake        &   12 (N$_2$), 11 (CH$_4$)                   & \cite{orti12} \\
Quaoar          &   21 (CH$_4$)                               & \cite{brag13} \\
                & 1000 (N$_2$), 700 (CO), 138 (CH$_4$)        & \cite{fras13} \\     
                &    6 (CH$_4$)                               & \cite{arim19a} \\
                &   85 (CH$_4$)                               & \cite{morg22} \\
Titania\footnotemark[2] & 22 (N$_2$), 8 (CH$_4$), 17 (CO$_2$) & \cite{wide09} \\           
Umbriel\footnotemark[2] & 13 (CO$_2$)                         & \cite{assa23a} \\
\bottomrule
\end{tabular}
\footnotetext[1]{Limits are given at 1$\sigma$-level. For each limit the assumed composition is given in parentheses.}
\footnotetext[2]{Although not Centaurs nor TNOs, these objects are given for comparison.}
\end{table}

\subsection{Central flashes}

The term $r/\mid z \mid$ in Eq.~\eqref{eq_phi_focusing} stems from the focusing effect of a circular limb
and causes a strong surge of flux near the shadow center where $\mid z \mid \ll r$. 
In principle the flux goes to infinity for $z=0$. In practice, however, 
it is limited by the finite diameter of the star project at the distance of the body, 
$\theta_\star$ (Eq.~\eqref{eq_star_diam_km}). 
The flash then reaches a maximum height of $\phi_{\rm max} \sim 4(H/\theta_\star)$ \citep{sica23}.
Since $H$ is typically 20--50 km, while $\theta_\star$ is of the order of a kilometer,
$\phi_{\rm max}$ can still reach values of more than one hundred at the very center of the
shadow for a transparent and spherical atmosphere.

Examples of central flashes observed during a Triton occultation are given in Fig.~\ref{fig_Triton_05oct17}.
More than 40 stations distributed around the shadow center detected this flash,
that is caused by the atmospheric layer near the 8~km altitude \citep{marq22}.
All these flashes are consistent with a spherical and transparent atmosphere at this altitude, 
implying an oblateness of less than 0.0019 for that layer, i.e. a difference between its equatorial and
polar radii of less than 3~km.
This imposes equatorial zonal winds at 8~km altitude of less than 
46 m s$^{-1}$ and 80 m s$^{-1}$ for prograde and retrograde regimes, respectively (Ibid.).

Various central flashes caused by Pluto's atmosphere have been recorded in the last two decades. 
For instance, a central flash was detected in two bands (red and blue) during an occultation 
recorded on 31 July 2007. No difference were detected in the two bands, and from this observation,
\cite{olki14} conclude that the flash is consistent with a transparent atmosphere with 
a temperature gradient of 5~K~km$^{-1}$around 10~km above the surface. This is consistent with 
multi-wavelength occultation light curves analyzed by \cite{gulb15}. 
However, combinations of a thermal gradient and a haze mechanism are not excluded by these works.

Another flash observation on 29 June 2015 is also consistent with a transparent atmosphere,
from a ground-based observation \citep{sica16}. However, it was simultaneously observed at 
0.57, 0.65, 0.81 and 1.8 $\mu$m  from the Stratospheric Observatory for Infrared Astronomy airplane 
(SOFIA, see \citealt{pers21}). Although the flash itself does not show marked wavelength dependence, 
the residual stellar flux around does have some color dependence, 
suggesting the presence of hazes just above the flash layer, i.e. just above the 5-km altitude level.
This color dependence could be explained by haze particles 
with sizes between 0.06 and 0.1 $\mu$m \citep{pers21}.
Both the line-of-sight optical depth and the particle size derived from the SOFIA Earth-based 
observations are consistent with the measurements made by the New Horizons spacecraft 
during its flyby of July 2015 \citep{glad16,chen17}.

The shape of the Pluto flashes are generally consistent with a spherical Plutonian atmosphere. 
The flash of 31 July 2007 exhibits some structure, though, that could be explained by 
a prolate atmosphere with an ellipticity of 0.09 \citep{olki14}. 
However, this would imply a variation of altitude of more than 100~km of the flash layer, 
an unrealistic occurrence for a layer that lies at a few kilometers above the surface. 
Consequently, this structure is probably better explained by patchy haze layers along the limb, 
or by changing thermal gradient near Pluto's surface.

\subsection{Seasonal evolution of atmospheres}

%
As the vapor pressure depends very steeply upon temperature
(for instance an increase of one Kelvin cause \bs{a} doubling of the vapor pressure of 
N$_2$ near 30~K), these atmospheres are prone to strong seasonal effects 
when insolation, sub-solar latitude and heliocentric distances vary significantly.

The monitoring of Pluto's and Triton's atmospheres using 
Earth-based occultations over several decades has provided strong constraints
on seasonal changes in Pluto's and Triton's atmospheres.
Both objects suffer large variations of the sub-solar latitude over centuries,
from 57$^\circ$S to 57$^\circ$N for the former, and 50$^\circ$S to 50$^\circ$N for the latter.
In the case of Pluto, these variations are also combined with large variations 
of heliocentric distance, from 29.7~au at perihelion to 49.3~au at aphelion.

\begin{figure}[htbp]
\centering
\includegraphics[width=0.7\textwidth,trim=50 50 50 40]{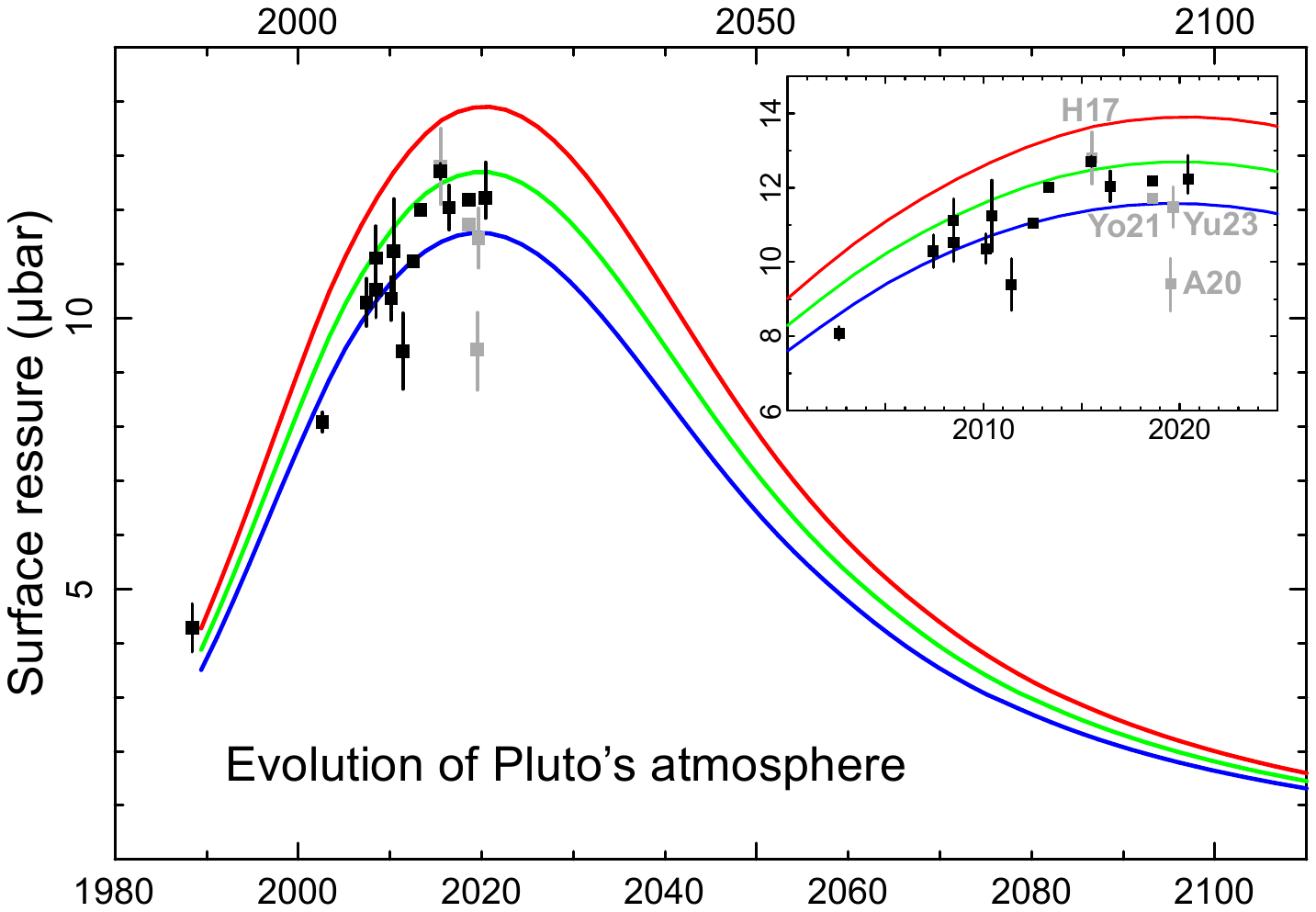}
\includegraphics[width=0.7\textwidth,trim=50 30 50 50]{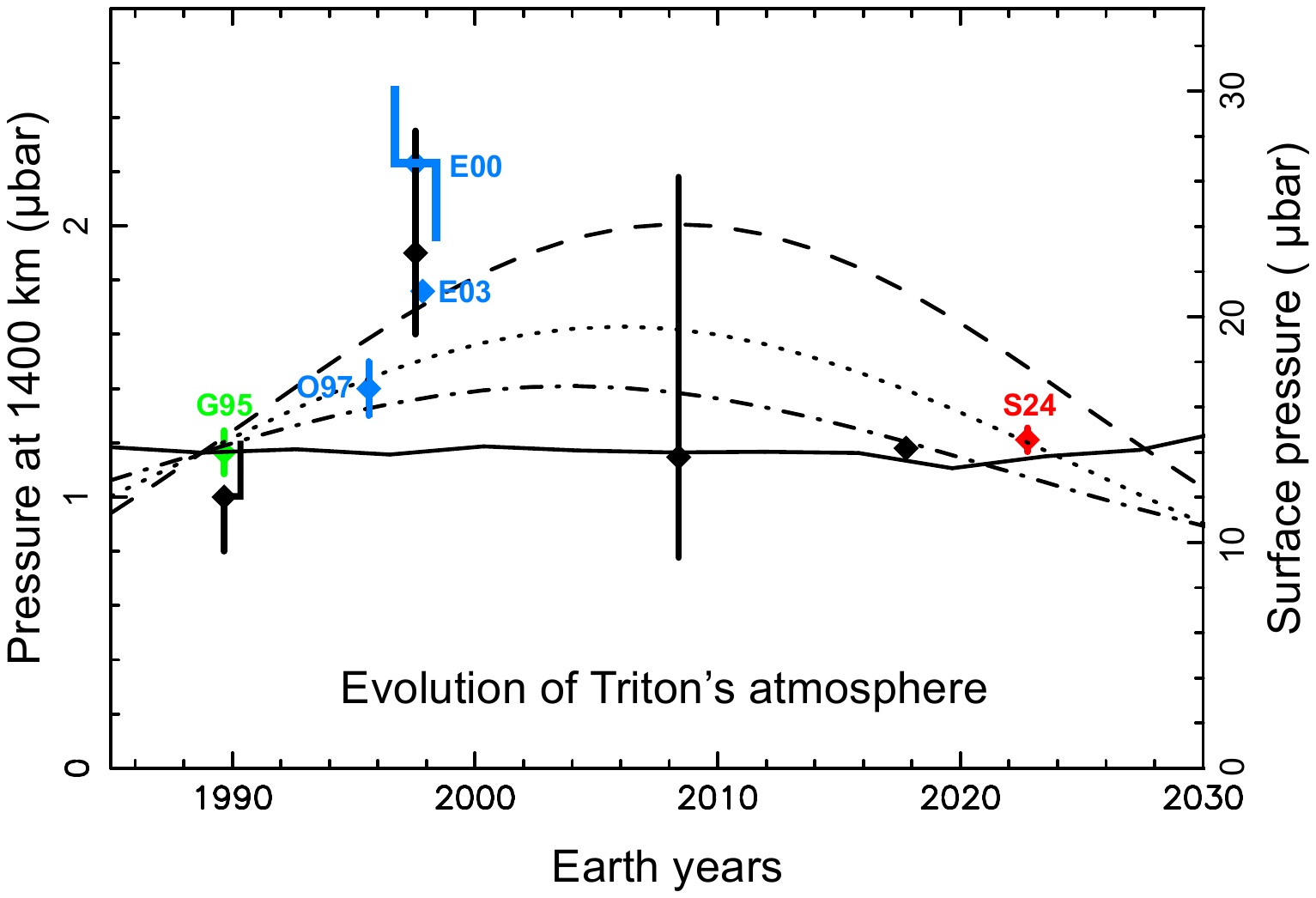}
\caption{%
Upper panel -
The evolution of Pluto's surface pressure between 1988 and 2020.
The black points are taken from \cite{meza19} and \cite{sica21a}.
The gray points are taken from Earth-based occultations using results from
Yo21 \citep{youn21} and Yu23 \citep{yuan23}.
The A20 point is from \cite{arim20}. It indicates a possible pressure drop. 
However, this remains to be confirmed since it is detected at a moderate level of 2.4$\sigma$ only.
The H17 point is from the radio occultation by the REX instrument on board 
of New Horizons spacecraft \citep{hins17}. 
The colored lines are evolution models based on the VTM described in \cite{meza19}. 
The blue, green, and red curves correspond to N$_2$ ice albedos 
of 0.73, 0.725, and 0.72, respectively.
Lower panel - 
The evolution of Triton's atmosphere between 1989 and 2022.
The left vertical axis gives the pressure at a reference radius of 1400~km (47~km altitude), 
while the right axis gives the pressure at Triton's surface.
The black points are taken from \cite{marq22}.
The blue points from O97 \cite{olki97}, E00 \citep{elli00a}, E03 \citep{elli03b}.
The red S24 point is from \cite{sica24}.
All points are from Earth-based observations, 
except for the G95 green point, obtained from the Voyager 2 radio experiment \citep{gurr95}.
Various VTMs of \cite{bert22} are shown as follows:
dashed-dotted line: 
southern cap within 90$^\circ$S-30$^\circ$S and northern cap within 45$^\circ$N-90$^\circ$N;
Dotted line: 
southern cap within 90$^\circ$S-30$^\circ$S, northern cap within 60$^\circ$N-90$^\circ$N;
Dashed line: 
southern cap within 90$^\circ$S-30$^\circ$S, no northern cap;
solid line: 
a simulation where the N$_2$ ice freely evolves over millions of years
in which local frosts deposits can form self-consistently.
All error bars are at 1$\sigma$-level.
}%
\label{fig_t_p_pluto_triton}
\end{figure}

Between an occultation observed near perihelion in 1988 \citep{yell97} 
and another one recorded in 2002, 
Pluto receded from the Sun by 3\%, meaning a drop of about 6\% in average insolation. 
Thus, the first-thought expectation was that the atmosphere should contract due to the ensuing temperature drop.
Observations showed the exact opposite, with a doubling of pressure between the two dates
\citep{elli03a,sica03}.
This trend was confirmed in the following years, using more occultation events
\citep{pasa05,elli07,youn08,bosh15,olki15,sica16}.
By 2020, the atmosphere reached a plateau value that represents a three-fold increase compared 
to its perihelion value, see \cite{meza19,sica21a,yuan23} and Fig.~\ref{fig_t_p_pluto_triton}.

In fact, models using global Volatile Transport Models (VTMs) did predict 
this seasonal effect, among other possible scenarios 
due to effects from changing sub-solar latitude
\citep{binz90,hans96}.
Subsequently, the New Horizons flyby revealed in July 2015 
the Plutonian topography and the N$_2$ ice distribution.
This triggered more accurate VTMs \citep{bert16,forg17,bert18}. 
In particular, the large Sputnik Planitia, a depression filled with N$_2$ ice,
appeared as the main ``engine" that controls the atmosphere 
as its insolation conditions vary.
Fig.~\ref{fig_t_p_pluto_triton} shows that the updated VTM basically 
captures the trend observed between 1988 and 2020. 
Two outliers points are seen in the insert of Fig.~\ref{fig_t_p_pluto_triton}.
One is from an occultation by a faint star recorded in 2011, which resulted 
in a very low contrast event. The other one (labeled ``A20") is from an
single-chord, grazing occultation observed in 2019 \citep{arim20}.
Both observations result in pressures that depart from the overall trend,
but at debatable levels of a bit more than $\sim$2$\sigma$.

The plateau observed since 2015 corresponds to the maximum value of pressure 
reached as Sputnik Planitia is illuminated under more and more grazing conditions, 
while Pluto is receding from the Sun.
A good test of the Pluto VTM presented in Fig.~\ref{fig_t_p_pluto_triton} 
would be the detection of a gradual decrease of pressure in the next decades,
something that Earth-based occultations will be able to provide.

The lower panel of Fig.~\ref{fig_t_p_pluto_triton} shows the same kind of analysis for Triton,
see \cite{sica24}.
Unfortunately, the Neptune system has been moving in front of depleted stellar
field since the 1990's, resulting in a scarce number of occultations.
Taken at face value, Fig.~\ref{fig_t_p_pluto_triton} shows that none of the proposed VTMs 
satisfactorily explains the observed trend.
Either the models account for the surge of pressure observed in the 1990's, 
but then overestimate the pressures in 2017 and 2017,
or they explain the basically constant pressures observed in 1989, 2017 and 2017,
but then fail to explain the surge in 1997.
However, some caution is called for regarding the increase observed in 1995 and 1997 \citep{sica24}.
While the black and red points in the lower panel of Fig.~\ref{fig_t_p_pluto_triton}
are obtained using the same methodology of \citealt{marq22}, 
the blue points come from other teams, and may not be fully consistent with this methodology.
In any instance, future occultations will be necessary to discriminate the Triton VTMs proposed in Fig.~\ref{fig_t_p_pluto_triton},
and in particular to see if the pressure remains constant in the forthcoming decades, 
or will show on the contrary the steady decrease predicted by some models.

\subsection{Atmospheric waves}
\label{subsec_atmo_waves}

Besides providing global information on atmospheres between pressure levels of 
a fraction of millibar to a few nanobars, Earth-based occultations may also reveal local fluctuations in the density, pressure and temperature profiles of these atmospheres.
These fluctuations are ubiquitous in occultations light curves, 
for instance in the case of the atmospheres of 
Jupiter (\citealt{rayn03}),
Uranus (\citealt{fren82}), 
Neptune (\citealt{roqu94}) and
Titan \citep{sica99}.
More recently, such fluctuations were also detected in Pluto's and Triton's atmospheres,
see below.

The common interpretation of these fluctuations is the presence of internal gravity waves, for which the restoring force is buoyancy. They usually have a strongly layered structure and long oscillation periods (tens of minutes to hours),  contrarily to acoustic modes, for which the restoring force is pressure.

Let us denote $k$ the wave number of the wave along the line of sight 
(i.e. essentially along the local horizontal direction) and 
$m$ its vertical wave number (perpendicular to the line of sight).
Then, if $\delta n$ is the density fluctuation,  
the rapid fluctuations of flux (or ``spikes") caused by $\delta n$ is \citep{sica99}
\begin{equation}
\frac{\delta \phi}{\phi} \sim (1-\phi)(mH)^{3/2} 
\exp \left[ -\frac{r}{4H} \left( \frac{k}{m} \right)^2 \right] 
\left( \frac{\delta n}{n} \right).
\label{eq_dphi_phi}
\end{equation}
The exponential factor represents the ``filtering" of the fluctuations along the
line of sight. If $k$ is very large, i.e. if there are many horizontal wavelengths 
along the ray trajectory, these fluctuations will be averaged out to zero from 
the observer's point of view, that is $\delta \phi/\phi$ will be very small.
For Pluto and Triton, 
we typically have $r \sim 1200$~km and $H \sim 25-65$~km (Table~\ref{tab_p_half}),
so that $r/H \gg 1$.
Consequently, Earth-based occultations will average out the horizontal fluctuations 
unless $k/m$ is very small, i.e. if $\lambda_{\rm h}/\lambda_{\rm v} \gg 1$, 
where $\lambda_{\rm h}= 2\pi/k$ (resp. $\lambda_{\rm v}= 2\pi/m$) is the
horizontal (resp. vertical) wavelength.

The equation~\ref{eq_dphi_phi} also shows that, from an observational point of view,
occultations favor waves with large $mH$'s ($\lambda_{\rm v} \ll H$), 
as they amplify the fluctuations $\delta n/n$.
Since the scale height $H$ is of the order of some tens of kilometers, 
occultations are well suited to detect waves with $\lambda_{\rm v}$ of the order of a few kilometers.
This said, waves with sub-km vertical wavelengths tend to become unnoticed due to the
smoothing effect of the stellar diameter (Fig.~\ref{fig_Delta_diam}).

Finally, Eq.~\eqref{eq_dphi_phi} shows that $\delta \phi \propto \phi(1-\phi)$. This means that the spikes are best seen near the half-light times ($\phi \sim 0.5$), and difficult to observe near the edge of the occultation ($\phi \approx 1$) and at the bottom of the occultation ($\phi \approx 0$). 

\begin{figure}[ht]
\centering
\includegraphics[width=0.9\textwidth,trim=0 0 0 0]{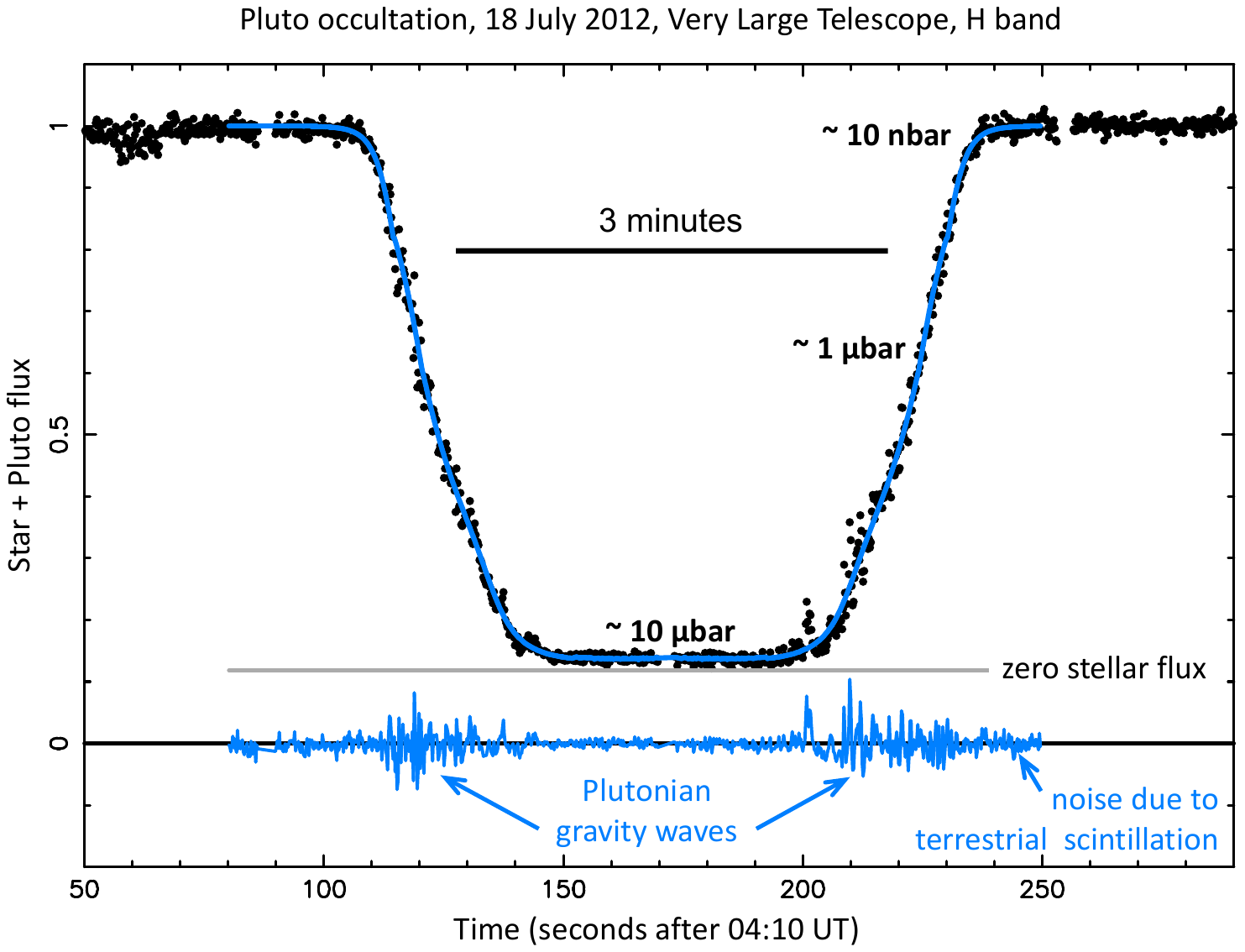}
\caption{%
The same as in Fig.~\ref{fig_Triton_05oct17}, 
but for an occultation by Pluto observed on 18 July 2012 in H-band from the 8.2-m 
Very large Telescope (VLT) of the European Southern Observatory (ESO) at Paranal.
Typical pressure levels are indicated along the occultation light curve.
Note that the residual to the fit (bottom blue curve) has significant fluctuations 
caused by gravity waves in Pluto's atmosphere, that exceeds the fluctuations caused
by the Earth atmosphere.
}
\label{fig_Pluto_18jul12}
\end{figure}

Occultations have revealed such spikes in both  Pluto's and Triton's light curves.
An illustration is given in Fig.~\ref{fig_Pluto_18jul12} in the case of Pluto, 
see also \cite{pasa05,youn08,macc08,pers08,hubb09,pasa17,silv22}.
%

As discussed above,
Earth-based occultations mainly detect waves around half-light levels,
favoring highly anisotropic stratified structures 
that have large aspect ratios $\lambda_{\rm h}/\lambda_{\rm v}$.
Such waves were already noticed in Uranus' atmosphere,
with aspect ratios well above 60 \citep{fren82}.
Similar aspect ratios (25 to 100) were also reported by \cite{nara88} in the case of Neptune,
while even larger aspect ratios -- up to 200 -- were reported in Titan's atmosphere \citep{sica99}.

Neglecting the Coriolis acceleration, the dispersion relation of gravity waves is
\begin{equation}
\omega_{\rm gw} = \left( \frac{k}{m} \right) \omega_{\rm BV},
\label{eq_disp_rela_gw}
\end{equation}
where $\omega_{\rm gw}$ is the frequency of the gravity wave and 
$\omega_{\rm BV}$ is the Brunt–V\"ais\"al\"a frequency
$$
\omega_{\rm BV}= \sqrt{\frac{(\gamma-1)g}{\gamma H}},
$$
$\gamma$ being here the ratio of 
the specific heat of the gas at constant pressure to
the specific heat at constant volume \citep{houg02}.

For Pluto and Triton, 
we have $g \sim 0.5$~m~s$^{-2}$, $H \sim 25-65$~km and $\gamma=1.4$ for N$_2$.
Thus, $\omega_{\rm BV} \sim 2 \times 10^{-3}$~s$^{-1}$,
corresponding to a period of almost one hour. Since $k \ll m$, Eq.~\eqref{eq_disp_rela_gw}
tells us that $\omega_{\rm gw} \ll \omega_{\rm BV}$,
so that the gravity waves have periods of many hours.
Consequently, as Pluto and Triton occultations last for a few minutes only,
they provide a ``frozen snapshot" of the gravity waves.

\begin{figure}[ht]
\centering
\includegraphics[width=0.9\textwidth,trim=0 0 0 0]{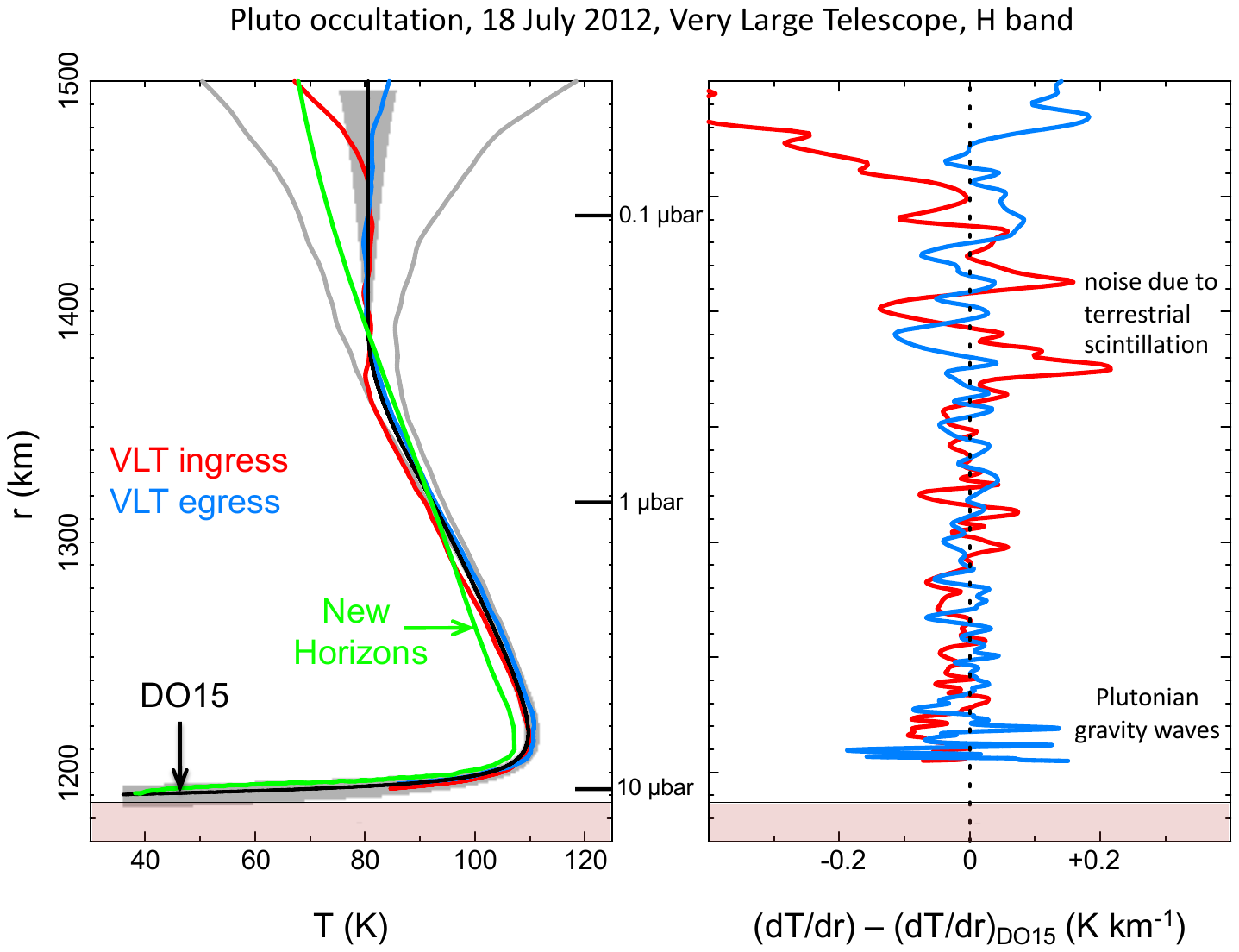}
\caption{%
Left panel - 
Red and blue curves: 
the thermal profiles $T(r)$ obtained from the Pluto occultation displayed in Fig.~\ref{fig_Pluto_18jul12}.
The dark gray areas mark the uncertainties due to the noise in the light curve.
Light gray profiles: 
possible solutions using other boundary conditions $T_0$ at some prescribed radius $r_0$,
see the dicussion after Eq.~\eqref{eq_ideal_gas_eq}.
Black profile ``DO15": 
a smooth thermal model by \cite{dias15} that fits the red and blue profiles.
Green profile:
the solution obtained from the New Horizons flyby \citep{ster18}.
Right panel - 
The temperature gradient $dT/dr$ corresponding to the red and blue profiles at left. 
The smooth gradient of the DO15 model, $(dT/dr)_{\rm DO15}$, 
has been subtracted to $dT/dr$ in order to better display 
the temperature fluctuations of Pluto's atmosphere.
While the fluctuations at the bottom of the profiles are real features associated with 
Plutonian waves, the fluctuations at the top are spurious effects caused by the terrestrial 
scintillation in the light curve (Fig.~\ref{fig_Pluto_18jul12}). 
The pale rose rectangles at the bottom of both panels indicate Pluto's surface.
}
\label{fig_T_dTdr_Pluto_18jul12}
\end{figure}


An example temperature profile for Pluto's atmosphere, obtained from the ground-based occultation of 18 July 2012 is displayed in Fig.~\ref{fig_T_dTdr_Pluto_18jul12}. It is compared with the profile obtained by the New Horizons instruments in 2015. It illustrates the excellent agreement between the two results and thus validates the Earth-based approach.
Above the radius 1400~km (altitude $\approx$200 km), the ground-based result suffers
from both an increasing sensitivity to noise and to the initial condition problem 
discussed after Eq.~\eqref{eq_ideal_gas_eq}.

The right panel of Fig.~\ref{fig_T_dTdr_Pluto_18jul12} displays the variations of the 
temperature vertical gradient $dT/r$, after subtracting the smooth contribution of
the general model ``DO15" of \cite{dias15}, 
revealing wavy fluctuations below a typical altitude of 100~km.
These structures have been interpreted as gravity waves excited by diurnal 
sublimation on N$_2$ frost patches that behave as ``actuators" to push up or pull down
parcels of air above Pluto's surface. The response of the atmosphere is then a system of
gravity waves with high aspect ratio $m/k$ propagating upward \citep{toig10} and
preferentially excited near Pluto's equator \citep{fren15}.
Another potential source of Pluto's gravity waves are stationary orographic waves forced
by winds flowing above topographic features (4 km-high mountains), 
based on images returned by the New Horizons spacecraft \cite{glad16,chen17}.

An earlier Pluto occultation observed on 18 March 2008 did in fact reveal waves 
in Pluto's atmosphere at higher altitudes (typically 100-200~km) 
showing a cutoff of the high wave number $m$ with increasing altitude,
consistent with viscous-thermal dissipation of gravity waves \citep{macc08,hubb09}.
An alternative interpretation of these features is not unique, though, as they can also be 
attributed to larger scale Rossby planetary waves \citep{pers08}.

The New Horizons missions actually confirmed the existence of high aspect ratio gravity waves 
in the 0--100~km altitude range, through the imaging of haze layers
that act as tracers of the waves \citep{glad16,chen17}.
%
The general picture that now emerges is that of Plutonian gravity waves
with smaller vertical wavelengths (typically 10~km) at lower altitudes (typically 0--150~km), and
larger wavelengths (some 20~km) at higher altitudes of 200--400~km, 
consistent with the expected effect of viscous-thermal dissipation.

Waves in Triton's atmosphere are more elusive. Some occultations do reveal spikes in the light curves \citep{elli00b}. However, Triton occultations remain scarce, and  none of the observations obtained so far have high enough timing resolution \bs{to permit} the detection of Pluto's atmospheric waves. 

\section{Rings}
\label{sec_rings}

Ground-based stellar occultations have proved to be a powerful tool 
for discovering and characterizing planetary rings.
More precisely, they are well suited to detect dense rings (optical depths $\tau \gtrsim 0.01$), 
whose physics is mainly controlled by inter-particle collisions.
Conversely, ground-based occultations have difficulties detecting tenuous rings ($\tau \lesssim 10^{-3}$)  
that are best revealed through direct imaging, especially in forward scattering geometries from spacecraft. 
These kinds of rings are mainly controlled by non-gravitational processes such as radiation pressure
or electromagnetic forces.

As of today, out of the eight dense ring systems known in the solar system, 
six have been discovered using ground-based stellar occultations.
After the Uranus ring discovery in 1977 \citep{elli77,mill77,bhat77},
Neptune's incomplete ring (arcs) were detected in 1984 \citep{hubb86,sica91}.
More recently, occultations have revealed the unanticipated presence of rings around
the Centaur object Chariklo in 2013 \citep{brag14}, 
then around the dwarf planet Haumea in 2017 \citep{orti17} and more recently,
around the large TNO Quaoar in 2018--2022 \citep{morg23,pere23}.
Dense material has also been detected around the Centaur Chiron during an occultation observed since 1995. 
Its nature is still debated, but it could be a ring system embedded in a cometary shell or a disk that evolves over timescales of a few years, see the discussion in Sect.~\ref{subsec_chiron}.

Although more than 60 Centaurs or TNOs have been explored using occultations, no more than 20 of them
have been scanned with sufficient quality to reveal narrow and dense rings. 
Thus, having detected four systems with rings or dense material in this modest sample 
means that 20\% of small bodies in the outer solar system might host rings,
announcing many new ring discoveries in the near future.

The main asset of Earth-based occultations is their high spatial resolution, down to sub-kilometric levels.
As such, they provide radial optical depth profiles of the rings, they reveal the sharp edges of some of them and through repeated campaigns, yield their orbital elements and reveal possible free or forced oscillation modes.
An illustration of the accuracy of this method is provided by Uranus' rings, 
whose keplerian elements are retrieved through occultations with accuracies of about 0.1~km, 
displaying numerous free or forced oscillation modes, 
see the overviews by \cite{fren23} and \cite{fren24}, and references therein.

Another asset of occultations is that they reveal structures that cannot be imaged directly
from Earth, even with state-of-the-art instruments. For instance, Chariklo's rings lie at less
than 40 mas from the central body and they are much fainter than the central object as seen from Earth. This precluded any usable imaging, even with space instruments like JWST. 
Only next-generation telescopes like the European Extremely Large Telescope (E-ELT) 
may break this limit and offer an opportunity to image rings around small bodies.

\begin{table}[!ht]
\caption{Physical properties of rings around small bodies.}
\label{tab_rings}
\begin{tabular}{@{}llllll@{}}
\toprule
Ring & Radius & Radial width $W_{\rm r}$       & Normal optical & Eccentricity $e$ & Pole (J2000) \\
     &  (km)  & (Equivalent width $E_{\rm p}$) & depth $\tau_{\rm N}$         &    & RA  (deg.)   \\  
     &        & (km)                           &                &                  & DEC (deg.)   \\ 
\midrule
\multicolumn{6}{c}{Chariklo\footnotemark[1]} \\
\midrule
C1R & 385.9$\pm$0.4 & Variable from & Average: 0.4  &  0.005$\leq$$e$$\leq$0.022 & 151.03$\pm$0.14  \\
    &               & 4.8 to 9.1    & Opaque edges  &                            &  +41.81$\pm$0.07 \\
    &               & ($\sim 2$)    &               &                            &                  \\
C2R & 399.8$\pm$0.6 & Unconstrained & Unconstrained & $e \leq 0.017$             & 150.91$\pm$0.22  \\
    &               & from 0.1 to 1 & from $+\infty$ to 0.1 &                    &  +41.60$\pm$0.12 \\
    &               & ($\sim 0.2$)  &               &                            &                  \\
\midrule
\multicolumn{6}{l}{%
Visible reflectivity and surface composition\footnotemark[2]: $(I/F)_{\rm V}=0.07$, 20\% water ice, 40\%--70\% silicates,
} \\
\multicolumn{6}{l}{%
10\%--30\% tholins, some amorphous carbon
} \\
\midrule
\multicolumn{6}{c}{Haumea\footnotemark[3]} \\
\midrule
H1R & 2287$^{+75}_{-45}$ & $\sim$70      & $\sim$0.1 &  Assumed  & 285.1$\pm$0.5 \\
    &                    & ($\sim 7$)    &           &  circular & -10.6$\pm$1.2 \\
\midrule
\multicolumn{6}{c}{Quaoar\footnotemark[4]} \\
\midrule
Q1R     & 4057$\pm$6  & Dense\footnotemark[5]: 5--6 (1.7--2.0) & 0.20--0.25  & Assumed  & 259.82$\pm$0.23  \\
        &             & Faint: 40--300 (0.4--0.7) & 0.005--0.02              & circular & +53.45$\pm$0.30  \\
Q2R     & 2520$\pm$20 & 10--15 (0.05--0.1)        & 0.005--0.01              & Assumed  & Assumed          \\
        &             &                         &                         & circular & coplanar to Q1R  \\
\midrule
\multicolumn{6}{c}{Chiron} \\
\midrule
\multicolumn{6}{l}{The 29 November  2011 occultation is consistent with two narrow rings  of widths 2--3.5~km each,} \\
\multicolumn{6}{l}{separated by a gap of 9~km, centered around the radius 305~km, with normal optical depth} \\
\multicolumn{6}{l}{of 0.2--0.4 \citep{sick20,brag23}. The 15 December 2022 occultation is} \\
\multicolumn{6}{l}{consistent with the material near 325~km, but also shows narrowly confined material near 423~km} \\
\multicolumn{6}{l}{and a broad $\sim$580 km disk around Chiron \citep{orti23} with J2000 pole RA=$160 \pm 10$~deg} \\
\multicolumn{6}{l}{and DEC=$28 \pm 10$~deg. No consistent permanent ring system can explain simultaneously the} \\
\multicolumn{6}{l}{occultations of 1994, 2011, 2018 and 2022} \\
\bottomrule

\end{tabular}
\footnotetext[1]{\cite{bera17} and \cite{morg21}.}
\footnotetext[2]{Derived from spectroscopic variations of the Chariklo system over time \citep{duff14}.}
\footnotetext[3]{\cite{orti17}.}
\footnotetext[4]{\cite{morg23} and \cite{pere23}.}
\footnotetext[5]{The dense part of Q1R is resolved and has a Lorentzian profile. The width given here is the 
full width at half maximum and the optical depth corresponds to the peak value at the deepest
point of the profile \citep{pere23}. 
The faint part of Q1R has a profile compatible with a square well model \citep{morg23,pere23}.}
%
\end{table}

The Table~\ref{tab_rings} provides the typical values of the width $W_{\rm r}$, normal optical depth $\tau_{\rm N}$ and equivalent width  $E_{\rm p}= W_{\rm r} p_{\rm N}$ of the rings detected around Chariklo, Haumea and Quaoar (see Appendix~\ref{app_physics_ring_occ} for more details on how these quantities are obtained).
This table also provides a summary of our current knowledge of the dense material detected around Chiron.
These four systems are now described in turn.


\subsection{Chariklo's rings}

\begin{figure}[ht]
\centering
\includegraphics[width=0.48\textwidth,trim=0 0 0 0]{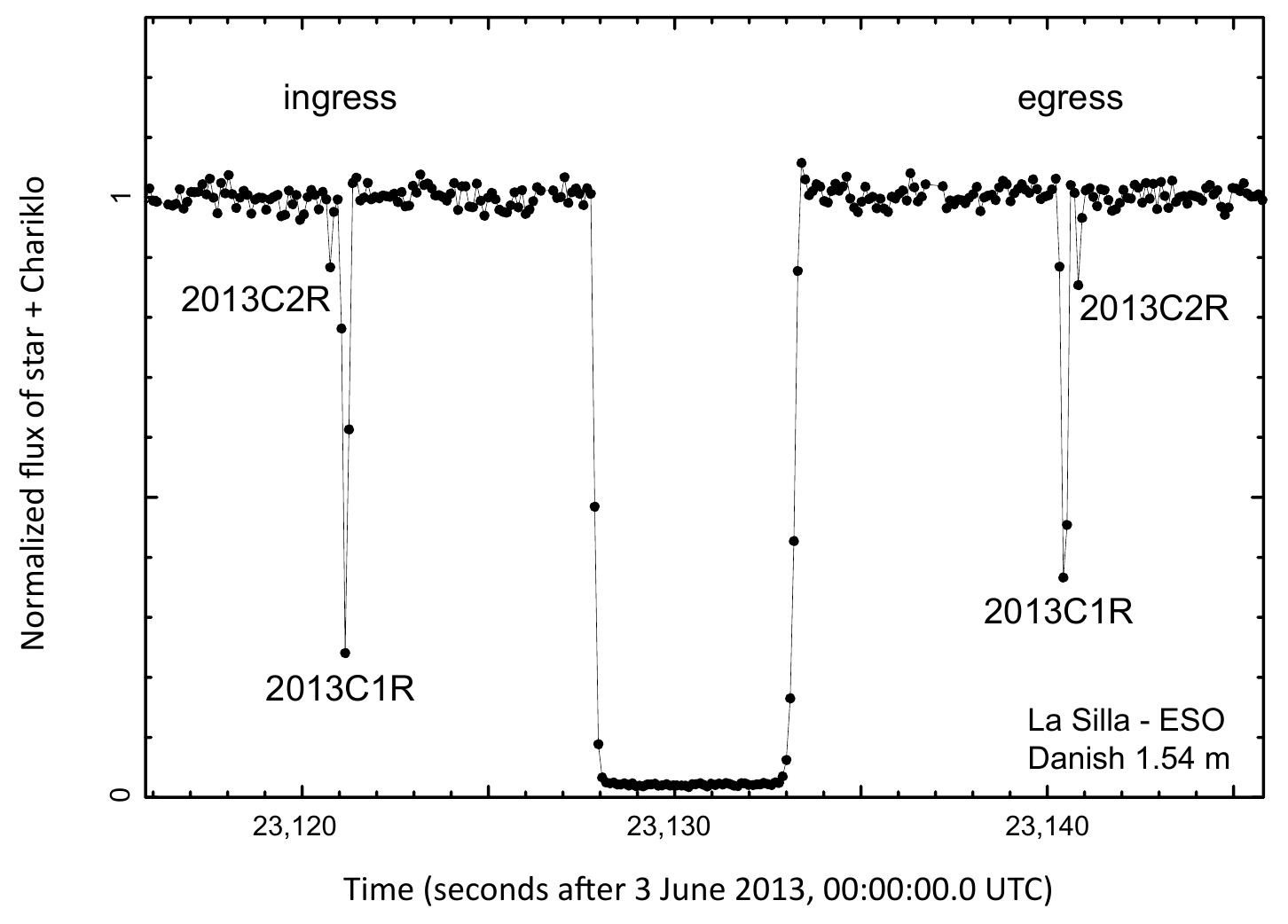}
\includegraphics[width=0.48\textwidth,trim=0 0 0 0]{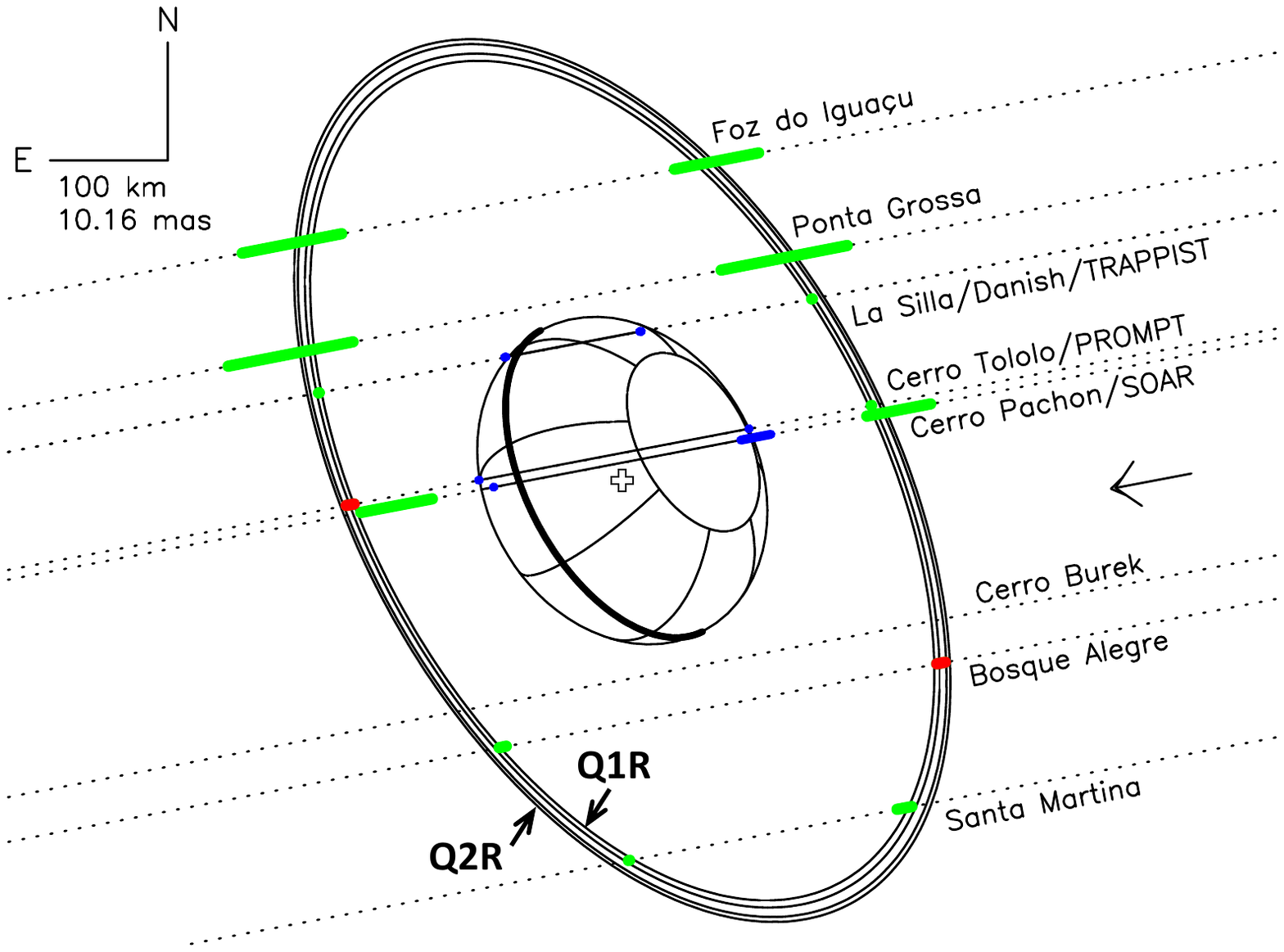}
\caption{%
Left panel - The occultation light curve recorded at the 1.54-m Danish telescope at La Silla (Chile)
during the 3 June 2013 occultation by Chariklo. Besides the disappearance of the star 
caused by the body seen at the center of the plot, two partial drops of stellar flux are detected
symmetrically around the body, revealing two semi-transparent narrow rings.
Right panel - The geometry of the rings reconstructed from the detections of 
secondary occultation events at various stations.
Each colored segment indicates the position of ring material as seen from these stations,
their lengths indicating the error bar on the location.
The black arrow at the right shows the motion of the star relative to the Chariklo system.
Adapted from \cite{brag14}.
}%
\label{fig_Chariklo_Danish_2013}
\end{figure}

Chariklo is the largest Centaur object currently known, with a volume-equivalent radius
of about 130~km \citep{leiv17}. It orbits between Saturn and Uranus, at heliocentric 
distances varying from 13.1~au to 18.7~au.
Its rings were discovered during a stellar occultation observed from South America on 3 June 2013. 
Besides the drops caused by the body itself, various sites in Argentina, Brazil and Chile 
detected secondary events in the form of partial stellar drops.
When combined, they were interpreted as caused by two narrow and dense rings,
2013C1R and 2013C2R (C1R and C2R for short) surrounding Chariklo, 
see \cite{brag14} and Fig.~\ref{fig_Chariklo_Danish_2013}.

At the time of ring discovery, Chariklo was moving in front of the Galactic Center.
Consequently, many occultations were planned, leading to about twenty campaigns between 2014 and 2022. 
These campaigns involved both amateur and professional instruments, including an observation
by the James Webb Space Telescope (JWST) in October 2022.
They confirmed the existence of C1R and C2R and refined various physical properties of the rings.
These also constrained Chariklo's shape, see \cite{bera17,leiv17,sica18,morg21} 
and Table~\ref{tab_dyn_param_rings}.

The occultations observed so far are consistent with two circular and co-planar 
rings with a fixed pole, orbiting near 386 and 400~km from Chariklo's center, respectively (Table~\ref{tab_rings}).
This is well outside the synchronous orbit at $\sim$200~km (where the orbital period of particles would match Chariklo's rotation), and a bit outside the estimated \gre Classical \bla Roche limit of the body, see the discussion by \cite{hedm23}.
The radial optical depth profile of the main ring C1R has been resolved on several occasions.
It displays a W-shaped structure reminiscent of the $\epsilon$ ring of Uranus.
Its edges are essentially opaque and very abrupt, and they remain unresolved 
down to the resolving power of the observations, about one kilometer.
Another property of C1R shared with Uranus' $\epsilon$ ring is the variation 
of its radial width $W_{\rm r}$, with values ranging from 4.8 to 9.1~km \citep{bera17,morg21}.

The equivalent width $E_{\rm p}$ of C1R is typically 2~km. 
This quantity essentially measures the width of an opaque ring that would block the same 
integrated flux as the actual ring, and as such, is a measure of the total amount of material 
along a particular radial scan of the ring.
The more tenuous C2R ring remains unresolved, and has $E_{\rm p} \sim 0.2$~km,
meaning that it contains about tens times less material than C1R.
The $\sim$10~km gap between C1R and C2R appears void of material,
down an upper limit of about 0.006 for its normal optical depth \citep{morg21}.

\subsection{Haumea's ring}

\begin{figure}[ht]
\centering
\includegraphics[width=0.63\textwidth,trim=0 0 0 0]{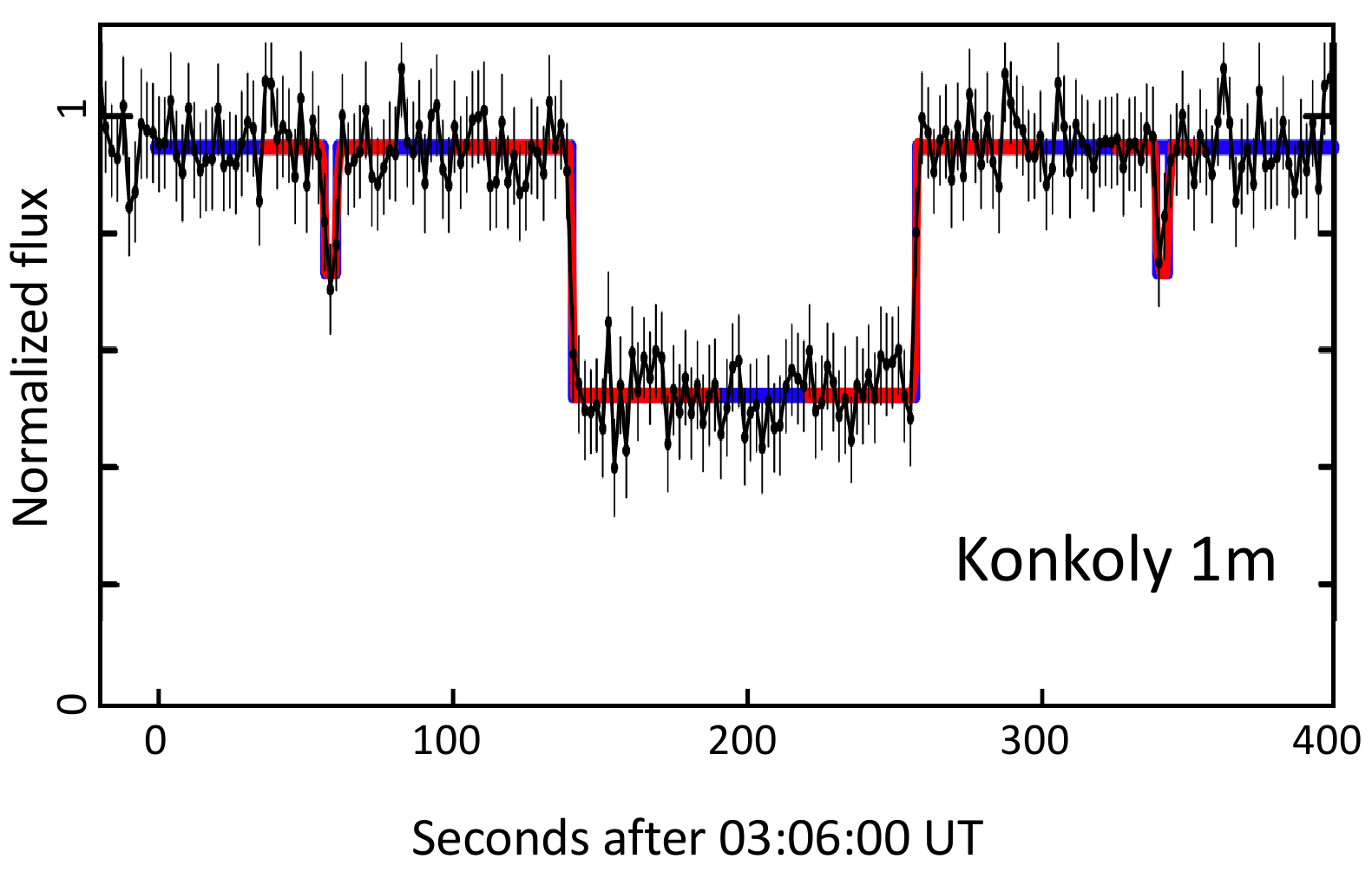}
\includegraphics[width=0.35\textwidth,trim=0 0 0 0]{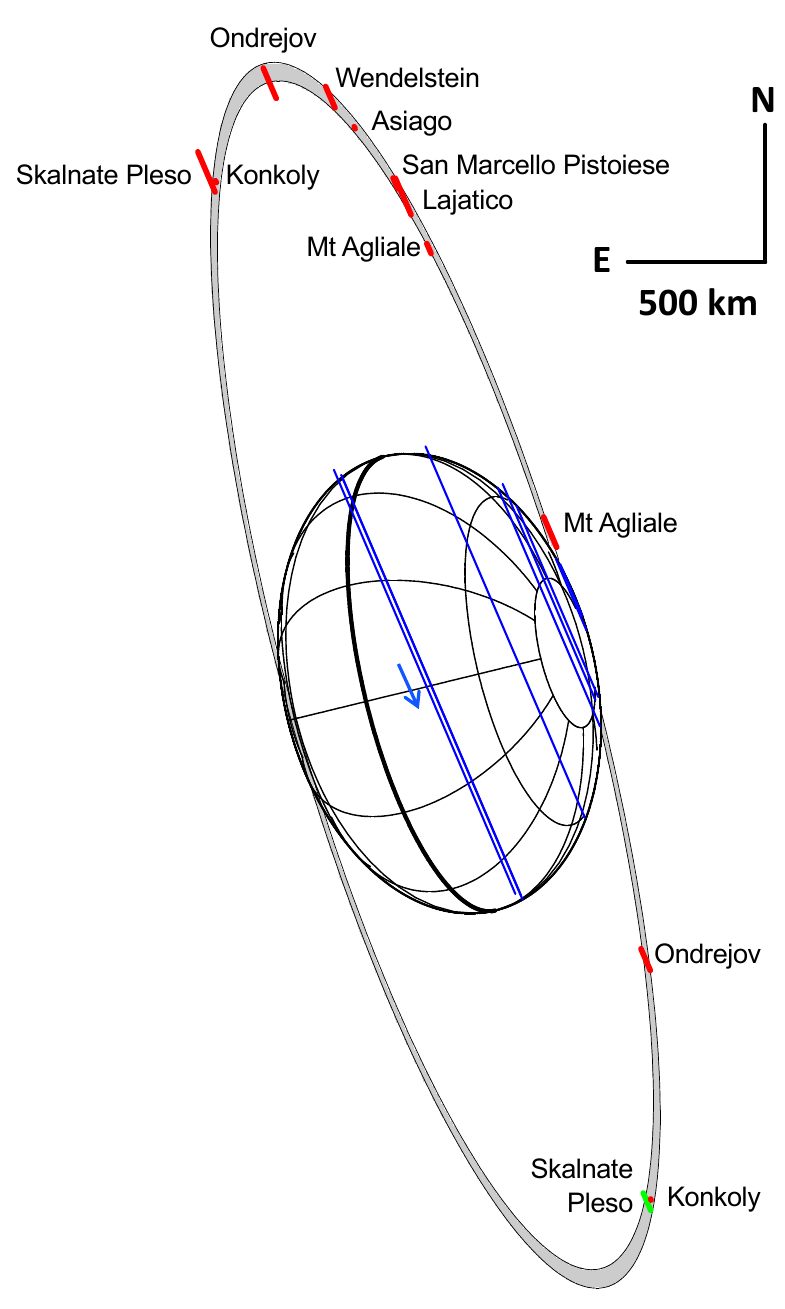}
\caption{%
Left panel - 
One of the light curves showing the presence of a ring around the dwarf planet Haumea 
during the 21 January 2017 occultation, as observed from the 1m telescope at the Konkoly site (Hungary).
The ring caused two partial drops of the stellar flux left and right of
the occultation by the main body at the center of the plot.
Right panel - 
The ring geometry and width retrieved from the various secondary events, 
whose locations are shown as small green or red segments.
The blue segments are the occultation chords by the main body,
from which the apparent limb of Haumea was derived (Table~\ref{tab_dyn_param_rings}).
The gray arrow shows the motion of the star relative to the Haumea system.
Images adapted from \cite{orti17}.
}%
\label{fig_Haumea_konkoly_2017}
\end{figure}

Haumea is a large TNO classified as a dwarf planet that moves from 34.6~au to 51.6 ~au from the Sun. 
It is a fast rotator with a rotation period of 3.9~h \citep{lell10}.
As of today (May 2024), only one occultation by Haumea has been recorded, on 21 January 2017.
This multi-chord event was observed from various countries in Europe. 
It provided an accurate shape for the body, a very elongated triaxial ellipsoid 
(\citealt{orti17} and Table~\ref{tab_dyn_param_rings}), a point discussed later.
As for Chariklo, secondary events revealed the presence of a narrow and
dense ring of width $\approx 70$~km orbiting at $2287^{+75}_{45}$~km from Haumea's center, 
see Table~\ref{tab_rings} and Fig.~\ref{fig_Haumea_konkoly_2017} and details in \cite{orti17}.
Again this is well outside the synchronous orbit, near 1400~km, and a bit outside the Roche limit \citep{hedm23}.

To be in line with the nomenclature used for Chariklo's and Quaoar's rings, we refer to it as H1R hereafter, 
although this name has not been used in previous publications. 
Assuming a circular ring, 
the apparent orientation and aspect angle of H1R provide its pole position. 
This position coincides with the orbital pole of Haumea’s main satellite 
Hi'iaka to within $2 \pm 1.5$ degrees.
Similarly, the position angle of the ring projected ellipse coincides with
the orientation of Haumea's limb to within $2 \pm 2$ degrees.
\citep{orti17}.
This strongly suggests a coplanar system with a circular H1R ring and a Hi’iaka's orbit
both contained in Haumea's equatorial plane. 

\subsection{Quaoar's rings}

\begin{figure}[ht]
\centering
\includegraphics[width=1.\textwidth,trim=0 0 0 0]{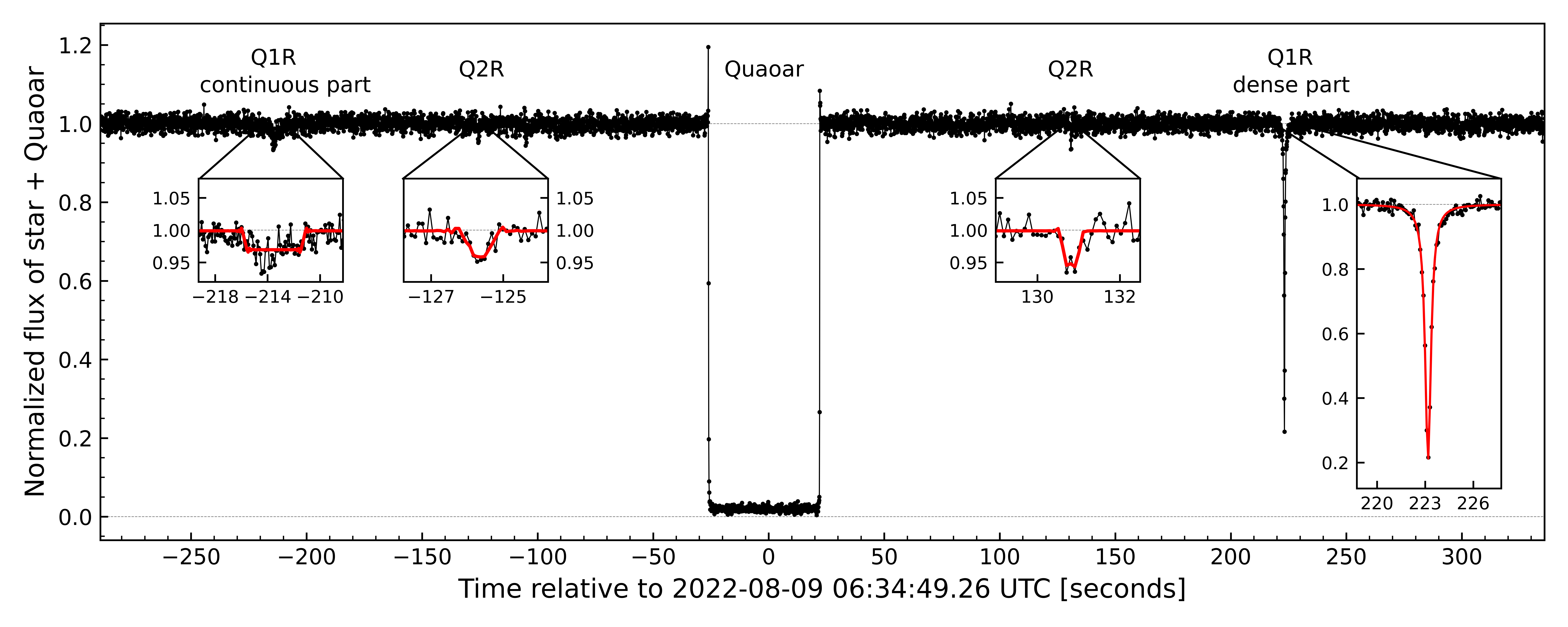}
\caption{%
Detection of the two rings (Q1R and Q2R) of Quaoar and the main body during the
9 August 2022 occultation observed from the 8.1m Gemini telescope at Mauna Kea (Hawaii) in the
z' band (947 nm).
The inserts show details of the profiles of each detection. The red curves are fits to the data,
using square well models for Q1R (ingress) and both Q2R detections and a Lorentzian model for
the dense part of Q1R at egress.
The spikes observed at the beginning and end of the occultation by Quaoar are
caused the Fresnel diffraction fringes due to the sharp, opaque edge of the body.
Image reproduced with permission from \cite{pere23}, copyright be the author(s).
}%
\label{fig_Quaoar_Gemini_z_2022}
\end{figure}

Quaoar is a large TNO with an area-equivalent radius of about 555~km \citep{brag13},
which makes it a candidate for being a dwarf planet.
It moves between 41.8~au and 44.8~au from the Sun and has a rotation period of 8.8 or 17.7~h, 
the latter value being preferred \citep{orti03}.
Its main ring Q1R was discovered by combining data from both professional and amateur teams obtained between 2018 and 2021, including data from space obtained by the ESA CHaracterising ExOPlanet Satellite (CHEOPS), see \cite{morg23}.
The more tenuous ring Q2R was seen during an occultation recorded at large telescopes in Hawaii in 2022, see below and \cite{pere23}.

Some physical parameters of Q1R and Q2R are listed in Table~\ref{tab_rings}. 
The main ring Q1R is unusual in two aspects.
First, it orbits at more than seven Quaoar's radii, which is well beyond the Classical Roche limit, 
usually estimated to lie at some 2.5-3 planetary radii. 
Second, Q1R is highly inhomogeneous, with a faint continuous part with $\tau_{\rm N} \sim 0.005-0.02$ 
and a dense part with $\tau_{\rm N} \sim 0.20-0.25$ 
(Table~\ref{tab_rings} and Fig.~\ref{fig_Quaoar_Gemini_z_2022}).  
The statistics of detections of this dense part indicate that it covers 
somewhere between 20 to 70 degrees of the entire orbit. 
As such, \bs{it is reminiscent of the neptunian Adams ring} that hosts a system of several dense arcs with $\tau_{\rm N} \sim 0.1$ spanning about 50 degrees in longitude, the rest being occupied by a faint continuous component with $\tau_{\rm N} \sim 0.003$ \citep{depa18}.

The ring Q2R has been detected only once, using large instruments at the Mauna Kea site in Hawaii
during an occultation on 9 August 2022 \citep{pere23}. 
The single chord obtained on Q2R during this event does not provide a unique orbital solution for this ring. 
However, this chord is consistent with a circular ring centered on Quaoar and coplanar to Q1R.
Its radius of 2520~km (Table~\ref{tab_rings}) places it at about 4.3 Quaoar's radii, 
i.e. well outside the classical Roche limit of the central body, like Q1R.

\begin{figure}[ht]
\centering
\includegraphics[width=0.9\textwidth,trim=0 0 0 0]{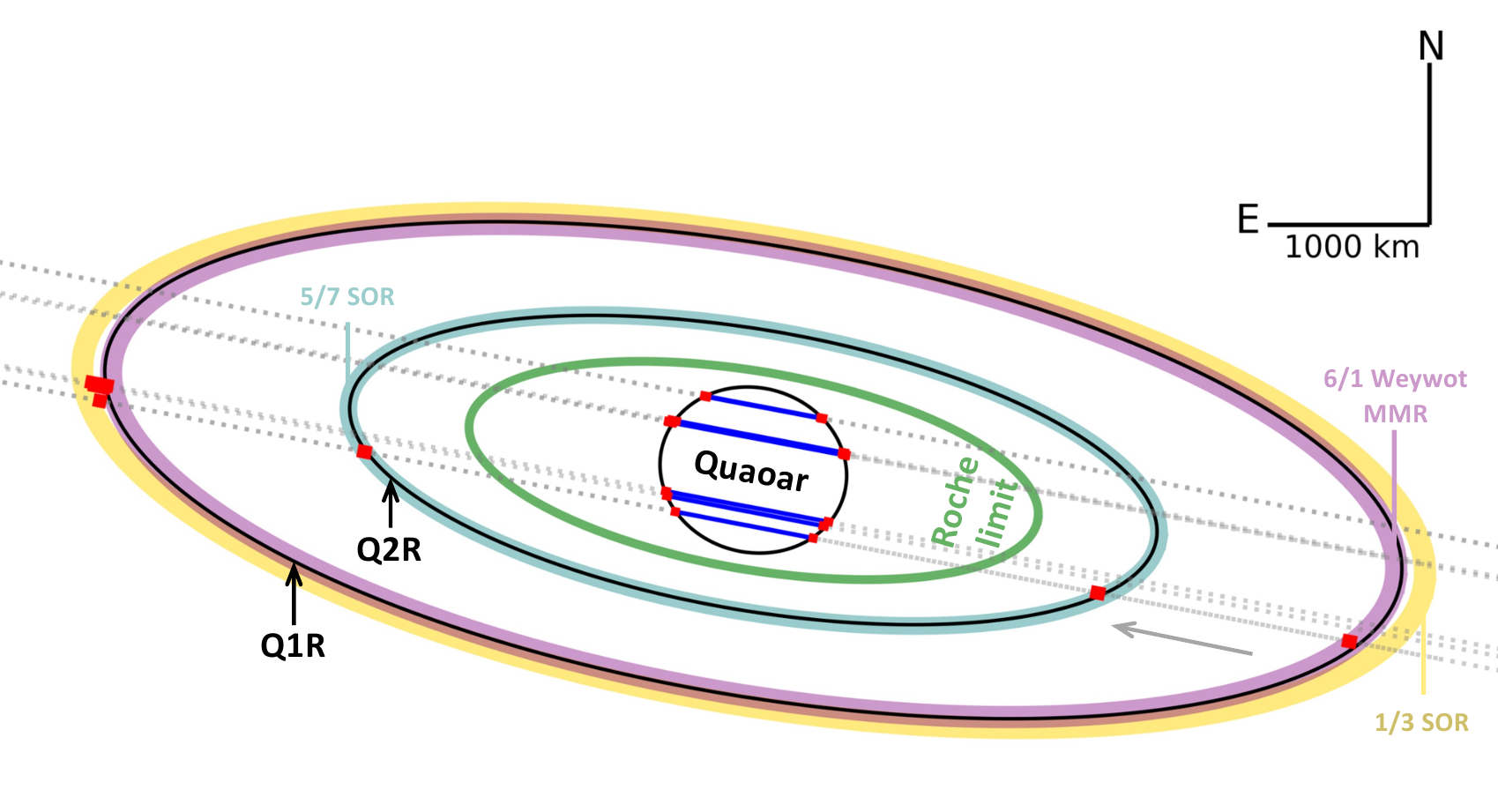}
\caption{%
The reconstructed geometries of Quaoar's rings (the black ellipses labeled QR1 and QR2) derived 
from the 9 August 2022 occultation, where the ring detections are shown as red squares.
%
The gray arrow gives the direction of the star relative to Quaoar along the occultation chords
(gray dotted lines).
The central ellipse is a fit to the blue chords obtained during this event, 
showing Quaoar's oblateness.
Both rings are outside the classical Roche limit (green ellipse) for particles with
bulk density $\rho=400$~kg~m$^{-3}$.
The inner ring is close to 
the 5/7 Spin-Orbit Resonance (SOR) with Quaoar (teal ellipse),
while the outer ring is close to both 
the 1/3 SOR with Quaoar (yellow ellipse) and 
the 6/1 Mean Motion Resonance (MMR) with Weywot (purple ellipse).
Image adapted from \cite{pere23}.
}%
\label{fig_Quaoar_two_rings}
\end{figure}

\subsection{Material around Chiron} \label{subsec_chiron}

Chiron is one of the largest Centaurs after Chariklo, with a volume-equivalent radius 
of about 100~km \citep{brag23}, versus 130~km for Chariklo.

Contrarily to Chariklo, Chiron has sporadic episodes of cometary activity 
that last for several months, separated by quiescent periods. 
%
During the outbursts, images show a tenuous coma extending thousands of kilometers away from Chiron. 
Although a dense inner coma was suspected near the body on Hubble Space Telescope (HST) images \citep{meec97}, 
it is currently impossible to resolve structures inside 
the first few hundred kilometers surrounding Chiron using classical imaging. 

Due to the depleted stellar fields that Chiron has crossed, 
very few stellar occultations have been observed in the last three decades.
However, they have revealed the presence of dense material 
inside a region of typical radius $\sim$500~km.
The first of these events was detected on 9 March 1994 by the \bs{Kuiper Airborne Observatory} (KAO). 
No occultation by the main body was observed then, but a sharp drop of the stellar flux followed by 
a smoother and shallower dip was recorded at distances of 100-200~km from Chiron \citep{elli95}. 
They were interpreted as cometary jet-like features originating from discrete active spots of Chiron's surface.

On 29 November 2011 event, two sharp secondary drops were recorded close to each other
on each side of the main body, resembling the sharp drops caused
by the Q1R and Q2R rings of Chariklo \citep{rupr15}.
These drops were interpreted either as arcs or shells by \cite{rupr15}, 
or as rings similar to Chariklo's near a radius of some 300--320~km 
(\citealt{orti15,sick20} and Table.~\ref{tab_rings}), 
with the possible presence of incomplete features or a shell from 900 to 1500~km \citep{sick20}.

Subsequent events on 28 November 2018 and 18 September 2019 pinned down Chiron's shape, 
with a possible triaxial shape (assuming hydrostatic equilibrium) given in Table~\ref{tab_dyn_param_rings}.
The location of the sharp drops seen in 2011 are consistent with features also seen in a light curve 
of the 2018 occultation. However, their optical depths have half of the previous values seen in 2011,
implying an evolution between the two dates \citep{sick23}.

\begin{figure}[ht]
\centering
\includegraphics[width=0.58\textwidth,trim=0 0 0 0]{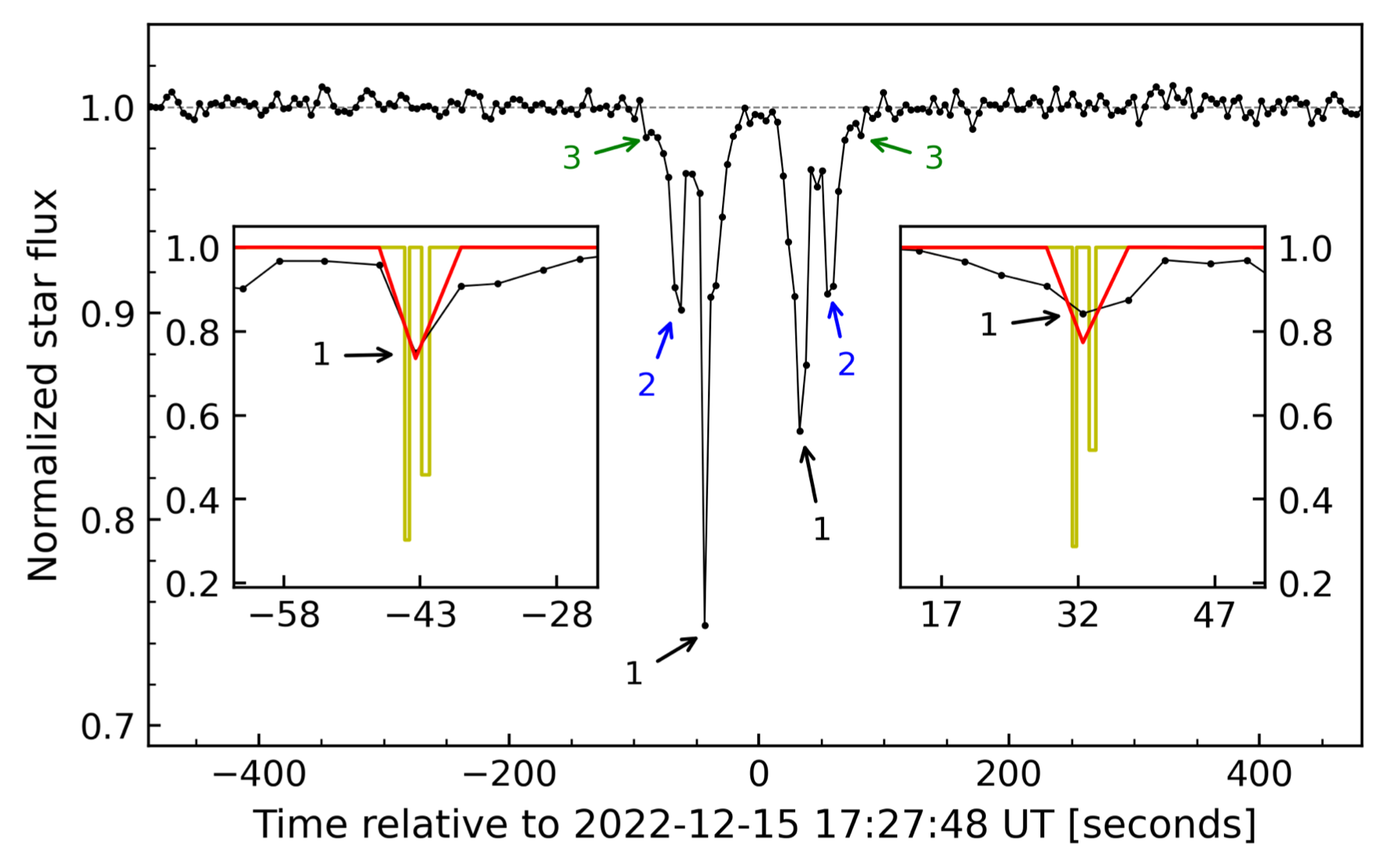}
\includegraphics[width=0.4\textwidth,trim=0 0 0 0]{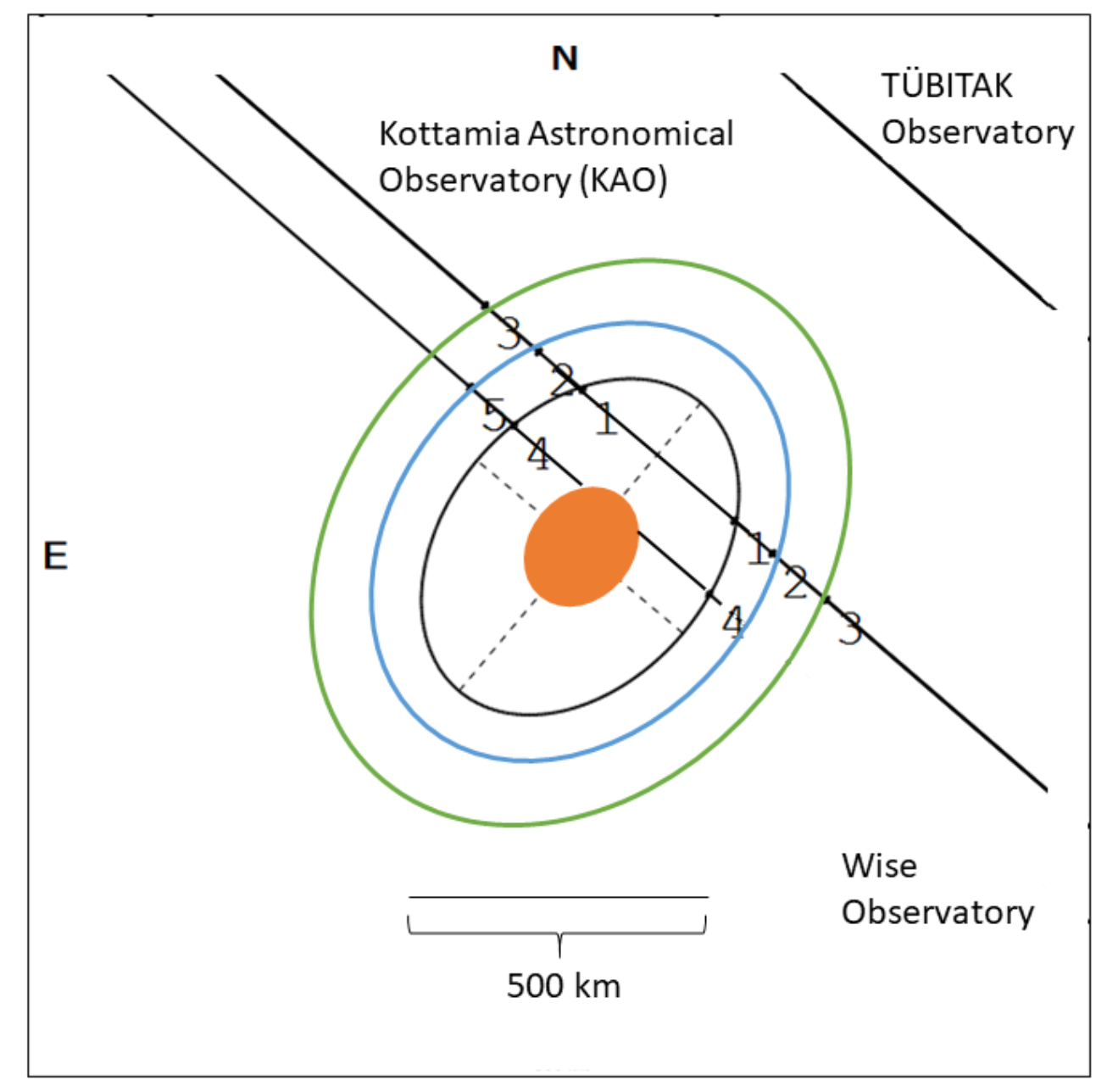}
\caption{%
Left panel - 
the light curve obtained from the Kottamia observatory (Egypt) during the Chiron 
occultation of 15 December 2022. The body did not occult the star, but dense material
caused stellar drops symmetrically positioned with respect to the time of closest
approach to Chiron, taken here as the origin of time.
The inserts show the modeling (red) of the deepest events 1 (black), assuming that it was
caused by the double events observed during the 29 November 2011 occultation (gold).
Right panel - 
Combining the Kottamia results with observations made at Wise observatory (Israel),
the events 1, 2 and 3 could be interpreted as rings drawn here in black, blue and green.
The orange ellipse at the center is a possible solution for Chiron's shape.
Images reproduced with permission from \cite{orti23}, copyright by the author(s).
}%
\label{fig_Chiron_occ_2022}
\end{figure}

Finally, an occultation observed on 15 December 2022 from Egypt and Israel clearly showed two deep stellar drops around 325 and 423~km from Chiron's center, embedded in a more diffuse disk of radius $\sim$580~km (\citealt{orti23} and Fig.~\ref{fig_Chiron_occ_2022}).
The sharp feature at 325~km labeled 1 in Fig.~\ref{fig_Chiron_occ_2022} is consistent with the double events observed during the November 2011 occultation. The event 2, corresponding to a ring at 423~km was also seen in 2011. However, the event~3 and the broad stellar drop surrounding them cannot be reconciled with the observations of 1994, 2011 and 2018. 

\subsection{Dynamics} 

Some basic parameters relevant to dynamics of rings around Chariklo, Haumea, Quaoar and Chiron
are given in Table~\ref{tab_dyn_param_rings}.
They show that all the four objects considered here depart significantly not only from a sphere, 
but also from an axisymmetric body.
This departure from axisymmetry creates strong spin-orbit resonances between the central body and the rings.
These resonances should clear the synchronous orbits, where the orbital period of the particles
matches the spin rate of the body, in a matter of years \citep{sica19}.
In the extreme case of Haumea, the body is so elongated (Table~\ref{tab_dyn_param_rings})
that any ring material inside a radius of about 2000~km is ejected or falls down onto the body
in a matter of a few revolutions \citep{ribe23}.

\begin{table}[!ht]
\caption{Dynamical parameters of Chariklo, Haumea and Quaoar.}
\label{tab_dyn_param_rings}
\begin{tabular}{@{}lllll@{}}
\toprule
& Chariklo\footnotemark[1]  & Haumea\footnotemark[2] & Quaoar\footnotemark[3] & Chiron\footnotemark[4] \\
\midrule
Mass (kg)                         & 
$(7 \pm 1) \times 10^{18}$        & 
$(4.006 \pm 0.04) \times 10^{21}$ & 
$(1.2 \pm 0.5) \times 10^{21}$    & 
$(4.8 \pm 2.3) \times 10^{18}$   \\

Rotation (h)           & 
7.004$\pm$0.036        & 
3.915341$\pm$0.000005  & 
17.6788$\pm$0.0004     & 
5.917813$\pm$0.000007 \\

$A$ (km)        
& $143.8^{+1.4}_{-1.5}$ 
& $1161^{+30}_{-30}$     
& $\sim 580$    
& $126^{+22}_{-22}$ \\

$B$ (km)        
& $135.2^{+1.4}_{-2.8}$ 
& $852^{+4}_{-4}$        
& $\sim 513$    
& $109^{+19}_{-19}$ \\

$C$ (km)        
& $99.1^{+5.4}_{-2.7}$  
& $513^{+16}_{-16}$      
& $\sim 471$    
&  $68^{+13}_{-13}$ \\

$R$ (km)           
& 121
& 712
& 516 
& 91     
\\
$C_{22}$           
& 0.00819            
& 0.0614                 
& 0.0138        
& 0.024             \\ 

$J_2$              
& 0.132              
& 0.305                  
& 0.0586        
& 0.224             \\ 

$a_{\rm syn}$ (km) 
& 196                
& 1104                   
& 2018          
& 154               \\

\midrule
\multicolumn{5}{c}{Ring and resonance radii (km)} \\
\midrule
Rings           
& Q1R: $385.9\pm 0.4$     
& H1R: $2287^{+75}_{-45}$ 
& Q1R: $4057 \pm 6$  
& $325 \pm 16$ \\
                    
& Q2R: $399.8\pm 0.6$ 
&                         
& Q2R: $2520 \pm 20$ 
& $423 \pm 11$ \\

Resonances      
& 1/3: $408 \pm 20$       
& 1/3: $2285 \pm 8$       
& 1/3: $4021 \pm 58$ 
& 1/3: $312 \pm 54$ \\
                
&                         
&                         
& 5/7: $2525 \pm 35$ \\
&                    \\


\bottomrule
\end{tabular}
\footnotetext{%
The semi-axes $A$, $B$ and $C$ give the best-fitting ellipsoids derived from occultations, 
see references below. The quantity
$R$ is the reference radius given by $3/R^2 = 1/A^2 +1/B^2 +1/C^2$, 
$C_{22}= (A^2 - B^2)/20R^2$ is the elongation, 
$J_2= (A^2 + B^2 - 2C^2)/10R^2$ is the dynamical oblateness, and
$a_{\rm syn}$ is the radius of the synchronous orbit, where the particle revolve at the same rate as the body.
For simplicity, the error bars on the mass, the rotation period and $A$, $B$ and $C$ have not been propagated
to $R$, $C_{22}$, $J_2$ and $a_{\rm syn}$.
}%
\footnotetext[1]{From \cite{leiv17} and \cite{morg21} and references therein.}
\footnotetext[2]{From \cite{orti17} and references therein.}
\footnotetext[3]{From \cite{vach12,morg23,pere23} and references therein. 
The values of $A$, $B$ and $C$ are estimations derived from the occultation of 9 August 2022 and 
from the amplitude of the Quaoar rotational light curve.}
\footnotetext[4]{From \cite{brag23}, \cite{orti23} and references therein.}
\end{table}

Because the bodies are irregular and the occultations observed so far provided few chords,
it is difficult to locate the center of mass of the bodies, and thus, 
to measure the orbital eccentricities of the rings.
The best case is obtained with Chariklo's rings, for which about fifteen occultations have been monitored.
Even so, only 3$\sigma$ upper limits of 0.022 for the eccentricity of C1R, and of
0.006 for the relative eccentricity of C1R and C2R have been obtained. 
These values can be compared to the orbital eccentricity of Uranus' $\epsilon$ ring, $e=0.0075$.

Another difficulty, particularly for Chariklo, which has no known satellites, 
is the lack of an accurate description of their gravitational fields. 
Thus, the apsidal precession rates of C1R and C2R are unknown, and it is impossible to
constrain the longitudes of the ring pericenters for a given occultation.
Consequently, it is not possible to establish whether a linear relation exists between 
the width and the radius of C1R, as it is the case for the $\epsilon$ ring of Uranus.
The variations of $W_{\rm r}$ for C1R mentioned earlier might indicate an eccentricity 
of about 0.005 for this ring, but this remains model dependent and is by no means 
a firm result based on a geometry reconstructed from occultation observations.

A noteworthy result derived from occultations is the proximity of all the rings detected 
so far around small bodies to second-order Spin-Orbit Resonance (SORs) resonances.
They correspond to a ratio between the mean motion $n$ of particles and the spin rate $\Omega$ 
of the central body of the form $n/\Omega = m/(m-2)$.
From Table~\ref{tab_dyn_param_rings}, we see that the rings C1R, C2R, H1R and Q1R all coincide to within error bars with the 1/3 SOR with the central body, meaning that a ring particle completes one revolution while the body completes three rotations. Meanwhile, the ring Q2R lies close to the 5/7 SOR.
The situation is not so clear for Chiron because its mass is poorly known and because the ring nature of its material is still to be established and characterized. This said, the dense material seen near 325~km can in principle be associated with the 1/3 SOR, that lies somewehere between 260 and 370~km.

This suggests that second-order resonances are a key ingredient to confine and maintain 
rings around small irregular bodies. Theoretical studies and N-body simulations seem 
to back up this hypothesis \citep{salo21,sica21b}.

Besides SOR resonances with Quaoar, we note that Q1R is also close to the 6/1 mean motion resonance with the satellite Weywot \citep{morg23}, which may have an important dynamical effect on the ring depending on Weywot's orbital eccentricity \citep{rodr23}.

Another interesting result concerning Quaoar is the unexpected location of its rings 
well outside its Roche limit, see Fig.~\ref{fig_Quaoar_two_rings}.
This is the first clear example of such behavior observed in planetary rings, 
which could be explained by the fact that collisions of cold icy particles could be more elastic 
than has been adopted for Saturn's rings, using laboratory experiments \citep{brid84}.
In fact, using rebound coefficients prevailing at low temperatures \citep{hatz88}, 
simulations show that a sufficiently high velocity dispersion is maintained in these rings, 
so that the ring particles can move away from each other after a collision with a velocity 
that exceeds their mutual escape velocities, thus preventing accretion \citep{morg23}.
In any instance, this discovery shows that search for rings using occultations 
should not be focused on the immediate vicinity of the bodies, 
but rather encompass a vast region around the body. 

Finally, our knowledge of Chiron's immediate surrounding is still fragmentary. 
The dense material detected around the body could be involved in some ongoing processes,
where a cometary-type shell is actually a ``ring system in the making".
In that context, Chariklo's rings would be the end product of such processes, 
that is coplanar, concentric and circular rings permanently surrounding a now inactive body.



\section{Trojans}
\label{sec_trojans}

A significant population of objects are resident in the L$_4$ and L$_5$ Lagrange points 
in the Jupiter-Sun system, at an average heliocentric distance of about 5~au.
%
There are compelling theories that the source region for these objects 
is the early TNO population and not drawn from the main-belt asteroids \citep{emer15}. \bla
Thus, it makes sense to consider 5~au to be the boundary where the outer solar system begins. 
In this section, a reference to ``Trojans'' is meant to refer specifically to Jupiter Trojans.

Much work remains to be done to fully characterize the Trojans.  Until recently, these objects have
been cataloged for their existence, but physical observations have been limited.  Despite this limitation,
there are some interesting attributes of the Trojans to consider that will be important when connecting
this population to the more distant TNOs.

\begin{itemize}
    \item The population of Trojans is considered to be complete at the bright (large) end, with the
    largest object, (624) Hektor, having a diameter of about 200 km.  The largest are comparable to the
    typical size of the cold-classical Kuiper Belt Objects (CCKBOs) but none are large enough to be in the dwarf planet category.  There are enough objects in both Trojan and \bs{CCKBO} populations 
    in similar size ranges to permit a comparison of their respective properties.
    
    \item The known population reaches to a smaller size range than the TNOs and is complete down to a size of roughly 20~km \citep{emer15}. This is not due to differences in the populations, \bs{but just due to the fact} that the apparent brightness of Trojans is higher than for TNOs due to being closer. This situation provides an opportunity to explore 
    a smaller size range than is currently feasible for TNOs.
    
    \item A large fraction of the Trojans have surface colors that are very similar to CCKBOs,
    suggesting a compositional linkage between the two populations.
\end{itemize}

The NASA Lucy spacecraft \citep{olki21} was launched on October 16, 2021 on a twelve-year mission to explore
Trojan asteroids \citep{levs21}.  In light of this new mission and the increasing appreciation of the
population, new attention is being paid to collecting observations.  The Lucy target
sample was chosen for its diversity.  The sizes range from roughly 20 to 120 km, covering the upper
range of the size distribution while also overlapping in size with the New Horizons target, Arrokoth 
(which is less than 40~km across along its longest dimension).

In surface composition, the sample covers C, P, and D-types 
and includes the largest member of the only recognized collisional family in the Trojan population. Four objects (Eurybates, Polymele, Leucus, and Orus) are in the L$_4$ cloud and will be visited in 2027-28. The last object is the Patroclus-Menoetius binary that will be reached in 2033.  
The mission objectives include 
the determination of the bulk densities, 
characterizing the cratering record and 
thus constraining the impactor population at 5~au, search and characterization of non-impact related surface processes, 
resolved spectral imaging, and thermal observations for adding to the knowledge of surface properties. 

The rotation periods and lightcurve amplitudes in magnitude cover a wide range: 
Eurybates (8.7~h, 0.20~mag), 
Polymele (11.5~h, 0.09~mag), 
Orus (13.5~h, 0.3~mag), 
Patroclus (102.8~h, 0.07~mag), and Leucus (445.7~h, 0.61~mag).

Observations with HST have surveyed the outer range of where stable satellites could be found Eurybates was discovered to have a small satellite, Queta, that is in a somewhat excited orbit and seems to be consistent with being a collisional fragment rather than a primordial object \citep{noll20,noll23}. No additional new satellites were found by HST around the other objects, noting that HST by itself cannot cover the entire region of stable orbits.

Earth-based occultations are playing a pivotal role for the Lucy Mission 
and, in so doing, are revealing new insights into these specific objects. 
They actually give us a glimpse of the wealth of information that could be obtained on the larger Trojan population by surveying the population with this technique, see Appendix~\ref{app_deployment}. 
As of May 2024, sixty-four different Trojans have already been detected in at least one stellar occultation, but only about one-third of them had a multi-chord event \citep{brag19}. 

With the second and later occultations (Appendix~\ref{app_deployment}),
the data from multi-chord detections are able to reveal any contact or very close binaries, as well as 
constraining the three-dimensional shape of the objects.  
Getting 3D shapes is especially important as one can now distinguish between 
spherical, prolate, and oblate shapes that are ambiguous from light curve inversion approaches.  
Also, the orbital period of twelve years means that we can see a useful range of observing geometries 
in as little as three years. Similar studies with TNOs will take far longer. 
In this case, having a good set of well-measured Trojans can be used as a reference basis set 
from which to make statistical inferences on the properties of TNOs 
where we have to wait as long as centuries to get useful changes in viewing geometry.

\begin{figure}[ht]
\centering
\includegraphics[width=1.0\textwidth,trim=0 150 0 150]{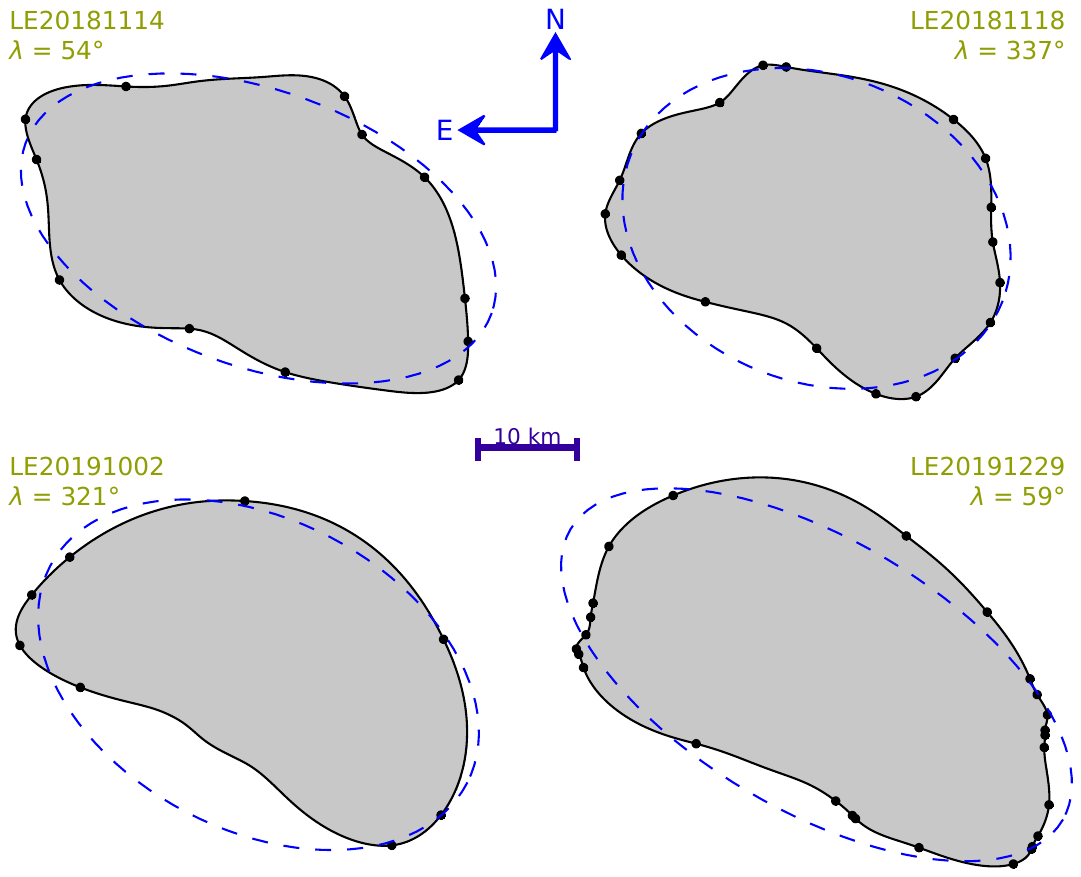}
\caption{%
Shape profiles of (11351) Leucus.  
Shown are four profiles of Leucus from as many separate occultation observations
\bs{(the date of each observation in encoded above each plot under the format ``LEYYYMMDD").}
The dots represent the extremities of the occultation chords.  
All four figures are drawn with the same scale (see the bar) 
and the J2000 sky-plane orientation is indicated by the arrows.
The dashed curve is an elliptical profile for the shape based on a weighted fit to the data.
The heavy curve is a somewhat artistic outer outline of the object drawn with a spline curve 
rather than just simple line segments.  
The outline is consistent with all of the constraints including the negative observations, 
while the ellipse fit only considers the positive detections. 
The 3D shape of Leucus is very irregular and 
the elliptical fits are poor approximations of the shape. 
Much of this structure and an understanding of the complete shape requires multiple 
measurements with densely spaced chords.
}%
\label{fig_leucus}
\end{figure}

Besides what has already been shown in Fig.~\ref{fig_trojan_3Dfit}, 
the results shown in Fig.~\ref{fig_leucus} demonstrate what can be achieved on small body shapes 
once the prediction reaches the Gaia catalog limit \citep{buie21}.  
The dots on the figure show the actual data, while the shaded outline is a smoothed shape 
consistent with both the positive and negative detections. 
The variation in shape seen in this figure is entirely due to the changing viewing geometry 
due to the rotation of the object.
Also shown are the fitted elliptical profiles for the same data.  
While the fitted ellipses have some use in the analyses, an ellipse is a poor representation 
of the actual shape of this small body.

It remains to be seen what the nature of the irregular shape of Leucus can be attributed to for this size of object.  The Lucy Mission will provide essential context from which these observations can be understood and in turn provide guiding context for objects that may be measured by occultation without assistance from a close-up spacecraft observation.  

As detailed in Appendix~\ref{app_deployment}, by the time one gets three or more occultations, one begins to have sensitivity on unseen satellite in the system through the wobble induced by the satellite.
For Trojans, the size of the barycentric wobble that is relatively easy to reach is about 3 km, peak to peak. 
This could reveal small unseen moons orbiting the primary.

The population of close and contact binaries in the Trojans will be important data for knowing how to
tie in the population with those of TNOs 
where such binarities are commonplace. 
From the occultation work for Lucy target a new moon was discovered 
orbiting the smallest target, Polymele \citep{buie22}.
This moon appears to be in a totally relaxed orbit (zero inclination and eccentricity)
and a semi-major axis of about 200 km.
The volume ratio and orbital properties are suggestive that 
Polymele is an accretional system, not the result of collisions as is expected for Eurybates. 

All five of the Lucy targets (or systems) show significant departures from a smooth tri-axial ellipsoidal shape.
This is already a decent sample of objects but it is now a straightforward task 
to extend this type of measurement to a very large sample of the population, using Earth-based occultations.  
Once the up-close imaging is obtained with Lucy, it will be an interesting investigation 
to see how to extend those detailed \bs{views into} a broader understanding.

\section{Centaurs and TNOs}
\label{sec_centaurs_tnos}


%
%


The primary and most straightforward scientific products from occultations are sizes and projected shapes, which, combined with absolute photometry of the bodies, allow us to determine geometric albedos as well. 
\bs{Currently\footnote{See \url{https://occultations.ct.utfpr.edu.br/results/}}}, 
about 195 occultations by Centaurs and TNOs involving 54 different TNOs 
(not counting Pluto and its satellite Charon) and 13 Centaurs \bs{have been observed},
although this depends on the exact definition that is used for the concept of Centaur \citep{brag19}.

\begin{table}[ht]
\caption{List of TNOs and Centaurs from which relevant results have been published using occultations (excluding Pluto and Triton).}
\label{tab_list_Published}
\begin{tabular}{@{}lllm{5.8cm}@{}}
\toprule
Object  & 
Type & 
Effective & 
References \\

&      
& 
Diameter  &
\\

&      
& 
(km) &
\\

\midrule
2002~TX$_{300}$   &
TNO   &
248$\pm$10\footnotemark[1]  &
\cite{elli10}, \cite{orti20a} \\

Eris   &
TNO   &
2326$\pm$12\footnotemark[1]  &
\cite{sica11}, \cite{orti20a}  \\

Makemake   &
TNO   &
1465$\pm$47   &
\cite{orti12}, \cite{orti20a}  \\

 Quaoar   &
 TNO   &
 1110$\pm$5   &
 \cite{brag13}, \cite{morg23},  \\
 &   &   & \cite{pere23} \\
 
Chariklo   &
Centaur   &
268$\pm$12   &
\cite{brag14}, \cite{leiv17}  \\
&      &       & \cite{bera17}, \cite{morg21} \\

2002~KX$_{14}$  &
TNO   &
369$\pm$25 & 
\cite{alva14} \\

Chiron   &
Centaur   &
196$\pm$34  &
\cite{rupr15}, \cite{orti15},  \\
&       &       & \cite{orti23}, \cite{brag23}\\

2007~UK$_{126}$   &
TNO   &
638$\pm$28   &
\cite{bene16}, \cite{schin17}, \\
&        &           & \cite{orti20a}\\

2003~AZ$_{84}$ &
TNO   &
764$\pm$6   &
\cite{dias17}  \\

Haumea   &
TNO   &
1392$\pm$26   &
\cite{orti17}  \\

Vanth &
TNO sat &
443$\pm$10\footnotemark[1] &
\cite{Sick19} \\

2003~VS$_{2}$  &
TNO   &
566$\pm$23   &
\cite{bene19}, \cite{vara23}  \\

2002~TC$_{302} $  &
TNO   &
499$\pm$14   &
\cite{orti20b}  \\

2002~VE$_{95}$ &
TNO   &
279.5$\pm$2.5\footnotemark[1]  &
\cite{romm20}  \\

2003~FF$_{128}$ &
TNO   &
131$\pm$45  &
\cite{romm20}  \\

Varda   &
TNO   &
753$\pm$20  &
\cite{soua20}  \\

Leleakuhonua &
TNO &
 112$\pm$12\footnotemark[1]&
\cite{buie20a} \\

Arrokoth &
TNO &
$\sim19$ &
\cite{buie20} \\

2014~WC$_{510}$ &
TNO  &
227$\pm$23\footnotemark[1]&
\cite{leiv20}\footnotemark[2]\\

2002~GZ$_{32}$   &
Centaur   &
211$\pm$12\footnotemark[1]  &
\cite{sant21}  \\

2014~YY$_{49}$ &
\coa{Centaur} &
32$\pm$4\footnotemark[1] &
\cite{strau21}\\

2013~NL$_{24}$ &
\coa{Centaur} &
132$\pm$10\footnotemark[1] &
\cite{strau21}\\

Huya   &
TNO   &
411$\pm$7   &
\cite{sant22}  \\

Bienor   &
Centaur   &
150$\pm$20  &
\cite{fern23}  \\

2002~MS$_{4}$   &
TNO   &
786$\pm$10  &
\cite{romm23}  \\
 
 Echeclus   &
 Centaur   &
 60$\pm$1  &
 \cite{pere24}  \\
 
\bottomrule
\end{tabular}
\footnotetext{Preliminary results on occultations by TNOs,  Centaurs, Satellites, and Trojans prior to publication can be found at \url{http://occultations.ct.utfpr.edu.br/results/}}
\footnotetext[1]{Spherical shape assumed}
\footnotetext[2]{This paper also reports the discovery of a satellite of 2014~WC$_{510}$}
\end{table}

%

The majority of the observed occultations have been single-chord events. 
They provide insufficient information to derive size, shape, and geometric albedo but often help refine the orbits (see Section \ref{sec_prediction} and Appendix \ref{app_deployment}). Events with only two chords also give rise to limited results because a unique ellipse cannot fit with only 2 chords as the mathematical problem is degenerate. 
In general, given the relatively large size of the TNOs involved in the detected occultations, we can safely assume that their shapes are ellipsoidal or spheroidal to a large degree, so their projected shapes are usually well fit with ellipses (Sect.~\ref{sec_shape_retrieval}). 

TNOs with \bs{diameters above around 400~km} are suspected to have enough mass to overcome rigid forces in the interior and adopt ellipsoidal or spheroidal equilibrium figures (\citealt{tanc08}). 
Most of the TNOs that have produced multichord events are above this size range. Clear exceptions are Arrokoth, which is in the 40~km size range and whose shape is highly irregular, 
2002 TX$_{300}$ and 2003 FF$_{128}$.
Also, the known Centaurs 
are all below the 300~km \bs{diameter} range.
A list of TNOs and Centaurs for which relevant results through occultation have already been published is given in Table~\ref{tab_list_Published}, their main highlights being available in the associated references.

\subsection{Geometric albedos}

Geometric albedos are derived using Eq.~\eqref{eq_albedo}. However, it is not always possible to obtain the instantaneous absolute magnitude of the object at the time of the occultation, as accurate values of the rotation period and good rotational light curves may be unavailable. This may result in uncertainties in the instantaneous absolute magnitude at the 20\% level. This can be made even worse if the solar phase curve is not accurately determined. 
Besides, some TNOs or Centaurs may have undiscovered large satellites or rings that contribute to the absolute magnitude, yielding overestimated geometric albedos. Therefore, at least some of the geometric albedo determinations will require updates whenever better values of the instantaneous absolute magnitudes become available or whenever rings or satellites are discovered. Within the uncertainties, the values determined in occultations generally agree with those from thermal observations except for two cases, as \bs{discussed below}.

\subsection{Occultation-derived vs. thermally-derived sizes}
\label{subsec_thermal}

Before comparing thermal observations with occultation observations, several intricate factors and nuances need addressing. Firstly, it is essential to compare equivalent diameters in the projected area sense, not volume, as determined by thermal models. Additionally, the presence of satellites must be considered. In certain studies, such as some Herschel ``TNOs are cool" papers \citep{mueller20}, satellite presence might not have been acknowledged or displayed in tables, leading to reported equivalent diameters combining both primary and satellite. However, occultations typically only record primaries. Furthermore, for triaxial bodies, the rotational phase at the time of occultation needs consideration. Moreover, differences in aspect angles during Herschel-Spitzer observations compared to occultations should be taken into account, albeit often minor, potentially only a few degrees. This calculation is advisable if information on spin axis orientation is available. On the other hand, when comparing occultation observations with thermal observations, it's crucial to ensure the thermal observations are analyzed similarly, using the same type of thermal models rather than thermophysical models. Thus, meticulous analysis is important.

\begin{figure}[ht]
\centering
\includegraphics[width=0.8\textwidth]{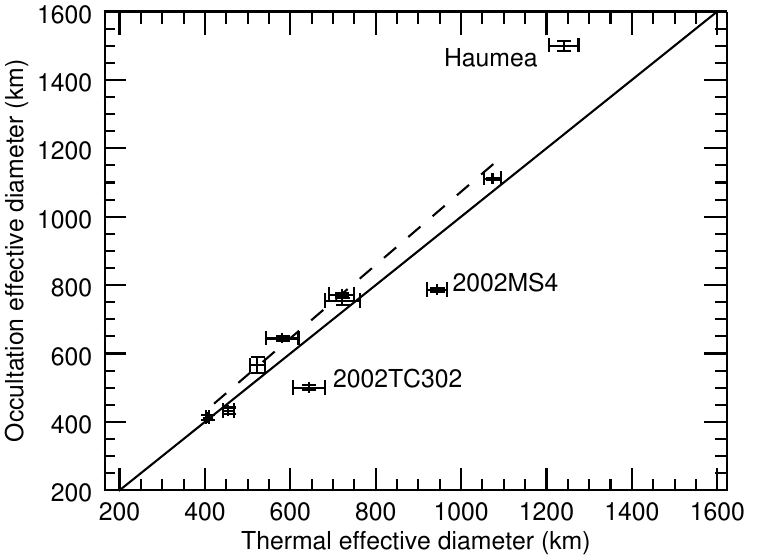}
\caption{%
Stellar occultation equivalent diameters versus diameters from thermal observations for the 10 published results that meet the requirements to make accurate comparisons (see text). The continuous line shows the y=x straight line. Most of the points are slightly to the left of this line, indicating that the thermal models tend to  underestimate the true sizes, at the six percent level on average. The dashed line shows a fit to the data excluding Haumea and the other two outliers 2002 TC$_{302}$ and 2002 MS$_4$.
}%
\label{fig_Occ_vs_Thermal}
\end{figure}

When all these aspects are taken care of, we come up with a good sample of 10 bodies from Table~\ref{tab_list_Published} suitable for the comparison and with more than two chords. The results are shown in Fig.~\ref{fig_Occ_vs_Thermal}. Except for two bodies (2002~TC$_{302}$ and 2002~MS$_4$), the thermal measurements tend to underestimate the true size, at only the 6 percent level in most cases. The two cases that are clearly outside this trend might indicate some sort of anomaly, such as the potential existence of a large satellite close in, not revealed in HST images, whose thermal emission could contribute. Enhanced thermal emission from a different reason does not seem likely, but cannot be discarded. Besides, problems with the data might also be a possibility.


\section{Serendipitous occultations by sub-km TNOs}
\label{sec_serendip}

\subsection{Principle}

Serendipitous occultations may occur when the uncertainty on the position of the occulting body is so large that planning campaigns anywhere on Earth becomes irrelevant. However, low-cost projects such as the Research and Education Collaborative Occultation Network (RECON, \citealt{buie16}) may detect the occultation by chance, thus pinning down the accuracy of the object ephemeris.

However, the serendipity method is used at its full potential for objects that are completely 
undetectable through direct imaging, but still expected from statistical arguments. 
In fact, the number of TNOs is expected to increase as their sizes decrease, thus increasing the probability of a serendipitous occultation. Meanwhile, the target stars must be bright enough to provide sufficient SNR, but not too bright so that their angular diameter remains comparable or smaller than that of the TNOs.
The best compromise occurs for km-sized objects, but even then, the probability of events remains small. This necessitates the simultaneous monitoring of many stars at cadence of tens of \bs{Hertz}. If successful, this method provides important statistical constraints on this ``invisible matter" of the solar system, composed of small size range of TNOs, including Oort cloud objects.

Two precursor articles proposed this approach. \cite{bail76} was the first to mention that TNOs that are undetectable through direct imaging can be glimpsed by monitoring stars in the ecliptic plane.
\cite{dyso92} elaborated on this work. He estimated a promising  occultation rate of one every 2 hours for km-sized  objects, assuming a population of 10$^{13}$ TNOs. He pointed out that the star must have an angular size smaller than the TNO, in practice stars bluer than the Sun with magnitude 13 or fainter.
There are some hundred of such stars per square degree in the direction of the galactic plane. This restricts optimal survey fields of view \bs{to} the intersections of the galactic and ecliptic planes. Dyson pointed out the problem of image quality, \bs{and suggested the use of} several close telescopes to discriminate between real and false events. He also mentioned the problem of data processing for hundreds of stars recorded at cadences of hundreds of \bs{Hertz}. He finally noticed that diffraction makes the occultation events smooth, and that the color dependency of diffraction can help to confirm an event.
%

Several teams and projects developed this method to probe the Kuiper Belt region and beyond.
\cite{roqu87} showed that diffraction is a fundamental phenomenon for interpreting the events. 
Subsequent theoretical and numerical studies have explored the key parameters of the serendipitous occultation technique and proposed observational approaches to detect small bodies in the outer solar system.

Simulation of occultation profiles that account for diffraction, stellar angular size and wavelength 
of observation provided an estimate the occultation rates in term of telescope size and assumption 
on the target populations \citep{roqu00}. 
\cite{nihe07} focused on the exploration of KBOs and Oort cloud comets for three instrumental configurations, small, and large ground-based telescopes, as well as space-borne telescopes.
\cite{coor03} showed that the serendipitous occultations may explore the turnover radius 
in the KBO size distribution between 0.1 and 1~km . 

The key parameters of the serendipitous occultation method are
the Fresnel scale $\lambda_{\rm F}$ (Eq.~\eqref{eq_Fresnel_Scale}) and 
the stellar diameter $\theta_\star$ projected at the body (Eq.~\eqref{eq_star_diam_km}).
Typical values of $\lambda_{\rm F}$ and $\theta_\star$ are given in Fig.~\ref{fig_Delta_diam}
for various geocentric distances, stellar apparent magnitudes and stellar types.
The diameter $W_{\rm sha}$ and the depth $D_{\rm sha}$ of the shadow caused by a small object of radius $r$ are then \citep{nihe07}
\begin{equation}
\begin{array}{l}
W_{\rm sha}= 
2 \left[ (\sqrt{3} \lambda_{\rm F})^{3/2} + r^{3/2} \right]^{2/3} + \theta_\star \\ \\
D_{\rm sha}= 
\left\{ 1 + \left[ 3 (r/\lambda_{\rm F})^2  \right]^{-3/2} \right\}^{-2/3}
\label{eq_width_depth_serendip}
\end{array}
\end{equation}

\begin{figure}[ht]
\centering
\includegraphics[width=1.0\textwidth,trim=0 0 0 0]{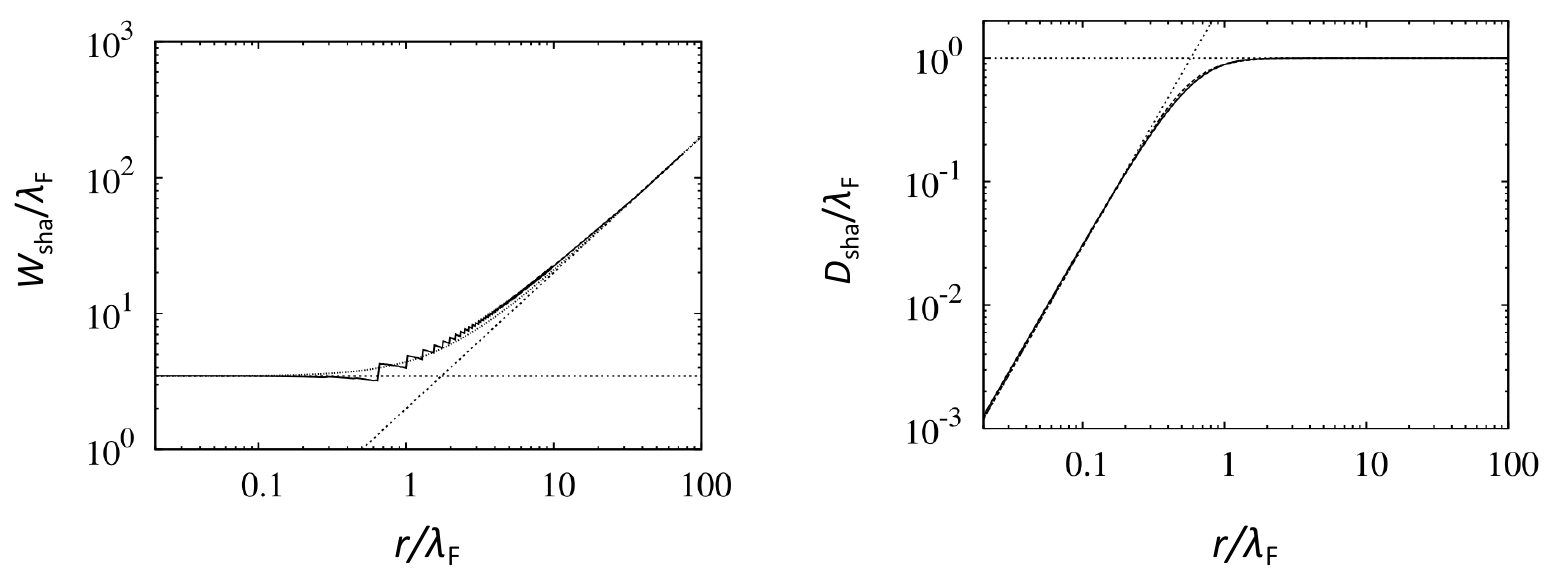}
\caption{%
Left panel - 
The diameter $W_{\rm sha}$ of occultation shadows caused by a
body of radius $r$, assuming a point-like stellar source (i.e. $\theta_\star=0$ 
in Eq.~\eqref{eq_width_depth_serendip}).
The quantities in the horizontal and vertical axes have been normalized 
to the Fresnel scale $\lambda_{\rm F}$.
Right panel - 
The same depth for the depth $D_{\rm sha}$ given in Eq.~\eqref{eq_width_depth_serendip}.
Image adapted from \citet{nihe07}.
}%
\label{fig_width_depth_serendip}
\end{figure}

Fig.~\ref{fig_width_depth_serendip} displays $W_{\rm sha}$ and $D_{\rm sha}$ as a function of radius $r$ of the occulting body. As expected, the geometric optics limits $W_{\rm sha} = 2r$ and $D_{\rm sha}= 1$ are reached for $r \gg \lambda_{\rm F}$. Conversely, $W_{\rm sha}$ remains of the order of a few times $\lambda_{\rm F}$  for $r \ll \lambda_{\rm F}$, while in that limit, the depth $D_{\rm sha}$ decreases as $r^2$.

The interpretation of serendipitous events is confronted \bs{with} a degeneracy between the TNO size and its distance. The degeneracy can be lifted if the occultation profile shows diffraction structure, which requires that $\theta_\star \lesssim \lambda_{\rm F}$. Note that for objects smaller than the Fresnel scale, $W_{\rm sha}$ is independent of $r$. This gives a measure of the Fresnel scale, and thus, the distance of the body, assuming a velocity of the occulting body projected in the sky plane.

In all cases, fitting potential events to models requires a good time resolution since synthetic profiles need several input parameters to be constrained: size, distance, and velocity of the occulting body, as well as the impact parameter of the encounter. 
As predicted by \cite{dyso92}, \bs{scintillation} can mimic diffraction profiles. A thorough examination of survey design shows that the Nyquist sampling rate for detection of a TNO at 40~au in the optical is 40 Hz, and that a detection threshold at some 8$\sigma$-level is necessary to disentangle real occultation events from false positives \citep{bick09}.

The ultimate way to confirm the reality of an occultation is the simultaneous detection of the profiles from two nearby telescopes separated by a distance smaller than $\lambda_{\rm F}$. Observation of occultations near opposition increases the relative velocity between the occulting body and the stars, increasing in turn the occultation rate. However, the lower apparent velocity at quadrature allows a better resolution of the diffraction pattern. Finally, control target stars with low expected occultation rate (i.e., far from the ecliptic plane or with very large sizes) can be used as useful ``witnesses".

\subsection{Research campaigns}



\begin{table}[!h]
\setlength{\tabcolsep}{1mm}		 
\renewcommand{\arraystretch}{0.6}   
\caption{Results of serendipitous occultation campaigns.}
\label{tab_serendip_search}

\begin{tabular}{@{}lllll@{}}

\toprule
Instrument/project & Frequency (Hz) & Number of                  & Surface density &  Index\footnotemark[3] \\
Reference          & Star-hours     & detections\footnotemark[1] & (objects deg$^{-2}$)\footnotemark[2] &  $q$ \\

\midrule
Pic du Midi\footnotemark[4] &
20                          &
1 (150)                     &
NA                          &
$\lesssim$4.5               \\
\cite{roqu03} & 15 & & & \\

\midrule
HST/FGS\footnotemark[5]            & 
40                                 & 
2 (500,45)                         & 
$1.1^{+1.5}_{-0.7} \times 10^{7}$  & 
3.7$\pm$0.3                       \\
\cite{schl09} & 31500 & (530,40) & $>$250 m  & \\
\cite{schl12} &       &          &           & \\

\midrule
CoRot\footnotemark[6]              & 
1                                  & 
13                                 & 
$1.4^{+4.2}_{-0.7} \times 10^{7}$  & 
4.5$\pm$0.2                        \\
\cite{liu15} & 144000 & (200--700,43\footnotemark[7])  & $>$400 m & \\

\midrule
RXTE\footnotemark[8]  & 
1                     & 
1    (150,40)         & 
NA                    & 
$\lesssim$3           \\
\cite{chan16} & 93 & & & \\

\midrule
TAOS\footnotemark[9]  & 
5                                    & 
0                                    & 
$<$10$^{10}$               & 
$<$3.34-3.82                \\
\cite{alco03}   & 836000 & & $>$250 m & \\
\cite{zhan13} &              & & & \\

\midrule
DAO\footnotemark[10]      & 
40                                    & 
0                                      & 
$< 3.5 \times 10^{10}$     & 
$\sim$3.5                          \\
\cite{bick08}   & 7 & & $>$430 m & \\

\midrule
MMTO\footnotemark[11]      & 
30                                    & 
0                                      & 
$< 2 \times 10^{8}$     & 
$\lesssim$4.6                          \\
\cite{bian09}   & 220 & & $>$500 m & \\

\midrule
IMACS\footnotemark[12]  & 
40                                     & 
\multicolumn{3}{c}{unpublished}   \\
\cite{payn15}   & 10000 & &  & \\

\midrule
Ultracam\footnotemark[13]  & 
65                         & 
3 (110,15)                 &
NA & NA \\
\cite{roqu06}   & 34 & (320,140) & & \\
\cite{dore13}   &    & (300,210) & & \\

\midrule
MIOSOTYS\footnotemark[14]  & 
20                                             & 
1 (380,41)                                 & 
$< 6.4 \times 10^{7}$               &   
                                                 \\
\cite{dore17}   & 9840 & & $>$380 m & \\

\midrule
AAO\footnotemark[15]                  & 
100                                                & 
1 (430,46)                                      & 
$9.8^{+3.5}_{-4.2} \times 10^{6}$  &
3.8\footnotemark[16]                     \\   
\cite{geor18}   & 6700 &  & $>$250 m & \\

\midrule
OASES\footnotemark[17]   & 
15.4                                    & 
1 (1150,40)                        & 
$6 \times 10^{5}$              &
3.9$^{+0.3}_{-0.5}$           \\   
\cite{arim19b}   & 60000 &  & $>$1200 m & \\

\bottomrule
\end{tabular}

\footnotetext{
$^1$The parentheses give the radii (m) and distance (au) of the detected objects;
$^2$The numbers just below the densities indicate the lower cutoff in size; 
$^3$The power-law size distribution index;
$^4$Using a 1-m and a 2-m telescopes;
$^5$Fine Guidance Sensors of the Hubble Space Telescope;
$^6$Convection, Rotation et Transits plan\'etaires (CoRot) satellite;
$^7$The distance 43 au is assumed;
$^8$Rossi X-ray Timing Explorer, using the single source Sco-X1;
$^9$Transneptunian Automated Occultation Survey, using four 50-cm telescopes at Lu-Lin Observatory, Taiwan.
The follow-up project TAOS-II operates three 1.3-m telescopes at San Pedro M\'artir Observatory (Mexico), and will monitor up to 10000 stars simultaneously at 20~Hz \citep{lehn23};
$^{10}$Dominion Astrophysical Observatory 1.8-m telescope;
$^{11}$Multi Mirror 6.5-m Telescope Observatory;
$^{12}$Inamori Magellan Areal Camera and Spectrograph, 6.5-m Magellan Clay telescope;
$^{13}$Triple-beam imaging photometer at the 8.2-m Very Large Telescope and the 4.2-m William Herschel Telescope;
$^{14}$Multi-object Instrument for Occultations in the SOlar system and TransitorY Systems, 1.9-m telescope (Observatoire de Haute Provence) and the 1.23-m telescope (Calar Alto);
$^{15}$Anglo-Australian Observatory, 1.2-m telescope;
$^{16}$The index $q$ is assumed;
$^{17}$Organized Autotelescopes for Serendipitous Event Surve, using two 28-cm telescopes separated by 52~m, Miyako Island, Japan.
}%

\end{table}

\begin{table}[!h]
\setlength{\tabcolsep}{1mm}		 
\renewcommand{\arraystretch}{0.6}   
\caption{Continuation of Table~\ref{tab_serendip_search}.}
\label{tab_serendip_search_contd}

\begin{tabular}{@{}lllll@{}}

\toprule
Instrument/project  & Frequency (Hz) & Number of    & Surface density       &  Index \\
Reference           & Star-hours     & detections   & (objects deg$^{-2}$)  &  $q$   \\

\midrule
Colibri\footnotemark[1]   & 
40                                      & 
0                                        & 
NA                                      &
NA                                      \\
\cite{pass18}   & 4000 &  & & \\

\midrule
CHIMERA\footnotemark[2]   & 
33                                            & 
0                                              & 
$\lesssim 10^{7}$                    &
NA           \\   
\cite{zhan23}   & some 10$^4$ &  & $>$500 m & \\

\midrule
W-FAST\footnotemark[3]   & 
10-25                                    & 
0                                           & 
$\lesssim 10^{6}$                 &
$<$3.93                                \\   
\cite{nir23}   & some 10$^6$ &  & $>$1000 m & \\

\midrule
McDonald\footnotemark[4]   & 
330                                       & 
0                                           & 
NA                                        &
NA                                        \\   
\cite{hitc24}   & 6900 &  &  & \\

\bottomrule
\end{tabular}

\footnotetext{
$^{1}$An array of three 50-cm telescopes separated by $\sim$150~m, Elginfield Observatory, Ontario, Canada;
$^{2}$Caltech HI-speed Multicolor camERA, 5.1-m Palomar Hale Telescope, simultaneous recording in i' and g' bands;
$^{3}$Weizmann Fast Astronomical Survey, 55-cm telescope, Neot Smadar, Israel;
$^{4}$A pilot study at the 2.1-m telescope of McDonald Observatory.
}
\end{table}

For more than two decades, campaigns aimed at detecting km-sized TNOs using serendipitous occultations have been conducted. The Tables~\ref{tab_serendip_search}-\ref{tab_serendip_search_contd} list the various projects undertaken and their main results. They provide, among others, the number of detections obtained by each project, the estimation of the cumulative surface density of TNOs and the constraints on the power-law size distribution index $q$ for km-sized TNOs.

\begin{figure}[ht]
\centering
\includegraphics[height=28mm,trim=0 0 0 0]{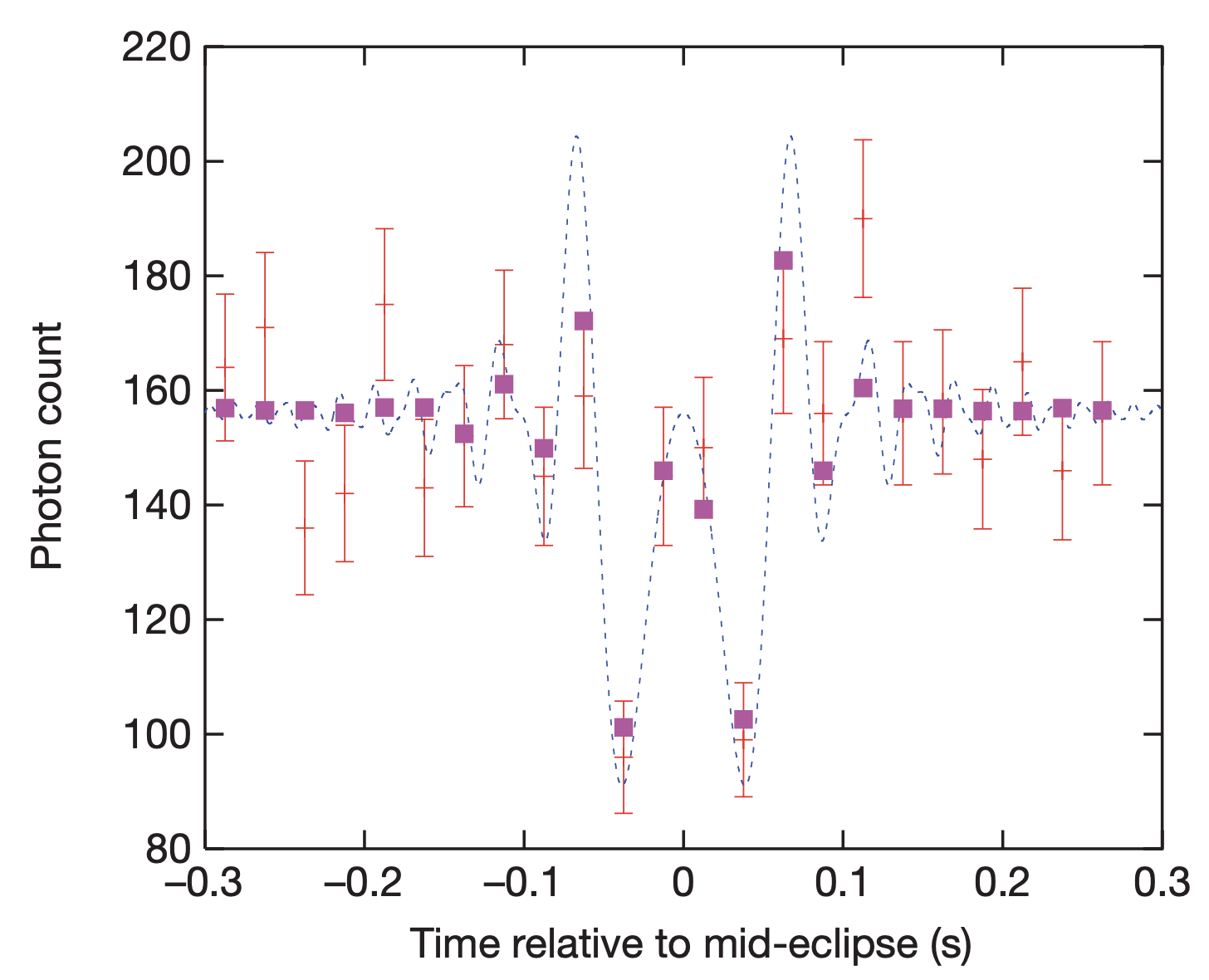}
\includegraphics[height=28mm,trim=0 0 0 0]{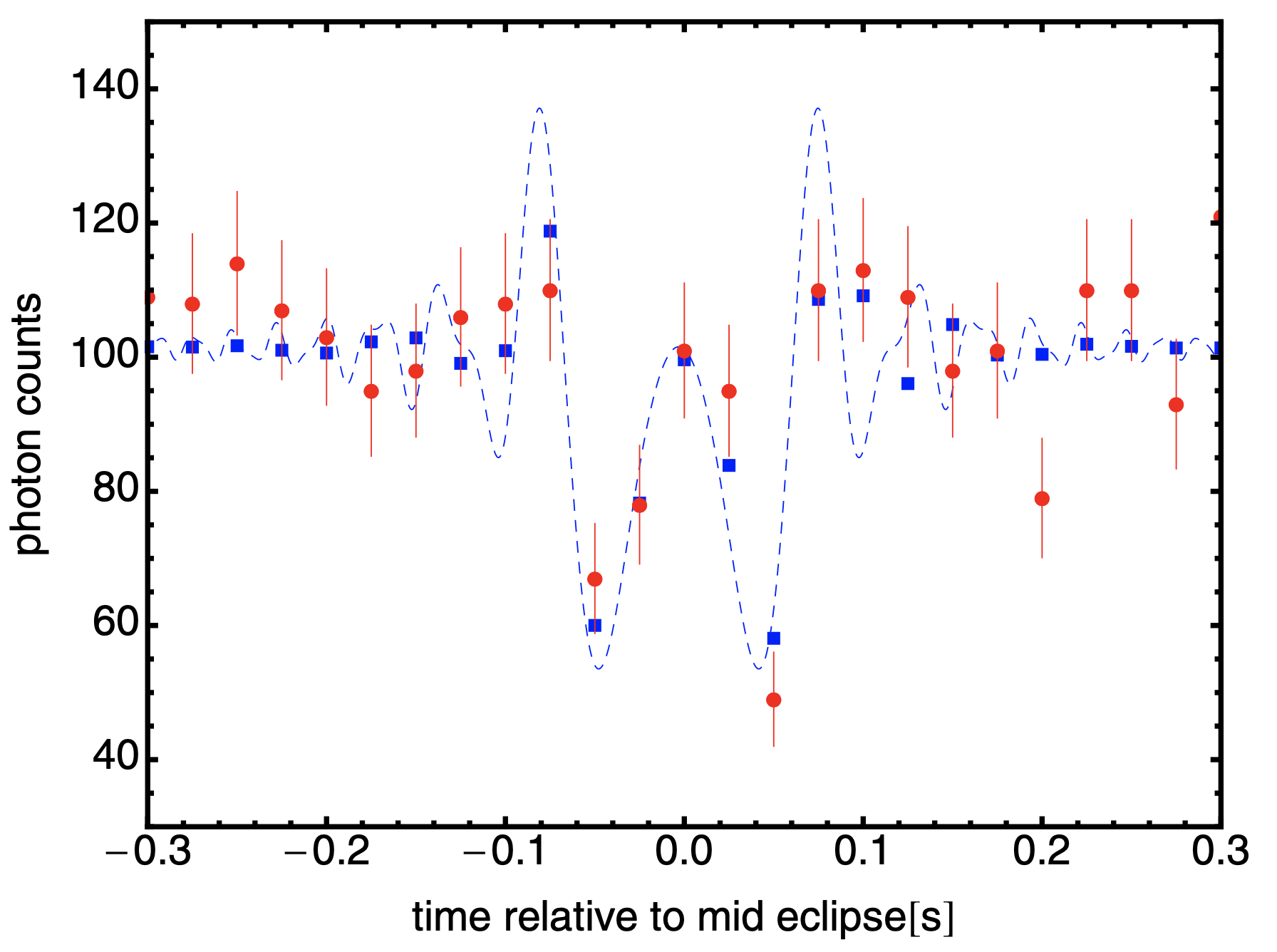}
\includegraphics[height=28mm,trim=0 0 0 0]{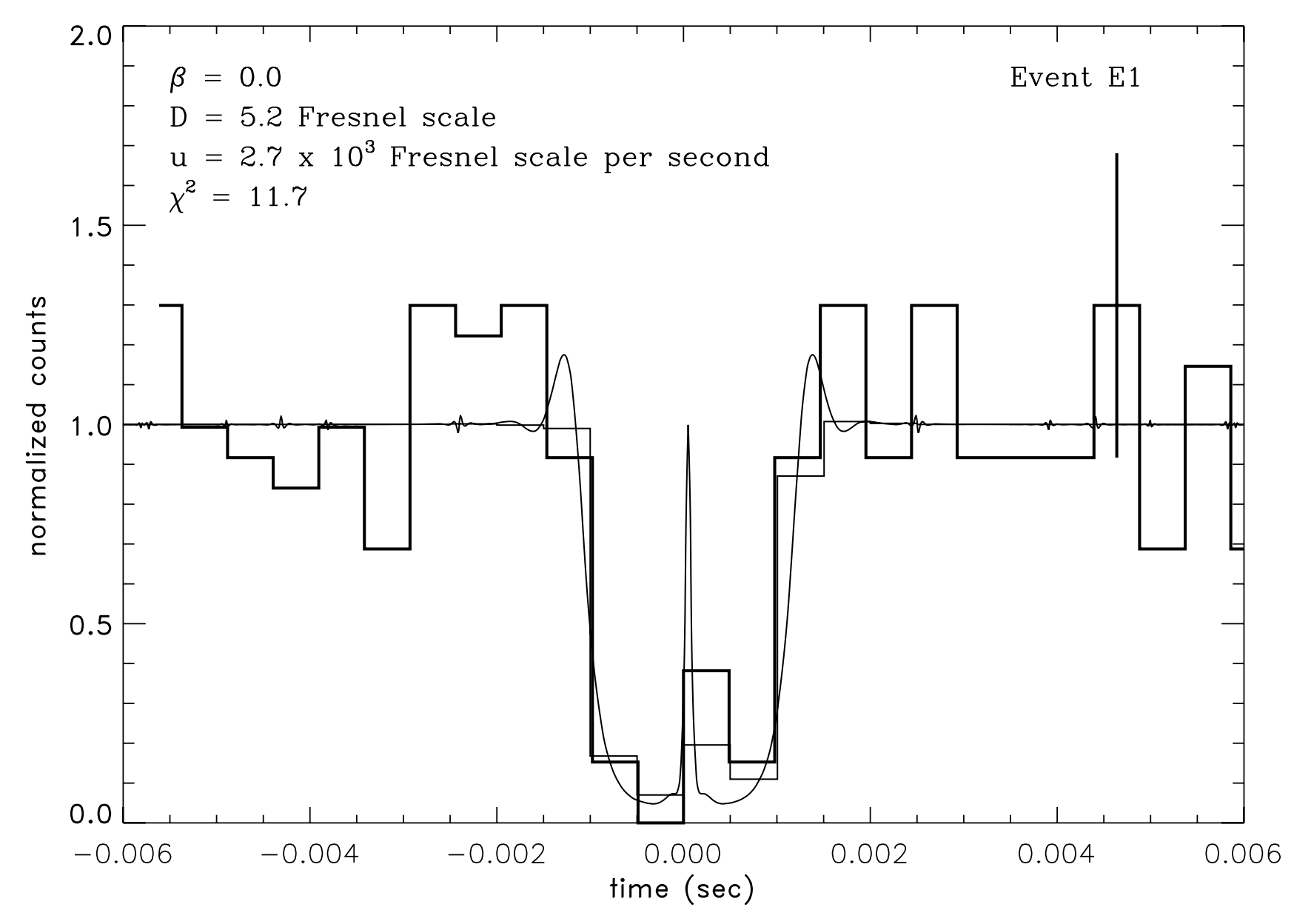}
\caption{%
Serendipitous occultation events detected from space borne instruments, see Table~\ref{tab_serendip_search} for details. In each plot, a synthetic profile has been superimposed to the data point.
Left and center panels: Two events detected in the visible using HST/FGS archival data \citep{schl09,schl12}.
Right panel: An X-ray event present in the RXTE archival data \citep{chan11}.
Images reproduced with permission; copyright by Macmillan, the author(s), AAS. 
}%
\label{fig_serendip_schlichting_chang}
\end{figure}

\begin{figure}[ht]
\centering
\includegraphics[height=0.4\textwidth,trim=0 0 0 0]{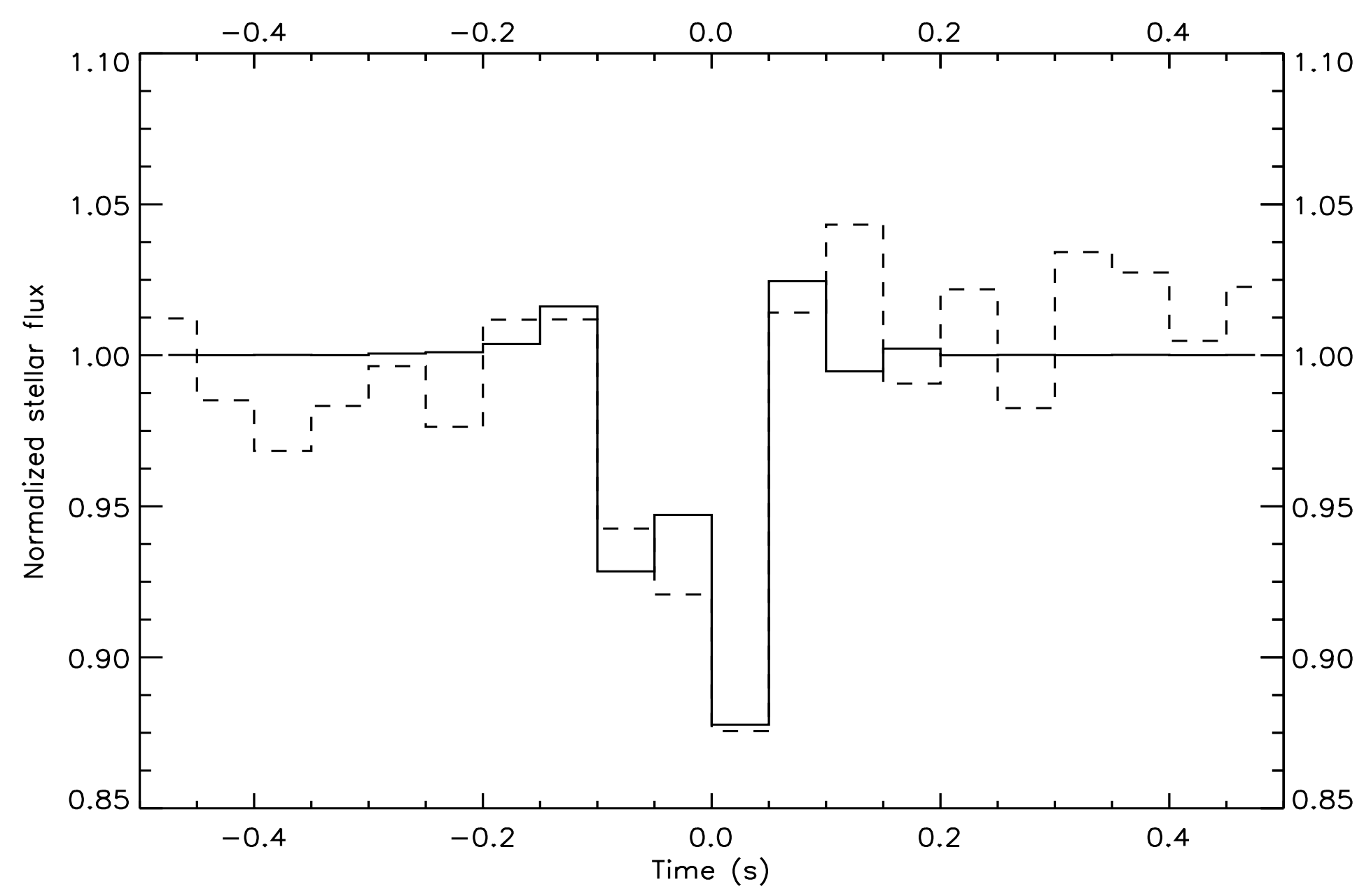}
\includegraphics[height=0.4\textwidth,trim=0 0 0 0]{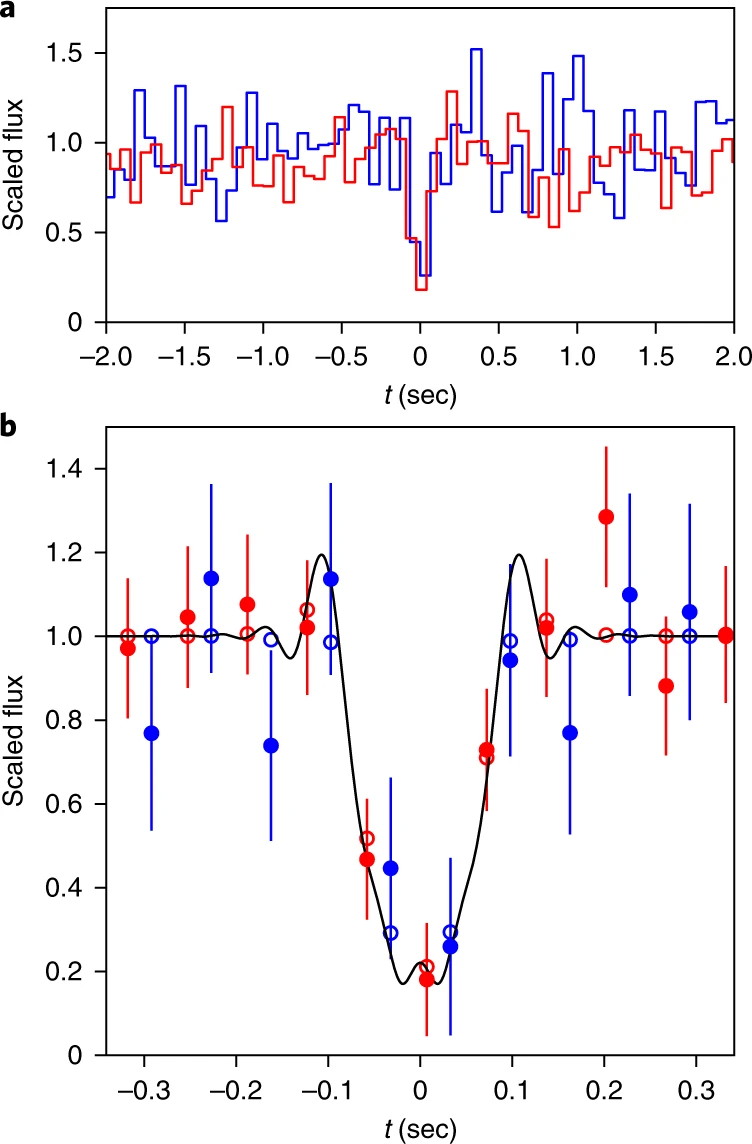}
\caption{%
Serendipitous occultation events detected from the ground, see Table~\ref{tab_serendip_search} for details.
Left panel:
A putative occultation event detected by the MIOSOTYS instrument, showing the data (dashed line) fitted by a model (solid line) consisting of a 380-m radius object at 41~au \citep{dore17}.
Right panel: 
An event recorded simultaneously at the two telescopes of the OASES project. The blue and red points are the data points obtained at the respective telescopes, while the solid line is a best-fitting model using an object of radius 1.3$^{+08}_{-01}$~km at 33$^{+17}_{-3}$~au \citep{arim19b}.
See text for details. Images reproduced with permission; copyright by the author(s).
}%
\label{fig_serendip_doressoundiram_arimatsu}
\end{figure}

The Fig.~\ref{fig_serendip_schlichting_chang} presents events obtained by the space-borne  HST/FGS \citep{schl09,schl12} and RXTE \citep{chan11} instruments, while Fig.~\ref{fig_serendip_doressoundiram_arimatsu} displays two other events obtained from the ground by the MIOSOTYS and OASES projects, respectively. We note that the RXTE results are unique, as they use X-ray data and are thus sensitive to 10-m-sized TNOs. This single detection gives an upper limit not only on the TNO population, but also on the 10-m-sized Main belt population \citep{chan13}. The RXTE data set also provides the first constraint without a priori hypotheses on the population of  300 to 900-m-sized comets in the Oort cloud \citep{chan16}.


\begin{figure}[ht]
\centering
\includegraphics[height=0.5\textwidth,trim=0 0 0 0]{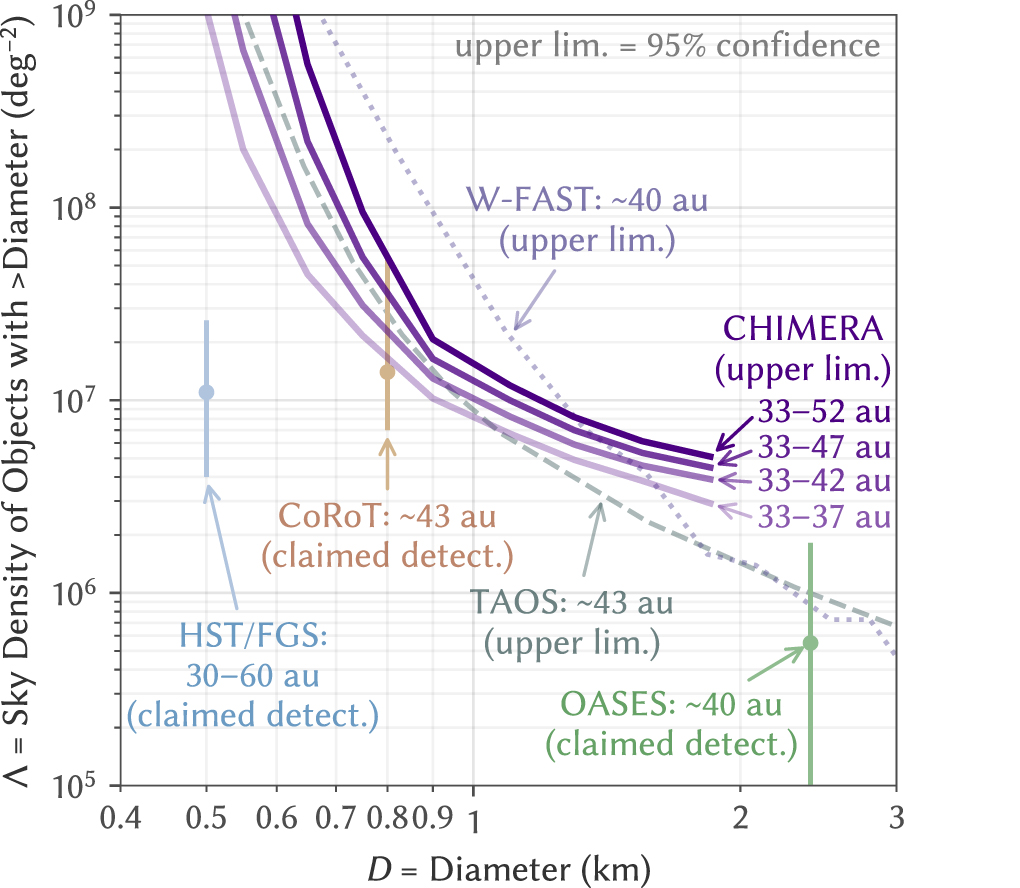}
\caption{%
Constraints from the CHIMERA survey for the ecliptic sky density of km-sized TNOs, compared with limits set by other previous searches, TAOS, W-FAST, HST/FGS, CoRoT, and OASES. Image reproduced with permission from \citet{zhan23}, copyright by the author(s).
}%
\label{fig_serendip_zhang}
\end{figure}

Besides the projects presented in Tables~\ref{tab_serendip_search}-\ref{tab_serendip_search_contd}, the serendipitous occultation technique is also proposed to detect extrasolar small bodies such as the interstellar comet 'Oumuamua. It consists in an all-sky monitoring program of $7 \times 10^{6}$ stars brighter than R=12.5, using 1-m telescopes at 10~Hz cadence. It is expected to discover about one interstellar object per year with such projects \citep{sira20}. 
The Tianyu project is dedicated to the exploration of the dynamic universe. It has two large-field-of-view 1-m telescopes equipped with high-speed CMOS cameras. It performs a high-precision photometric survey of about 100 million stars. The survey time is divided into three modes of second, day and week cadences, respectively. It is dedicated to various scientific cases. In particular, the second mode may improve the likelihood of capturing TNO occultations \citep{feng24}. 



Discriminating real events from false positives remains a delicate task, in particular because of the variability of the atmospheric noise. More reliable events are still needed to extract solid information about the trans-Neptunian population. This requires more data, and more importantly, light curves with better SNR combined with improved data-analysis tools in order to discriminate real occultations from instrumental and atmospheric noise.
Better SNR needs large telescopes and optimized target stars. An option is to mount cameras on otherwise unused areas in the focal planes of large, ground-based telescopes, as proposed by \cite{hitc24}. Data analysis requires massive comparisons between observations and synthetic profiles. Template banks of simulated occultations must in fact cover a large parameter space, resulting in time-consuming data processing. In particular the assumed geocentric distance of the objects is often limited to classical TNOs, while the photometric light curves probe the entire line of sight, from near-Earth objects to the Oort Cloud population. Recent progresses in big data science could help exploring these possibilities.

Meanwhile, the re-analysis of existing fast-photometry light curves with optimized tools could be a successful approach for exploring the ``hidden matter" of the solar system. An example of it is the Kepler data set, which may reveal the signatures of TNOs or Oort cloud objects \citep{gaud04}. Turning to the future, projects such as the Whipple space telescope could contribute to this search \citep{nihe07}, as well as future X-ray satellites \citep{chan11}.

Ultimately, the serendipitous occultation method constrains the density of km-sized objects beyond Neptune, as illustrated in Fig.~\ref{fig_serendip_zhang}.
These results can be compared with crater-counting on the surface of Pluto and Charon \citep{robb17}.
This yields contradictory results, as the serendipitous occultation counts taken at face value reveal about ten times the number of bodies inferred from cratering record.
This problem can be solved by developing models where the initial size of the bodies \it decreases \rm with increasing heliocentric distance, while the surface density of the bodies \it increases \rm beyond the 2:1 mean-motion resonance with Neptune \citep{shan23}. This overabundance of objects beyond 50~au, compared to previous models, seems to be supported by a recent survey conducted at the Subaru Telescope \citep{fras24}.

\bla

\section{Conclusions}
\label{sec_conclusion}

The last five years or so have seen an explosion of results returned by stellar occultations. Thanks to the Gaia mission, we can now predict events at the mas-level accuracy or better.
Thus, \bs{inaccurate predictions no longer represent a main source of failure when observing occultations, as was the case in the pre-Gaia era.}
In fact, the number of predicted events is now much larger than the capacity of the professional and amateur communities to observe them. Forthcoming large surveys, such as the Legacy Survey of Space and Time (LSST) at the Vera Rubin Observatory, will amplify this trend by providing more predictions for TNOs of various classes. 

In this context, we have to adopt a science-oriented approach, where objects are chosen for their specific interests. This can be done keeping in mind that the occultations technique does not suffer the usual limitation imposed by a finite telescope aperture since the measurement depends more on the brightness of the occulted star. Thus we can hope to learn nearly as much about an object at 1~au as we can about something at 100~au.

To give a few examples, we can now directly determine the three-dimensional shape, as well as the albedo and bulk density of many objects. This enables us to classify objects into different regimes, such as equilibrium vs. non-equilibrium states or homogeneous vs. differentiated internal structures. Occultations also allow us to detect topographic features and discover close binary systems while revealing and accurately measuring rings, providing their orbital elements, widths, and optical depths. Moreover, occultations can reveal and monitor seasonal changes of tenuous atmospheres down to the nanobar level. Taken together, all these results provide direct information on the formation and history of objects on a very large scale of size and heliocentric distance.

Moreover, stellar occultations are now observed in a context where space telescopes (e.g., CHEOPS, JWST, E-ELT) or ground-based 30-m class telescopes are or will be available. 
Meanwhile, the Lucy space mission will also return high-resolution images of some of the Trojans that are already characterized from Earth using occultations. This will serve to validate the method and then apply it to other Trojans or more remote bodies that will \it not \rm be the targets of space missions \bs{for} a long time.

The method of serendipitous occultations by km-sized TNOs 
has returned limited results so far. Most of the current searches only gave upper limits for the density of small TNOs in the sky plane. The few claimed events remain debatable, while the most convincing events are in tension with other surveys. A larger number of confirmed events must be obtained to better assess the TNO population in the kilometer range. As the capacity to process large volume of data is rapidly increasing, in parallel with improved automated data analyses, important results are expected with this method in the future.

In any instance, we have to keep an open mind and make ``random observations''. Even if no specific scientific goal is pursued for a given object, many surprises are in store. Diversity and unexpected behaviors are more the rule than the exception. For instance, before the discovery of Chariklo's ring in 2013, no one imagined that small objects may have rings. Considering that three (and possibly four) ring systems have already been detected in the outer solar system -- among a relatively modest sample--, many more are yet to be discovered. 
Similarly, the stability and seasonal evolution of Pluto's and Triton's atmospheres are far from being fully understood. In that context, it is important to continue these observations and push the occultation method to its limit (the nbar-level or below). In particular, large telescopes that provide high SNR may lead to the discovery of permanent or transient atmospheres that may be global or local.

Finally, we also have to keep in mind that occultations are and will be very complementary to what can be done with high-resolution imaging.
Imaging is efficient for covering the entire stable region for satellites, but it eventually fails to probe closer to the body, looking for tighter pairs. Extending the population and attributes of close binaries is an important goal 
that occultations are uniquely poised to provide. In doing so, they can answer questions such as, 
do equal mass binaries exhibit a minimum separation before becoming contact binaries?
Or is there more of a continuous distribution down to contact binaries?
And also, do contact binaries only exist up to a maximum size?
Or even, are all the Cold Classical Objects binaries \citep{fras17}? 
%
These questions can be addressed with either expensive spacecraft for a limited number of objects or more easily and extensively with occultation observations involving small ground-based telescopes.

\backmatter

%
%

\bmhead{Acknowledgments}

This study was financed in part by the Coordenação de Aperfeiçoamento de Pessoal de Nível Superior - Brasil (CAPES) - Finance Code 001 and  National Institute of Science and Technology of the e-Universe project (INCT do e-Universo, CNPq grant 465376/2014-2). FBR acknowledges CNPq grant 316604/2023-2. JLO acknowledges support by the Spanish projects PID2020-112789GB-I00 from AEI and Proyecto de Excelencia de la Junta de Andalucía PY20-01309. Financial support from the grant CEX2021-001131-S funded by MCIN/AEI/10.13039/501100011033 is also acknowledged.

\clearpage
\begin{appendices}

\section{Occultation deployment considerations}
\label{app_deployment}

Although a very efficient tool, stellar occultations have faced a major limitations in the past: 
the difficulty to predict accurately the shadow path.
This caused a waste of time, energy and money when organizing on-site campaigns 
(sometimes involving portable telescopes) that turned out to be negative.
It also restricted the access to large telescopes, for which the risk of failure was 
a potential negative point, given the high pressure put on these instruments. 
A useful parameterization of this risk was encapsulated in the dimensionless quantity $Q$ by \citet{mill79},  
\begin{equation}
    Q = 2 \sigma \Delta / D,
\label{eq_coeff_Q}
\end{equation}
where $\sigma$ is the astrometric uncertainty (in radians) of the object relative to the star 
at the time of occultation, 
$\Delta$ is the geocentric distance to the object, and 
$D$ is the diameter of the object, measured in the same units as $\Delta$.
We note that the probability of success varies as $1/Q$ and that
of the three parameters, only $\sigma$ is under our control. 

Let us consider a goal for an occultation to be the collection of observations 
from two stations separated by approximately the radius of the occulting body.  
If these two detections (or ``chords") are successful, reconstructing a spherically equivalent size is possible.  
From Eq.~\eqref{eq_coeff_Q}, this minimal goal is reached for 
a required number $N_{\rm T}$ of stations (or telescopes) given by
\begin{equation}
    N_{\rm T} = 3(Q+1),
    \label{eq_N_T}
\end{equation}
As $Q$ approaches zero (i.e. as predictions improve) the number of stations tend to three.  
Of these, most likely two would see the event and one would miss.
We note that the number of stations $N_{\rm T}$ grows linearly with $\Delta$, 
which illustrates the difficulty of detecting stellar occultations by remote TNOs. 

Of course, one can strive for more than just this minimal two-chord measurement. 
With a sufficient density of stations, an occultation can be used to retrieve not only a direct limb-profile 
of the object, but also limits on other material (dust, rings, moons) around the object.

Returning to the case of a minimal two-chord occultation goal, let us consider the following baseline: 
$\sigma=1$~arcsec, $\Delta=1$~au, and $D=1500$~km.  
Then $Q=1$, resulting in a total of $N_{\rm T}=6$ stations with a spacing of 750 km between stations.
This set of stations would cover the predicted track over $\pm1.5\sigma$, yielding a 71\% chance of getting two chords, assuming no weather losses or equipment failures. 
This strategy has a good chance of meeting the goal but implies that 4 out of the 6 stations will not see an event.  

For main belt asteroids with typical distance $\Delta=3$~au and diameter $D=100$~km, and 
with a star position accurate to the 10 milli-arcsec (mas) level easily obtained with Gaia,
we get $Q=0.4$ and $N_{\rm T}=4$. 
For objects on the inner edge of the Kuiper Belt region ($\Delta=30$~au) 
we need astrometry that is good to 1 mas for the same sized object 
to get a similarly deterministic prediction.
With positions good to only 10~mas, a deployment of 40 stations is required for the same 71\% chance of success for a two-chord TNO result.
Extending this even further, occultations at $\Delta$=300~au require 0.1 mas positions and $\Delta=3000$~au 
would require 10 micro-arcsecond ($\mu$as) accuracy on positions, again for a 100 km body. 

The consequence of having the Gaia catalog, with ultimate accuracy of 100~$\mu$as,
is that typical main belt asteroid occultations are now easy and do not require any special effort. 
Even just twenty years ago, this was not true.
With the improvement, we can now seriously consider investigations of objects that were previously nearly impossible. 

One example is that of near-Earth asteroids. With effort, successful observations of Apophis, Phaethon, Didymos, and even Dimorphos have been achieved. 
Most of these objects are km-scale bodies or even smaller. Another example is the study of outer solar system objects -- the subject of this review. With TNOs, nearly all successful occultations prior to Gaia were of the very largest objects, such as Pluto and Triton. The following discussion lays out some guiding concepts for the study of the smaller members of the outer solar system going down to sizes as small as 10~km.

\begin{figure}[ht]
\centering
\includegraphics[width=1.0\textwidth,trim=0 150 0 150]{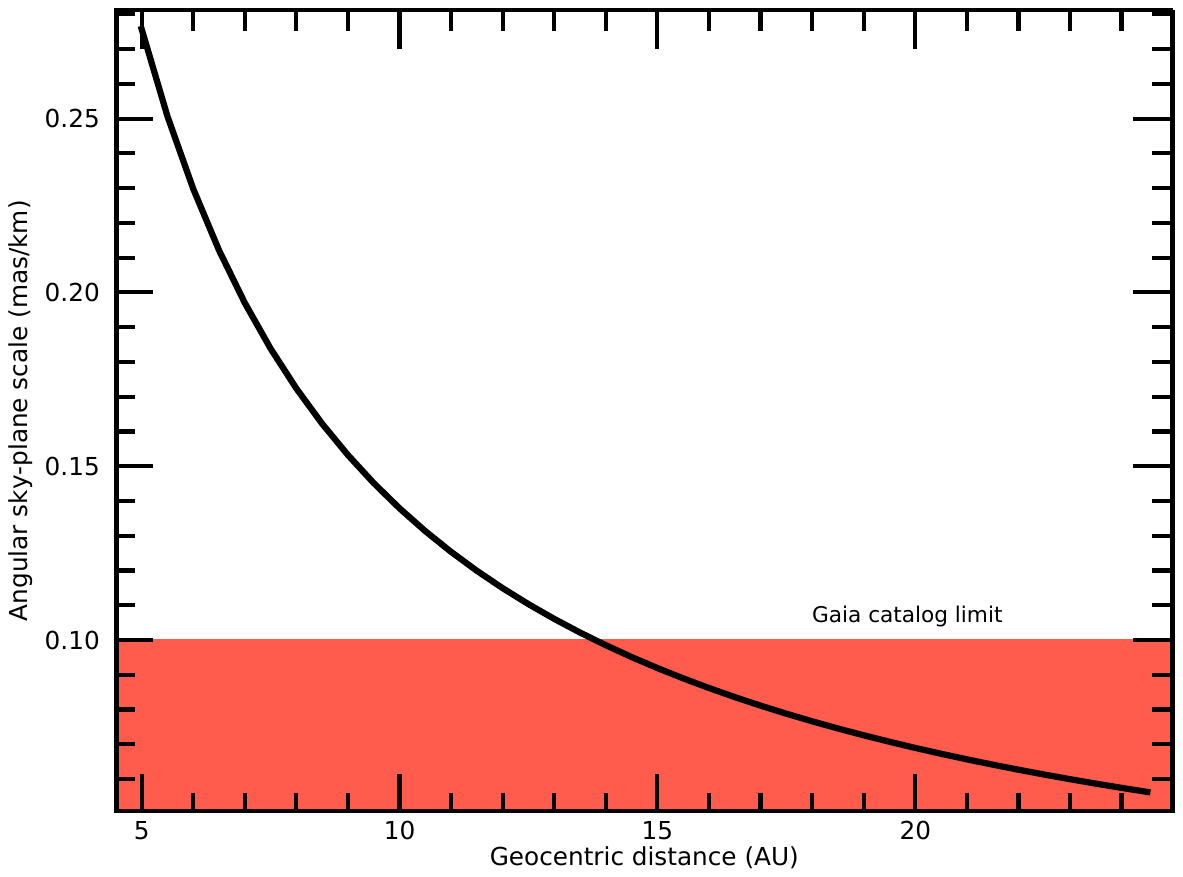}
\caption{%
Relationship of angular to physical scale with changing distance.  The black curve shows the
relationship between angular scale (mas) to distance (km) in the sky plane of an object over a
range of distances relative to the observer. A 1~km offset is equivalent to 100~$\mu$as, 
comparable to the typical precision of the Gaia catalog.  For a constant physical scale accuracy of 1 km, the measurement will be limited by the observation inward of 13.8~au and will be limited by the Gaia catalog at greater distances.
}%
\label{fig_sigplot}
\end{figure}

We see in Fig.~\ref{fig_sigplot} the relationship between angular and physical scales 
as a function of the distance to the observer. 
The ratio of the scales is given as mas/km to facilitate thinking about the constraints imposed 
by needing to place an observation point on a scale that is tied to the size of the object (in km).
The scale is also easy to read if one thinks of a fiducial case of 1~km.
This fiducial case is chosen because in practice it is now feasible to measure the position of an object 
relative to a star using multi-chord occultation data to a level of about 1~km.
Getting more accurate positions for objects of this size requires a very detailed knowledge of the three dimensional shape of the object as well as its internal mass distribution.

For this discussion, we adopt here a canonical uncertainty from a well-measured occultation 
to give a position good to 1~km. The red region shown in Fig.~\ref{fig_sigplot} shows an important transition zone,
as it represents the limit of the Gaia catalog precision, typically 100~$\mu$as.
Toward the smaller end of this range, occultation predictions are limited by the knowledge of the object.
Starting around 13.8~au, the catalog limit dominate the uncertainties.

The important point here is that a successful occultations in the outer solar system, 
will produce astrometry as good as the Gaia catalog limit,
\it regardless of the apparent brightness of the object. \rm
In contract, the difficulty of such measurements in the outer solar system scales as 
$\Delta^{-4}$ from direct imaging, while the occultation technique scales as $\Delta^{-1}$.  
An important consequence of this difference means small telescopes are sufficient to get Gaia-quality astrometry 
rather than the most expensive and access-limited telescopes such as HST or JWST.

Even with the Gaia catalog at hand, 
we do not yet have a catalog of orbits for all objects of interest that are fully based on that catalog.  
The resulting ephemeris uncertainties then dominate the occultation predictions.  
Each object has its own observational history that must be considered for specific campaigns. 
Here, we adopt a canonical uncertainty of 10~mas for the quality of a current orbit 
that is as yet unconstrained by any occultation data.  
This is a level that represents well-measured objects with a particularly good orbit.  
New occultation-derived astrometry will then be added to the observational record and 
will improve the prediction. It will be many years or even decades 
before traditional astrometry based on the Gaia catalog will begin to dominate. 

\begin{figure}[ht]
\centering
\includegraphics[width=1.0\textwidth,trim=0 150 0 150]{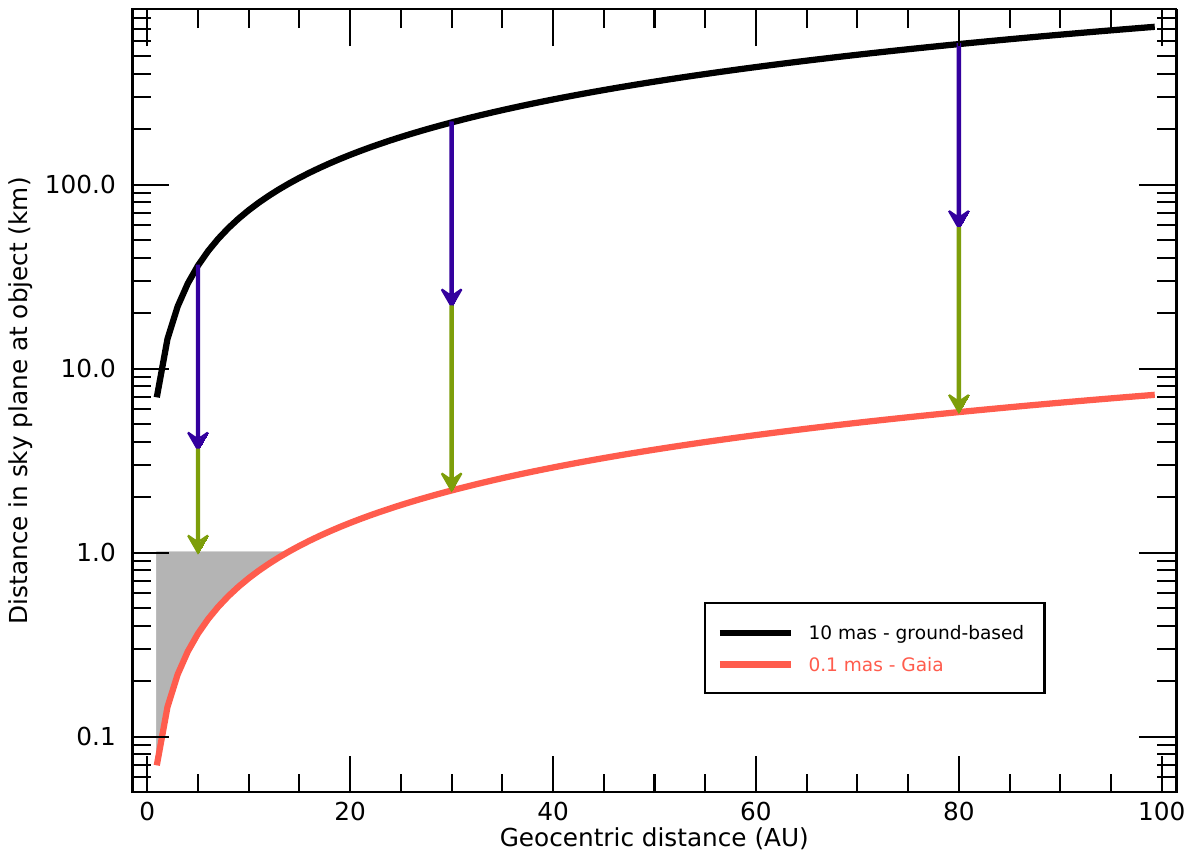}
\caption{%
Implication of astrometric uncertainty for a range of target distances.  The upper curve
represents the fixed positional uncertainty of a typical well-observed object constrained only
traditional astrometry.  The bottom curve shows the notional limit of how good the uncertainty will
be once it reaches the precision limit of the Gaia catalog.  The shaded gray region shows the consequence of
limiting an occultation astrometry measurement to no better than 1~km (see text).  Three test cases are
shown for a notional sequence of two occultations.  The first event is at the top where the uncertainty
will be due entirely to traditional astrometry.  After the first event, the uncertainty drops dramatically
and shown here as a 10x drop.  After the second event, the uncertainty future events will drop to the
Gaia catalog limit.  The case for a Jupiter Trojan will be limited by the 1 km uncertainty limit.
In the Kuiper Belt, the uncertainty floor is determined by the Gaia catalog.
}%
\label{fig_sigkmplot}
\end{figure}

Using the information from Fig.~\ref{fig_sigplot}, Fig.~\ref{fig_sigkmplot} shows the size of this 10~mas uncertainty mapped onto a physical scale.  This upper curve shows the uncertainty expected for an occultation prediction for an object not yet seen to occult a star.

The bottom curve in Fig.~\ref{fig_sigkmplot} represents the limit 
when the position of the object can be predicted with an accuracy comparable to the Gaia catalog.  
We cannot reasonably expect to ever get lower than the Gaia catalog limit anytime soon, 
but this limit represents anyway a 100-fold decrease in the uncertainty 
compared to classical astrometry.
%
The shaded region at small geocentric distances depicts the limit imposed by the notional 1-km noise floor 
from occultation measurements for our chosen size range for the discussion.  
In this region, a successful occultation can still improve an orbit but the improvement will not be quite as good, 
as seen in the three specific examples of Fig.~\ref{fig_sigkmplot}.  

The downward pointing arrows indicate qualitatively what happens as new occultation astrometry are obtained.  
After the first successful occultation, future uncertainties move downward by a factor of $\sim$10.  
After two successful events, future prediction are essentially at the lower limit.  
The first case shown at left is for Jupiter Trojan asteroids at 5~au\null.  
Prior to the first occultation the uncertainty is $\sim$35~km, comparable to the small end of the size range considered.  
After one occultation, the uncertainty drops to 3-4~km, already smaller than the objects of interest.  
After two occultations, further predictions will become limited by what we do not know about the object 
rather than uncertainties in the ephemeris.  

The second case is at 30~au, chosen to match the approximate position of Pluto.  
The initial uncertainty is around 200~km (corresponding to 10~mas), 
dropping to 20~km, and then reaches the limit at 2~km (0.1~mas).  
This situation makes it harder to get the smallest objects of interest ($D$ = 30~km).  
In the case of Pluto, we can now specifically target the central flash zone of the occultation which covers a width 
less than 100 km.  

The last case is for a much more distant TNO where the initial uncertainty is close to 700~km.  
In this case, only the largest objects are straightforward to obtain the first occultation.   
Either the first occultation will be a lucky success or some particular help will be required 
such as obtaining special observations (e.g. by HST) that bypass the need for the first occultation and
get us to the next level of orbit accuracy reaching $\sim$100~km.  
Even for such distant objects, reaching the Gaia limit for predictions enables 
deterministic multi-chord campaigns on very small objects. 
This situation is true for even more distant objects but at present our sample of known objects 
at distances of 100~au or greater does not extend to the smaller sized objects.

\begin{figure}[ht]
\centering
\includegraphics[width=1.0\textwidth,trim=0 150 0 150]{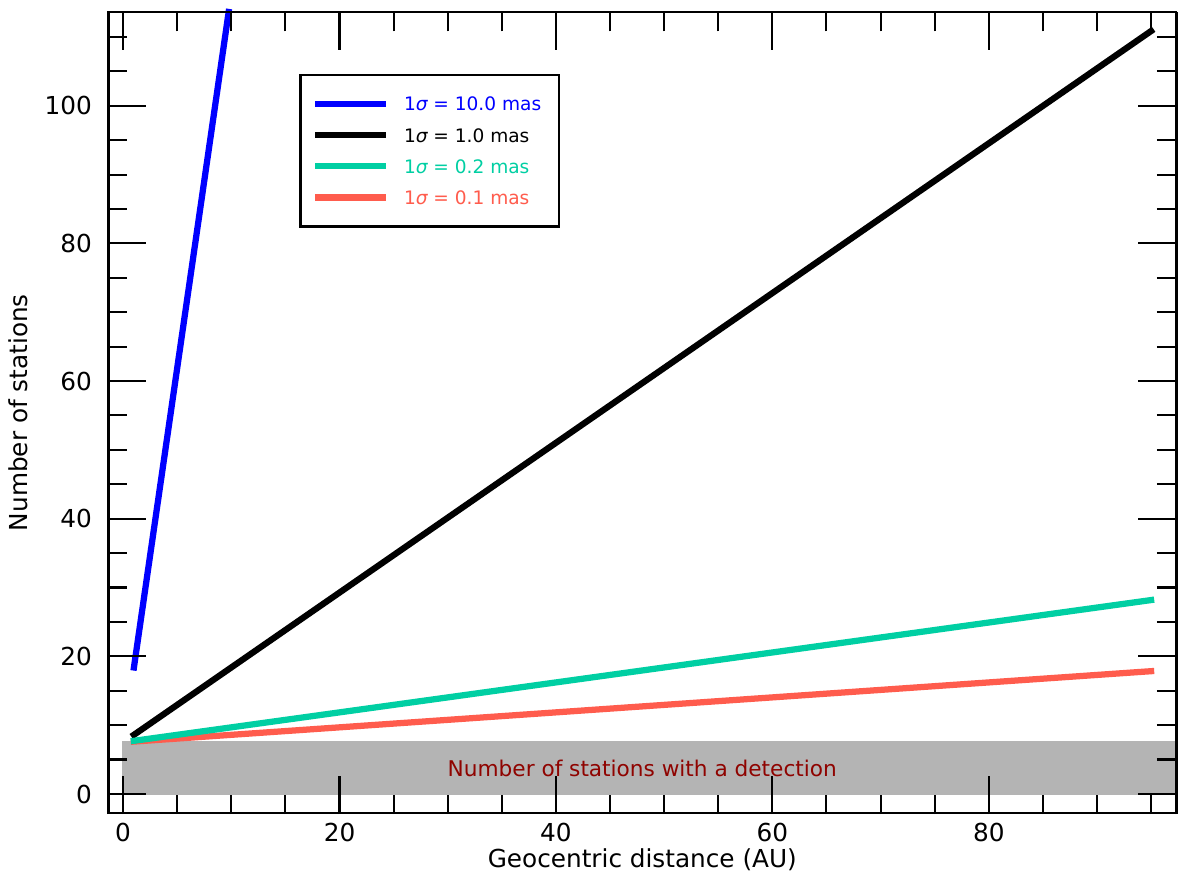}
\caption{%
Number of deployed stations required for different prediction uncertainties.
This figure illustrates the case of a 30~km diameter object and a 4~km spacing between
stations and assuming a coverage of $\pm3\sigma$ (i.e. with a formal chance of success of 98\%).  
Such a deployment would give seven chords across the body as indicated by the shaded zone.  
The cases shown are for typical ground-base prediction quality (10 mas), a notional HST-supported prediction (1 mas),
Gaia catalog limited prediction (0.1 mas), and twice the Gaia limit.  Reaching the Gaia limit for
predictions enables detailed observations of small and distant TNOs for a reasonable deployment size.
}%
\label{fig_numplot}
\end{figure}

A deterministic multi-chord campaign is one where a high-likelihood of success is assured 
with a spacing that is at most four times smaller than the diameter of the object,
ensuring both size and shape measurements. 
Figure~\ref{fig_numplot} illustrates the effort required to obtain such a result for an object with a 30~km diameter.  
Here, we choose a goal of 4~km spacing so that to return seven chords on the body. 
Other cases are easily treated by scaling from this example.  
The lines on the plot show the number of stations vs. geocentric distance required to meet the goals 
for a range of prediction uncertainty bounded by the same range shown in Fig.~\ref{fig_sigkmplot}. 
A single chord result rarely provides an astrometric constraint as good as 1~km, 
but it will still provide a significant reduction in future predictions.
For a 30-km object, even the best case first occultation still requires nearly 20 stations.  
In practice, working on this size range of object is most easily accommodated by relaxing 
the constraint on the number of desired chords.
Dropping the requirement by a factor of 7 to get just a single chord is almost as effective as 
what the first occultation brings for its astrometric contribution.

Reaching an uncertainty of 1~mas brings most of the Kuiper Belt into range of reasonable sized efforts.  
At this stage it makes sense to still avoid a campaign as demanding as a 7-chord plan 
but one can now reasonably attempt getting a few chords.  
After two occultations, the uncertainties collapse down into the range of the lowest two lines 
and detailed shape measurements are now practical.  
Note that in this description the term ``difficult" equates to noting the fraction of deployed stations 
that will not obtain positive detections of the target.  
The shaded bar at the bottom shows the expected number of positive chords and 
the gap from there to the appropriate line indicates the number of negative observations.

Applied too strictly, this approach risks missing unexpected discoveries of satellites and rings, among others.
The case of a close binary is even more important since we know that there a significant number of such systems in the TNO population
\citep{fras17}.
Consider a notional case of a CCKBO that is an equal mass binary 
with two components each of a 100~km diameter and orbiting each other with a 700~km distance.  
If the objects are oriented perpendicular to the shadow motion direction, there is a 600~km gap between the two bodies.  
A plan to tightly target a 140~km single body easily provides data that passes in between the components, missing both.  
The first stage high uncertainty effort serves to place direct constraints on companions 
provided much more value to the large number of inevitable negative observations.

\section{The physics of ring occultations}
\label{app_physics_ring_occ}

As the stellar flux is dimmed by a semi-transparent ring, the transmitted flux $I$ is related to the incident stellar flux $I_0$ by the apparent opacity $p'$ through
\begin{equation}
p'=1-\frac{I}{I_0}.
\label{eq_opacity}
\end{equation}
Thus, $p'= 0$ corresponds to a transparent ring while $p'= 1$ is for an opaque one. The apparent optical depth 
$\tau'$ is then 
\begin{equation}
 \tau'=-\ln{(1-p')}. 
 \label{eq_app_opticaldepth}
\end{equation}
These quantities are called apparent as they are measured in the plane of the sky. 

%
%

The apparent values described above can be calculated for a normal viewing so as to be independent of the particular event being analyzed. An individual ring particle of radius $r$ diffracts light of wavelength $\lambda$ in a cone with angular diameter $\phi_{\rm A}=\lambda~D/r$, the Airy scale. Thus, a meter-sized particle at distances greater than $10^9$~km, observed in the optical, will spread the light in a cone of hundreds of kilometers.
This is much larger than the widths of the rings considered here, leading to the extinction paradox \citep{cuzz85}, which makes the ring appear from Earth with twice the optical depth than it actually has. This paradox can be explained by the fact that the ring acts as a phase screen (instead of an intensity screen) when particles are much smaller than the Fresnel scale $\lambda_{\rm F}$ \citep{roqu87}. 

This effect must be accounted for to derive the actual ring opacity and optical depth, distinguishing two limiting cases, monolayer and polylayer rings, see details in \cite{bera17} and \cite{morg23}. 
For a monolayer ring, we obtain the normal opacity from
\begin{equation}
    p_{\rm N}= \mid \sin{B} \mid\cdot(1-\sqrt{1-p'}), 
    \label{eq_realopacity}
\end{equation}
while for a polylayer ring, the normal optical depth is
\begin{equation}
    \tau_{\rm N} = \mid \sin{B} \mid \cdot\frac{\tau'}{2}.
    \label{eq_realopdepth}
\end{equation}

The radial ring width ($W_{\rm r}$) of the ring can be combined with $p_{\rm N}$ to provide the \it equivalent width \rm
\begin{equation}
E_{\rm p} = p_{\rm N} W_{\rm r}.
\label{eq_equiv_width}
\end{equation}

For a monolayer ring, $E_{\rm p}$ is a measure of the total amount of material contained in a radial scan of the ring.
Similarly, $\tau_{\rm N}$ can be used to define the \it equivalent depth \rm
\begin{equation}
A_{\tau} = \tau_{\rm N} W_{\rm r},
\label{eq_equiv_depth}
\end{equation}
which is a measure of the total amount of material contained in a radial scan of a polylayer ring. 

If a ring occultation is resolved, e.g. with more than three data points in the profile, then 
$E_{\rm p}$ and $A_{\tau}$ are obtained from
\begin{equation}
    E_{\rm p}=\int_{W_{\rm r}}(v_{\rm r} p_{\rm N}) dt
    \label{eq_intequivwidth}
\end{equation}
and
\begin{equation}
    A_{\tau}=\int_{W_{\rm r}}(v_{\rm r} \tau_{\rm N})dt
    \label{eq_intequidepth}
\end{equation}
where $v_{\rm r}$ is the velocity of the star perpendicular to the ring, as observed in the ring plane.

Since the discovery of rings around Centaurs and TNOs, it makes sense to sound the close environment of these objects and put limits to the presence of dense materials on every stellar occultation \citep{sant21,pere24}. Detection limits are assessed using the apparent equivalent width
\begin{equation}
    E'_p(i)=[1-\phi(i)]\Delta r(i),
    \label{eq_apequivwidth}
\end{equation}
where $\phi(i)$ is the normalized stellar flux of a given data point and $\Delta(i)$ is the radial interval covered by it. Considering that the light curve dispersion follows a Gaussian distribution, the standard deviation outside the main occultation can be taken as the limiting opacity $p'$ (Eq. \ref{eq_opacity}). A suspicious flux drop can be considered as a possible detection of a secondary event if it is above the 3$\sigma$ level; see Fig. 11 of \cite{brag23} as an example. 





\end{appendices}


\phantomsection
\addcontentsline{toc}{section}{References}
\bibliography{bibliography}


\end{document}